\documentclass[a4paper,11pt,notoc]{article}
\usepackage{jheppub} 

\usepackage{verbatim}
\usepackage{graphicx}
\usepackage{slashed}
\usepackage{soul} 
\usepackage[usenames,dvipsnames,svgnames,table]{xcolor} 

\newcommand{\MET}{\slashed{E}_T}

\title{Towards the next generation \\ of simplified Dark Matter models}

\vspace{0.5em}

\author[1]{Andreas Albert,}
\affiliation[1]{III. Physikalisches Institut A, RWTH Aachen, Physikzentrum, 52056 Aachen, Germany}

\author[2]{Martin Bauer,}
\affiliation[2]{Institut f\"ur Theoretische Physik, Universit\"at Heidelberg, Philosophenweg 16, 69120 Heidelberg, Germany}

\author[3]{Jim Brooke,}
\affiliation[3]{H. H. Wills Physics Laboratory, University of Bristol, Bristol, BS8 1TL, U.K.}

\author[4]{Oliver Buchmueller,}
\affiliation[4]{High Energy Physics Group, Blackett Laboratory, Imperial College, Prince Consort Road, London, SW7 2AZ, U.K.}

\author[5]{David G. Cerde\~no,}
\affiliation[5]{Institute of Particle Physics Phenomenology, Durham University, U.K.}

\author[3]{Matthew~Citron,}

\author[3]{Gavin Davies,}

\author[6]{Annapaola~de~Cosa,}
\affiliation[6]{Physik-Institut, Universit\"at  Z\"urich,  Winterthurerstrasse 190, CH-8057 Z\"urich, Switzerland}

\author[7,8]{Albert De~Roeck,}
\affiliation[7]{Experimental Physics Department, CERN, CH 1211 Geneva 23, Switzerland}
\affiliation[8]{Antwerp University, BÐ2610 Wilrijk, Belgium. }

\author[9]{Andrea~De~Simone,}
\affiliation[9]{SISSA and INFN Sezione di Trieste, via Bonomea 265, I-34136 Trieste, Italy}

\author[7]{Tristan~Du~Pree,}

\author[4]{Henning Flaecher,}

\author[10]{Malcolm~Fairbairn,}
\affiliation[10]{Theoretical Particle Physics and Cosmology Group, Department of Physics, \\
King's College London, London WC2R 2LS, U.K. }

\author[10,11]{John Ellis,}
\affiliation[11]{Theoretical  Physics Department,  CERN, CH-1211 Geneva 23, Switzerland}

\author[12]{Alexander~Grohsjean,} 
\affiliation[12]{Deutsches Elektronen-Synchrotron, Notkestr. 85, 22607 Hamburg, Germany}

\author[13]{Kristian~Hahn,}
\affiliation[13]{Department of Physics and Astronomy, Northwestern University, Evanston, Illinois 60208, USA}

\author[10,14]{Ulrich~Haisch,}
\affiliation[14]{Rudolf Peierls Centre for Theoretical Physics, University of Oxford, Oxford, OX1 3PN, U.K.}

\author[7]{Philip~C.~Harris,} 

\author[5]{Valentin~V.~Khoze,} 

\author[15]{Greg~Landsberg,} 
\affiliation[15]{Physics Department, Brown University, Providence, Rhode Island 02912, USA}

\author[16]{Christopher McCabe,}
\affiliation[16]{GRAPPA Centre of Excellence, University of Amsterdam, Amsterdam, The Netherlands}

\author[3]{Bjoern Penning,}

\author[17]{Veronica Sanz,}
\affiliation[17]{Department of Physics and Astronomy, University of Sussex, Brighton BN1 9QH, U.K.}

\author[12]{Christian Schwanenberger,} 

\author[18]{Pat Scott,}
\affiliation[18]{Astrophysics Group, Imperial College, Blackett Laboratory, Prince Consort Road, London, SW7~2AZ, U.K.}

\author[7]{and Nicholas Wardle}

\abstract{~~\\
This White Paper is an input to the 
ongoing discussion about the extension and refinement of simplified Dark Matter (DM) models. It is not intended as a comprehensive review of the discussed subjects, but instead summarizes ideas and concepts arising from a brainstorming workshop that can be useful when defining the next generation of simplified DM models (SDMM). In this spirit, based on two concrete examples, we show how existing SDMM can be extended to provide a more accurate and comprehensive framework to interpret and characterise collider searches. In the first example we extend the canonical 
SDMM with a scalar mediator to include mixing with the Higgs boson. We show that this approach not only provides a better description of the underlying kinematic properties that a complete model would possess, but also offers the option of using this more realistic class of scalar mixing models to compare and combine consistently searches based on different experimental signatures. The second example outlines how a new physics signal observed in a visible channel can be connected to DM by extending a simplified model including effective couplings. In the next part of the White Paper we outline other interesting options for SDMM that could be studied in more detail in the future. Finally, we review important aspects of supersymmetric models for DM and use them to propose how to develop more complete SDMMs.
~~\\
{\it This White Paper is a summary of the brainstorming meeting ``Next generation of simplified Dark Matter models" 
that took place at Imperial College, London on May 6, 2016, and corresponding follow-up studies on selected subjects.} }

\begin{document}
\maketitle
\flushbottom

\section{Introduction} 

This White Paper summarises discussions during the brainstorming meeting ``Next generation of simplified Dark Matter models"
held at the Imperial College, London on May 6, 2016 \cite{Agenda} and expands on a few selected topics that were considered to be the most important for the near future. 

This informal brainstorming meeting followed thematically the one hosted at Imperial in May 2014, which focused on the interplay and characterization of Dark Matter (DM) searches at colliders and in direct detection experiments, summarised in a White Paper~\cite{Malik:2014ggr}, which in part builds upon work documented in~\cite{Buchmueller:2014yoa}.

Since then several important developments in the characterisation of DM searches at colliders have taken place, most notably the activities of the LHC DM forum (LHC~DMF)~\cite{Abercrombie:2015wmb} and the newly-founded LHC DM working group~\cite{Boveia:2016mrp}.

Central to this effort are simplified DM models (SDMM), which have replaced interpretations using a universal set of operators in an effective field theory (EFT)
\cite{Beltran:2010ww,Goodman:2010yf,Bai:2010hh,Goodman:2010ku,Rajaraman:2011wf,Fox:2011pm} as the main vehicle to characterise DM searches at colliders. However, as discussed in~\cite{Abercrombie:2015wmb}, EFT interpretations can still provide useful information and complement the SDMM approach for collider searches. Today, SDMM are also used for comparisons with other searches, such as those conducted by direct detection and indirect detection experiments (see~\cite{Boveia:2016mrp}).

The majority of these SDMMs  are derived from simple Lagrangians that are governed by four basic parameters: a mediator mass ($m_{\rm med}$), the DM candidate mass~($m_\chi$), the coupling of the mediator to Standard Model (SM) particles (usually quarks or gluons,~$g_{\rm SM}$), 
and the coupling of the mediator to DM particles ($g_{\rm DM}$).
While these simplistic models have been very useful to map out the general characteristics of DM searches at colliders, they are often too simple to capture fully the detailed physics of all relevant searches.

Therefore, a well-defined extension of these SDMMs is required in order to allow for a more refined characterisation and comparison of all relevant DM searches. This should also include resonance searches in the dijet, dilepton, diphoton and other channels with only~SM particles in the final state, which are not directly looking for the DM particles but can nevertheless be very powerful in constraining the mediator mass and couplings.
Furthermore, this next generation of SDMMs should ideally also address some of the theoretical shortcomings inherent to the simplistic first-generation SDMMs.

The scope of the brainstorming meeting was to discuss options for defining the next generation of  SDMMs and, if deemed relevant/possible, to contribute to the development of consistent, state-of-the-art SDMM extensions.

In Section~\ref{scalar} of this White Paper we discuss in detail a simplified scalar singlet mediator model, which includes mixing between the 
SM Higgs boson and another scalar. In contrast to the simplified scalar model recommended in~\cite{Abercrombie:2015wmb} this class of mixing models allows for a more consistent interpretation of missing transverse energy searches, such as monojet, mono-$V$, and VBF-tagged analyses that are sensitive to different production modes --- gluon fusion, associated, and vector boson fusion (VBF) production, respectively.

In Section~\ref{750} we use the example of the observed $750 \, {\rm GeV}$ excess in high-mass diphoton searches at ATLAS and CMS
with the 2015 data to outline how a hypothetical signal for the production of a new mediator can be connected to DM using simplified models. 
While this excess was not confirmed by the new data collected by both experiments up to  mid 2016,    
this exercise is an example of a  case study on how to correlate searches with different experimental signatures to characterise the properties of a new particle discovery in the context of DM studies.   

Following these two detailed examples comparing and combining different experimental searches using  SDMMs, in Section~\ref{otherSMS} we outline qualitatively other interesting options for simplified models that could be studied in more detail in the future, while in Section~\ref{SUSY} 
we review some aspects of supersymmetric (SUSY) models that are important for DM physics.

We summarise the White Paper and make recommendations for future work on the extension and refinement of SDMMs in Section~\ref{summary}.


\newcommand{\hOne}{\ensuremath{h_{1}}}
\newcommand{\hTwo}{\ensuremath{h_{2}}}
\newcommand{\mhOne}{\ensuremath{m_{h_{1}}}}
\newcommand{\mhTwo}{\ensuremath{m_{h_{2}}}}
\newcommand{\mchichi}{\ensuremath{m_{\chi\bar{\chi}}}}
\newcommand{\mchi}{\ensuremath{m_{\chi}}}
\newcommand{\chichi}{\ensuremath{\chi\bar{\chi}}}
\newcommand{\XsecHTwo}{\ensuremath{\sigma_{pp\rightarrow\ttb h_{2}}{\rm BR}(h_{2} \rightarrow X)}}
\newcommand{\XsecPhi}{\ensuremath{\sigma_{pp\rightarrow\ttb\phi}{\rm BR}(\phi \rightarrow X)}}
\newcommand{\XsecTThOnehTwoExcl}{\ensuremath{\sigma_{pp\rightarrow\ttb h_{1},\ttb h_{2}}{\rm BR}(h_{1},h_{2}\rightarrow\chi\bar{\chi})}}
\newcommand{\XsecPhiExcl}{\ensuremath{\sigma_{pp\rightarrow\ttb\phi}{\rm BR}(\phi\rightarrow\chi\bar{\chi})}}
\newcommand{\ttb}{\ensuremath{t\bar{t}}}
\newcommand{\XsecHttIncl}{\ensuremath{\sigma_{t\bar{t}H}{\rm BR}(H \rightarrow X)}}
\newcommand{\XsecSinglet}{\ensuremath{\sigma{\rm BR}_{singlet}}}
\newcommand{\XsecBR}{\ensuremath{\sigma{\rm BR}}}
\newcommand{\whOne}{\ensuremath{\Gamma^{tot}_{h_{1}}}}
\newcommand{\whTwo}{\ensuremath{\Gamma^{tot}_{h_{2}}}}
\newcommand{\pt}{\ensuremath{p_{T}}}
\newcommand{\ptchichi}{\ensuremath{p_{T}^{\chi\bar{\chi}}}}
\newcommand{\ttDM}{\ensuremath{t\bar{t}+\chichi}}
\newcommand{\yDM}{\ensuremath{y_{\rm DM}}}

\section{Scalar singlet model with mixing}
\label{scalar}

In this Section we discuss the simplest extension of the scalar-mediated DM model recommended by the LHC DMF~\cite{Abercrombie:2015wmb} that includes mixing with the SM Higgs boson. Extensions with a more complicated scalar sector have been discussed, for example, in \cite{Izaguirre:2014vva,Ipek:2014gua,Berlin:2015wwa,Baek:2015lna,Choudhury:2015lha,Duerr:2016tmh}, some of them are aiming to address the Fermi-LAT galactic center $\gamma$ ray excess~\cite{TheFermi-LAT:2015kwa}.

Besides the SM particles and interactions, the model considered here contains a scalar mediator $s$ and a~DM particle~$\chi$, which for concreteness is taken to be a Dirac fermion. The additional scalar interactions relevant for the further discussion are~\cite{Kim:2008pp,Kim:2009ke,Baek:2011aa,LopezHonorez:2012kv,Baek:2012uj,Fairbairn:2013uta,Carpenter:2013xra,Khoze:2015sra,Abdallah:2015ter}

\begin{equation} \label{eq:Linteractions}
{\cal L} \supset -y_{\rm DM} \hspace{0.25mm} s \hspace{0.25mm} \bar \chi \chi  - \mu \hspace{0.25mm} s \hspace{0.25mm} |H|^2 \,,
\end{equation}
where $y_{\rm DM}$ is a dark-sector Yukawa coupling and $H$ denotes the usual SM Higgs doublet. 

As a result of the portal coupling $\mu$, the Higgs field $h$ and the real scalar field $s$ mix, giving rise to mass eigenstates $h_1$ and $h_2$:
\begin{equation} \label{eq:mixing}
\begin{pmatrix} h_1 \\[2mm] h_2 \end{pmatrix} = \begin{pmatrix} \cos \theta & \hspace{1mm}   \sin \theta \\[2mm] -\sin \theta & \hspace{1mm} \cos \theta \end{pmatrix} \begin{pmatrix} h \\[2mm] s \end{pmatrix},
\end{equation}
where $\theta$ is the mixing angle. In terms of these mass eigenstates the trilinear couplings of the scalars to DM and to the massive SM gauge bosons and fermions take the following form 
\begin{equation}
\begin{split} \label{eq:smh1h2}
{\cal L} \supset &- y_{\rm DM}  \hspace{0.5mm} \big ( \sin \theta \hspace{1mm} h_1  + \cos \theta  \hspace{1mm} h_2  \big )  \hspace{0.5mm}  \bar{\chi}\chi \\[2mm]  & + \big(  \cos \theta \hspace{1mm} h_1 -  \sin \theta  \hspace{1mm} h_2  \big)   \left(\frac{2M_W^2}{v} \, W^{+}_\mu W^{-\,\mu} +  \frac{M_Z^2}{v} \, Z_\mu Z^{\mu}  - \sum_f \frac{m_f}{v}\, \bar{f}f\right) \,,
\end{split}
\end{equation}
where $v \simeq 246 \, {\rm GeV}$ is the Higgs field vacuum expectation value, $M_W$ and  $M_Z$ are the $W$ and $Z$ boson masses, respectively, and $m_f$ denotes the masses of the SM fermions. Since the mixing angle~$\theta$ is defined such that for $\theta \to 0$ the DM sector is decoupled from the~SM, the state~$h_1$  plays the role of the observed Higgs boson with $m_{h_1} \simeq 125 \, {\rm GeV}$, while the mass of the state $h_2$, along with $y_{\rm DM}$ and $\theta$, are free parameters of the model.\footnote{In (\ref{eq:smh1h2}) the trilinear scalar couplings and all quartic couplings have not been included. These couplings are all simple functions of $\sin \theta$ and $\cos \theta$ and uniquely fixed in the model (\ref{eq:Linteractions}). Apart from the $h_1 h_2^2$ and~$h_1^2 h_2$ vertices, we ignore them here because they do not play a role in the phenomenological applications discussed in this Section.} Note that, as far as the couplings between $h_2$  and fermions are concerned, the interactions~(\ref{eq:smh1h2}) resemble those of the scalar-mediated DM model recommended by the LHC DMF~\cite{Abercrombie:2015wmb} after identifying $g_{\rm DM} = y_{\rm DM} \hspace{0.25mm} \cos \theta$ and $g_{\rm SM} = -\sin \theta$.  Couplings between the SM Higgs~$h_1$ and~DM as well as $h_2$ and electroweak (EW) gauge bosons are, on the other hand, not present  in the latter model, while in the context of (\ref{eq:Linteractions}) such interactions and their precise form are  an unavoidable consequence of EW symmetry breaking.

In this paper we focus specifically on the possible collider signatures of this model, and how they differ from the LHC DMF scalar singlet case without mixing. Constraints on the model from non-collider DM experiments can be found in~\cite{Kim:2008pp,Baek:2011aa,LopezHonorez:2012kv,Baek:2012uj,Fairbairn:2013uta}. 

For $m_{h_1}  > 2 m_\chi$, the most obvious manifestation of the interactions (\ref{eq:smh1h2}) is through their contributions to the invisible decay of the Higgs boson. The corresponding decay width is
\begin{equation} \label{eq:h1DM2}
\Gamma (h_1 \to \chi \bar \chi) = \frac{y_{\rm DM}^2 \sin^2 \theta \, m_{h_1}}{8 \pi} \left ( 1 - \frac{4m_\chi^2}{m_{h_1}^2} \right )^{3/2} \,.
\end{equation}
After the transformation $\sin \theta \to \cos \theta$ and $m_{h_1} \to m_{h_2}$ the same expression holds in the case of $h_2$, if it is sufficiently heavy. To determine the invisible Higgs boson branching fraction from (\ref{eq:h1DM2}), one has to keep in mind that all partial widths of $h_1$ to SM particles are suppressed by~$\cos^2 \theta$ and that depending on the mass spectrum also $h_1 \to h_2 h_2$ may be allowed.

Another important feature of (\ref{eq:smh1h2}) is that the couplings between $h_1$ and the EW gauge bosons, as well as the SM fermions, receive  a universal suppression factor of $\cos \theta$ relative to the SM values. The mixing angle and hence (\ref{eq:Linteractions}) is therefore subject to the constraints that arise from the ATLAS and CMS measurements of the signal strengths in Higgs boson production and decay~\cite{HIG-15-002}. Global fits \cite{Farzinnia:2013pga,Belanger:2013kya} to the LHC Run 1 data find $\sin \theta \lesssim  0.4$, which implies that the state $h_1$ ($h_2$) is mostly Higgs-like (singlet-like). Constraints on $\theta$ also arise from the oblique parameters $T$  and $S$~\cite{Baek:2011aa,Lopez-Val:2014jva,Robens:2015gla,Robens:2016xkb}, but  are weaker than those that follow from the Higgs boson measurements.

\begin{figure}[!t]
\centering
\includegraphics[width=0.975\textwidth]{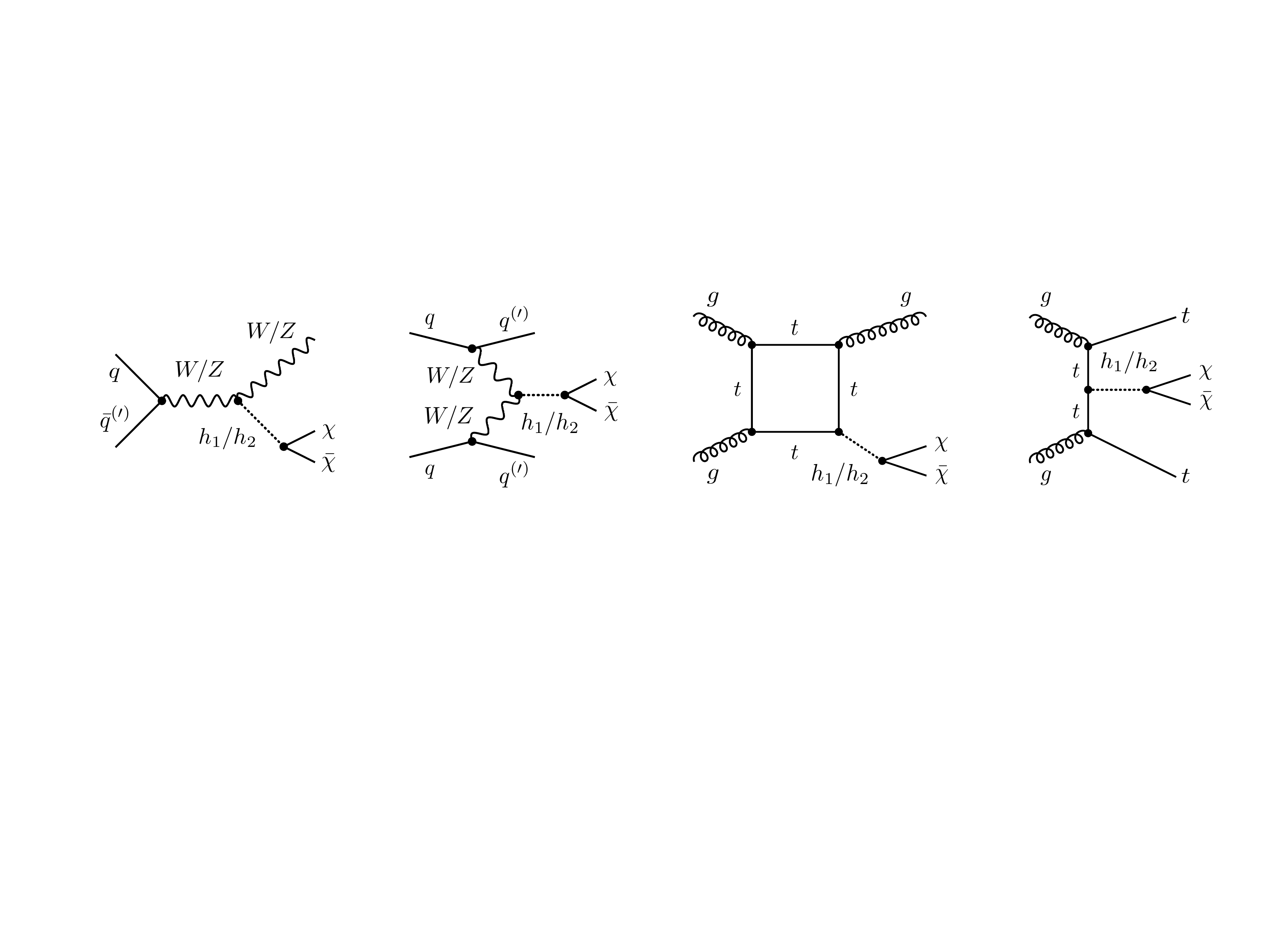}
\vspace{2mm}
\caption{\it Examples of diagrams with an exchange of a $h_1/h_2$ mediator that lead, respectively, to a mono-$W/Z$ signal, a $\MET + 2 j$ signature in vector boson fusion,  $\MET + j$ events from a top quark loop, and a $\MET + t \bar t$ signature.}   
\label{fig:scalardiagrams}
\end{figure}

Turning our attention to the $\MET$ signals, an important observation  is that the phenomenology of  the scalar singlet model with mixing (SMM) is generically richer than that of the scalar-mediated DM model recommended by the LHC DMF.  For instance, the couplings in  (\ref{eq:smh1h2})   that involve EW gauge bosons will give rise to mono-$W$ and mono-$Z$ signals  at tree level. The relevant diagrams are shown in the left panel of Fig.~\ref{fig:scalardiagrams}. The resulting amplitudes take the following schematic form:
\begin{equation} \label{eq:amplitude}
{\cal A} (pp \to \MET + W/Z) \propto y_{\rm DM} \sin (2 \theta) \left ( \frac{1}{s - m_{h_1}^2 + i m_{h_1} \Gamma_{h_1}} - \frac{1}{s  - m_{h_2}^2 + i m_{h_2} \Gamma_{h_2}}  \right ) ,
\end{equation}
where $s$ denotes the invariant mass squared of the DM pair, and  $\Gamma_{h_1}$ and $ \Gamma_{h_2}$ are the total decay widths of the two mass eigenstates in the scalar sector. Similar results hold in the case of  $\MET + 2j$ production through vector boson fusion (VBF), top quark loop induced $\MET + j$ signals, and $\MET + t \bar t$ production. Examples of diagrams that lead to these signals are also displayed in  Fig.~\ref{fig:scalardiagrams}. We note that the contributions from virtual $h_1/h_2$ exchange have opposite signs in (\ref{eq:amplitude}), which is a simple consequence of the mixing matrix (\ref{eq:mixing}) being orthogonal. 
The destructive interference of the two scalar contributions  is a  feature that is also well-known from the DM-nucleon scattering cross section relevant for direct detection~\cite{Kim:2008pp,Baek:2011aa,LopezHonorez:2012kv}. 

It is easy to understand from (\ref{eq:amplitude}) that  the parameter space of the model (\ref{eq:smh1h2}) can be divided into several cases with distinct phenomenologically:
\begin{itemize}
\item Scenario A: For  $m_{h_2} > 2 m_\chi > m_{h_1}$, only the second propagator in (\ref{eq:amplitude}) can go on-shell and, as a result, only diagrams involving $h_2$ exchange will contribute to the various~$\MET$ signals arising in the model (\ref{eq:Linteractions}). This feature implies, for instance, that the normalised kinematic distributions of the monojet signal in  the scalar  models with and without mixing are the same. In consequence, the $\MET + j$ cross sections in the two models can be obtained by a simple rescaling procedure. Working in the narrow-width approximation~(NWA) and taking into account only top quark loop induced diagrams, one obtains 
\begin{equation} \label{eq:rescale}
\frac{\sigma ( pp \to h_2 \hspace{0.5mm} (\to \chi \bar \chi) + j )}{\sigma ( pp \to \phi  \hspace{0.5mm}  (\to \chi \bar \chi) + j )} \simeq \left ( \frac{\sin (2 \theta)}{2 g_{\rm SM}} \right )^2 \frac{\Gamma_\phi}{\Gamma_{h_2}} \,,
\end{equation}
where $\Gamma_\phi$ denotes the total  width of the scalar mediator in the LHC~DMF spin-0 simplified model. We note that additional contributions to $\MET + {\rm jets}$ production arise in the context of (\ref{eq:smh1h2}) also from mono-$V$ or VBF topologies. Such contributions are not present in the LHC DMF model, but are consistently described in the~SMM. 

\item Scenario B: If $m_{h_1} > 2 m_\chi > m_{h_2}$, the roles of $h_1$ and $h_2$ are interchanged, which means that the interactions (\ref{eq:smh1h2}) can be mapped onto the simplified models that are employed in the context of direct and indirect searches for invisible decays of the SM Higgs boson \cite{Aad:2015uga,Aad:2015txa,Chatrchyan:2014tja,CMS-PAS-HIG-16-009}. Again, simple rescaling relations like the one given in (\ref{eq:rescale}) can be worked out to translate the signal strengths in a given $\MET$ channel between the different SDMMs. Unlike the LHC DMF model, the SMM again allows for a consistent description of searches for invisible Higgs boson decays across all channels. 

\item Scenario C: For $m_{h_2} > m_{h_1} > 2 m_\chi$, both scalars can be produced on-shell and, in principle, diagrams with $h_1$ and $h_2$ exchange can be relevant for describing correctly $\MET$ signals arising from (\ref{eq:Linteractions}). However, in large parts of the parameter space the state~$h_1$ will give the dominant contribution, due to a resonance enhancement associated to the first propagator in (\ref{eq:amplitude}). This is an immediate consequence of the fact that  $\Gamma_{h_1}$, being the width of the Higgs-like scalar, is experimentally observed to be small, while $\Gamma_{h_2}$ can receive sizable contributions from decays into~DM and, if kinematically allowed, into top quark pairs. The phenomenology of scenarios~B and~C can therefore be expected to be similar for searches with $\MET$ signatures. 

\item Scenario D: If $m_{h_1} > m_{h_2} > 2 m_\chi$, both scalars can be produced on-shell  like in scenario C, and both contributions can again be important if $\Gamma_{h_1} \simeq \Gamma_{h_2}$. As we will argue in the following, such cases can only be realised if $y_{\rm DM}$ is sufficiently small, and thus are not relevant for searches in $\MET$ signatures. 

\item Scenario E: If $m_{h_1}, m_{h_2} < 2 m_\chi$, the scalars cannot decay to DM, and the prospects for observing $h_2$ production in $\MET$ channels will be very challenging. To probe this scenario one thus has to exploit resonance searches in the SM final states. Depending on the mass and width of $h_2$, possible channels are $\gamma \gamma$, $\gamma Z$, $t \bar t$, $h_1 h_1$, and $t {\bar t }t {\bar t }$. 

\end{itemize}

We now quantify these general observations by studying $\MET$ signals for different mass hierarchies, values of the mixing angle $\theta$, and values of the dark-sector Yukawa coupling~$y_{\rm DM}$.  We compare predictions from the SMM  \eqref{eq:smh1h2} with those of the scalar models~\cite{Haisch:2013ata,DMFmodel_Buckley,DMFmodel_Hahn} used in earlier LHC DMF studies~\cite{Abercrombie:2015wmb}. The SMM and DMF models are used to produce leading order kinematic distributions and cross sections for the monojet and $t \bar t + \MET$ processes.  Monojet, $t \bar t + \MET$ and SM Higgs boson events are generated with {\tt MadGraph5\_aMC@NLO}~\cite{Alwall:2014hca} using the {\tt SMM} UFO model~\cite{Haisch:2016} for the SMM case and the {\tt DMSIMP} UFO model~\cite{Backovic:2015soa} for the LHC DMF and SM Higgs boson cases.  The widths of the $\hOne$ and $\hTwo$ mass eigenstates in the SMM and DMF models are determined automatically with {\tt MadGraph5\_aMC@NLO} as a function of the relevant masses and $y_{\rm DM}$ values.  

\begin{figure}[!t]
        \centering
        \includegraphics[width=0.49\textwidth]{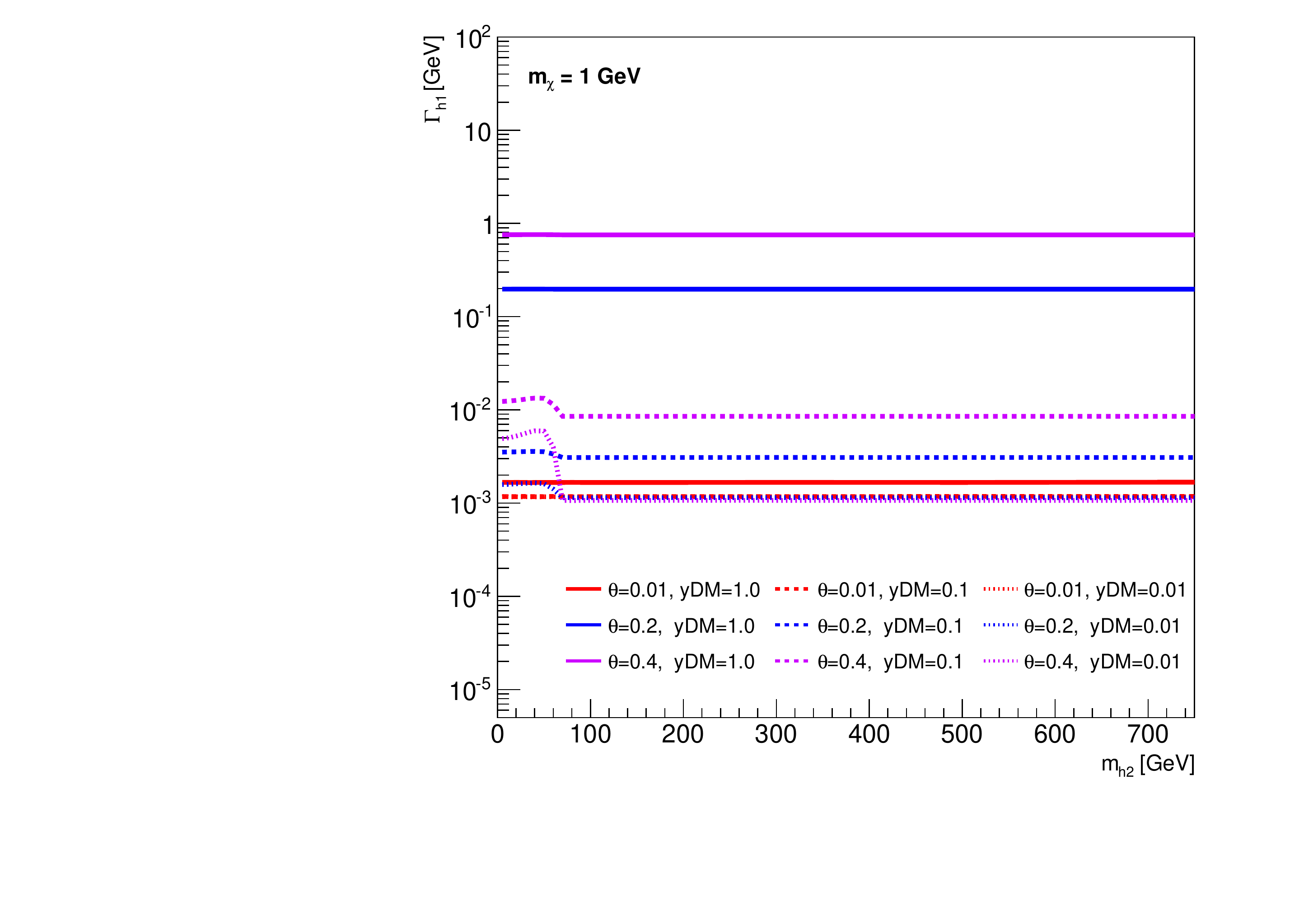}
        \includegraphics[width=0.49\textwidth]{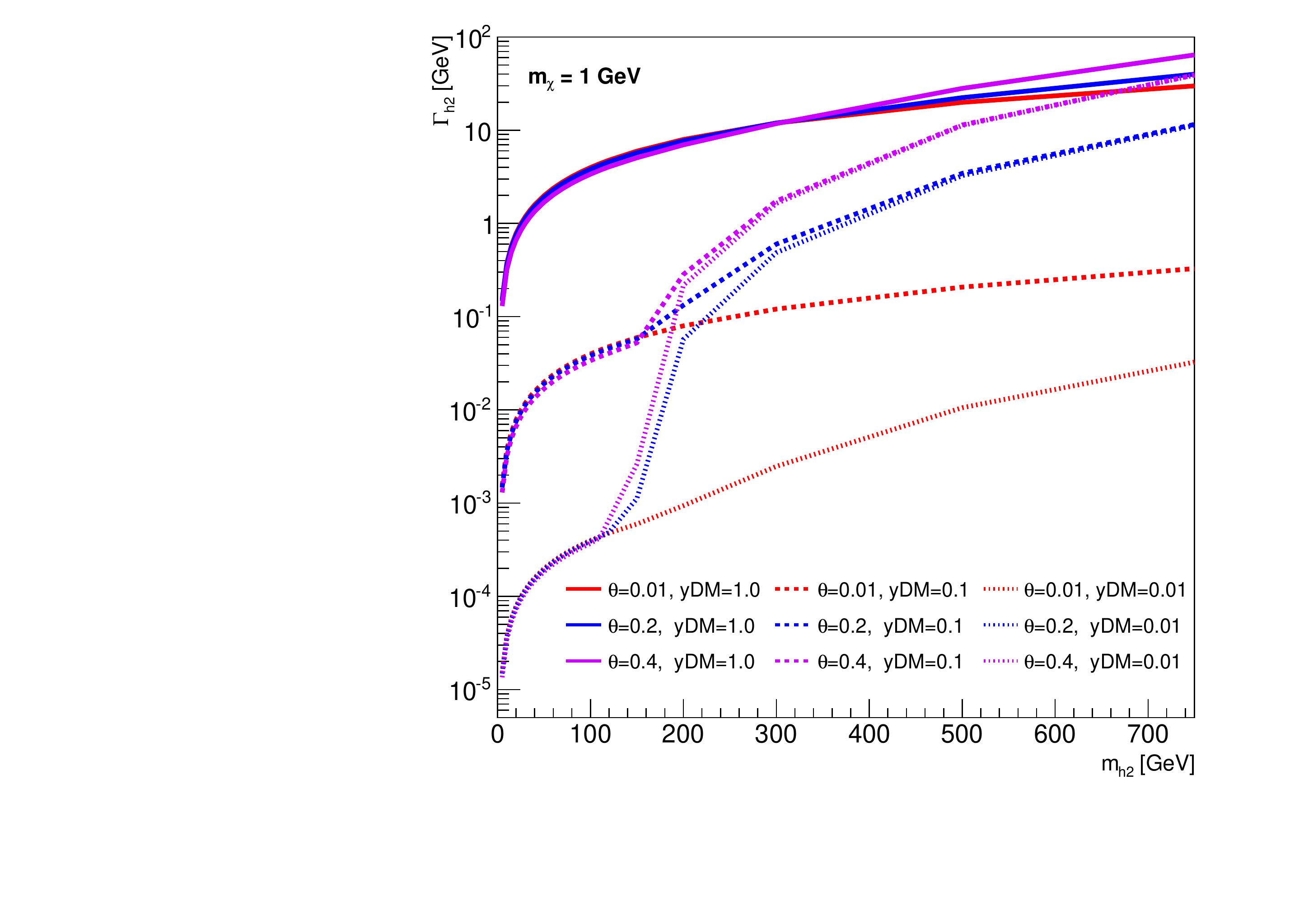}
        \caption{\it
	The $\hOne$ (left) and $\hTwo$ (right) total decay width in scenarios C and D for several values of $y_{\rm DM}$ and $\theta$. In both panels the DM mass is fixed to $1 \, {\rm GeV}$.  An increase in the total decay width of the $\hOne$ state at low masses and low values of DM coupling $y_{\rm DM}$ is due to the contribution of the $h_1 \to h_2 h_2$ decay channel. Note that the $\theta = 0.01$ lines for $y_{\rm DM} = 0.1$ and 0.01 nearly overlap and hence are seen as a single line in the left plot.}
        \label{fig:widths}
\end{figure}

In general, the total width $\Gamma_{h_1}$ in the SMM differs from the SM Higgs boson width due to the additional $h_1 \to \chichi$ and $h_1 \to h_2 h_2$ decay channels, and the $\cos^{2}\theta$ suppression of $\hOne$ decays into SM particles.  Similarly, $\Gamma_{h_2}$ includes decays both to DM and SM particles, and depends on~$\mhTwo$, $y_{\rm DM}$, and~$\theta$.  Figure~\ref{fig:widths} shows $\Gamma_{h_1}$ and $\Gamma_{h_2}$ as a function of these SMM parameters for scenarios C and D, for which both $\hOne$ and $\hTwo$ can potentially contribute to the different~$\MET$ signatures.  The kinematics in the SMM is expected to be driven by $h_1$ ($h_2$) exchange when $\Gamma_{h_1} \ll \Gamma_{h_2}$~($\Gamma_{h_2} \ll \Gamma_{h_1}$) --- we will demonstrate below that this expectation is indeed  correct. From the comparison of the two panels in Fig.~\ref{fig:widths} it also is evident that in order to have~$\Gamma_{h_1} \simeq  \Gamma_{h_2}$, the DM coupling $y_{\rm DM}$ has to be small. In the limit $y_{\rm DM} \to 0$, the decay rates of $h_1$ and~$h_2$ to SM particles will however become dominant and, as a result, mediator searches in SM final states will  typically provide the leading constraints on the parameter space.  One can thus conclude that in the parameter space where $\MET$ searches are strongest, depending on the mass hierarchy, either $h_1$ or $h_2$ exchange dominates the signals. 

Next, we study the exclusive $\chichi$ production cross section and its kinematics for the mass hierarchies corresponding to scenarios A, B, C, and D.  As mentioned, scenarios C and D correspond to the on-shell decays of both the $\hOne$ and $\hTwo$ mediators to DM particles.  Earlier studies have shown that kinematics and cross sections are independent of $\mchi$ in such scenarios~\cite{Abercrombie:2015wmb}.  Therefore, without loss of generality, we consider DM particles with a mass of $\mchi = 1\rm~GeV$ in scenarios C and D, and scan values for $\mhTwo$, $y_{\rm DM}$, and $\theta$.  In scenario A (B), the DM  particles are heavier than $\mhOne/2$ ($\mhTwo/2$), and $\hOne$ ($\hTwo$) decays to~$\chichi$ are prohibited.  Provided that $\mchi$ is smaller than $\hTwo$ ($\hOne$) in scenario A (B), SMM kinematics should also be independent of $\mchi$.  Consequently, we focus on $\mchi = 100 \, {\rm GeV}$~($\mchi = 10 \, {\rm GeV}$) for these scenarios.

We compare SMM and LHC DMF kinematics by means of the predicted transverse momentum of the $\chichi$ system, $\ptchichi$, which is a useful generator-level proxy for the $\MET$ observable typically used in collider-based DM searches.  Our treatment ignores experimental effects (e.g.,~selection efficiencies, energy resolutions, and detector effects) that would be relevant in an analysis at the reconstruction level. 

\subsection{Scenario A} 

\begin{figure}[!ht]
        \centering
        \includegraphics[width=0.49\textwidth]{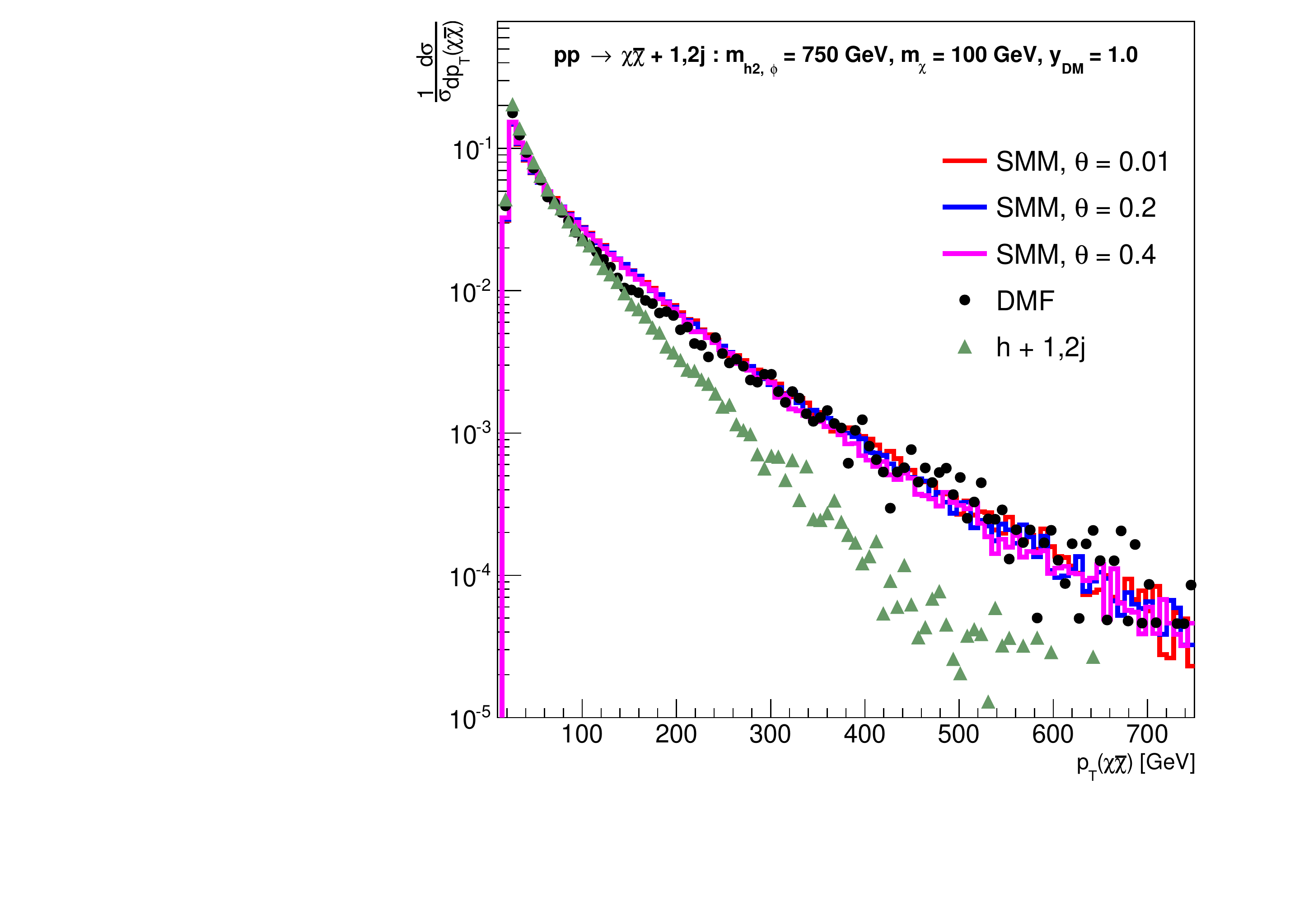}
        \includegraphics[width=0.49\textwidth]{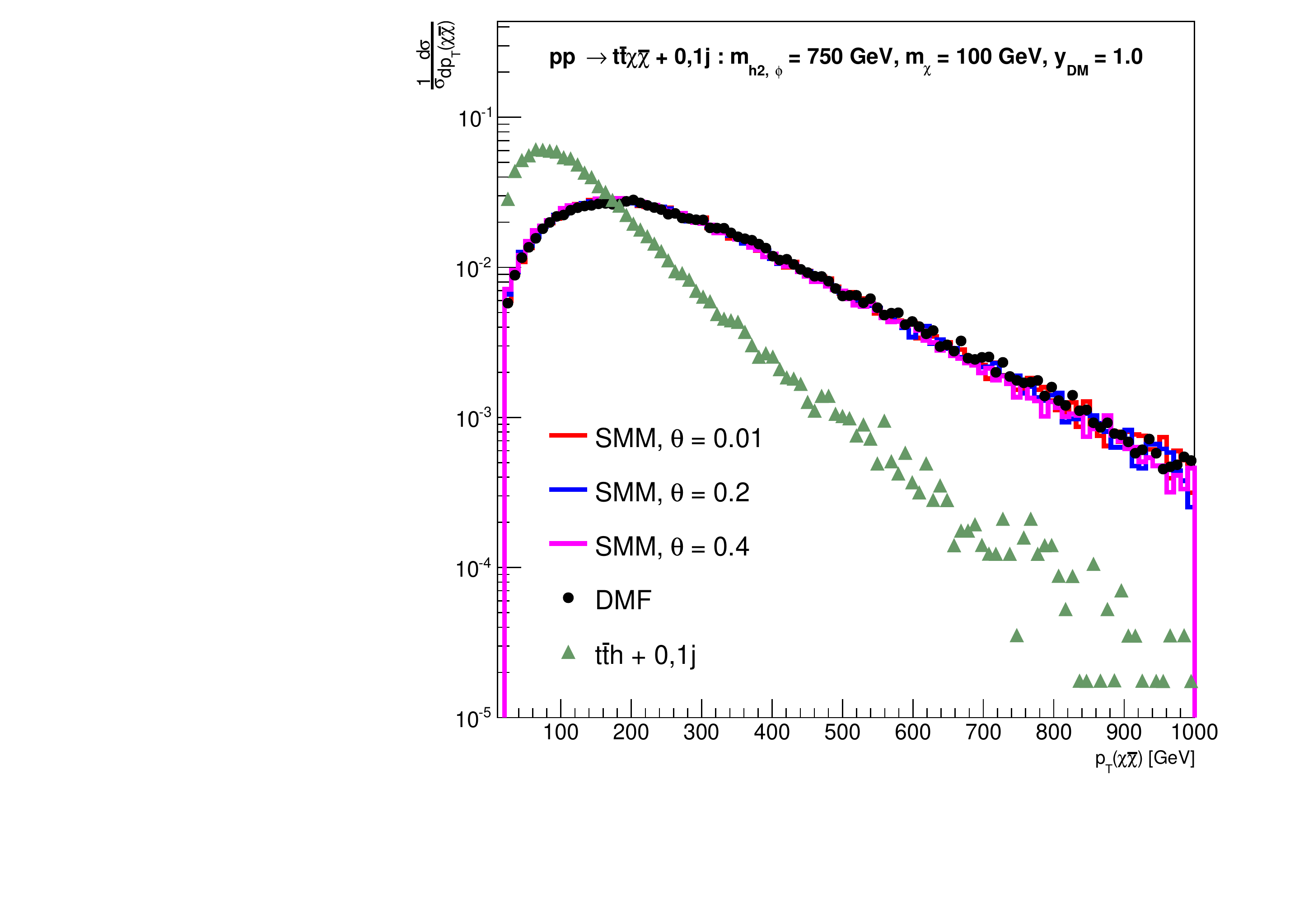}
        \caption{\it
	 Scenario A kinematics: $\ptchichi$ in the SMM and LHC DMF models for the monojet (left) and $t \bar t + \MET$ (right) channels.  Both plots correspond to $\mchi = 100 \, {\rm GeV}$, $\mhTwo = 750 \, {\rm GeV}$, $\yDM=1.0$.  The monojet plot includes a comparison with the SM Higgs boson production in association with one or two jets, while the $t \bar t + \MET$ plot includes a comparison with SM Higgs boson production in association with $\ttb$ (the Higgs boson $p_{\rm T}$ is displayed in these cases). The SMM kinematics for both monojet and $t \bar t + \MET$ generally agree with the LHC DMF model predictions in this scenario. 
        } 
        \label{fig:scenarioA_kin}
\end{figure}

Figure~\ref{fig:scenarioA_kin} compares SMM and LHC DMF model kinematics in scenario A.  In accordance with the expectation, we observe a close correspondence between the kinematics in these models.  The discrepancy observed in the monojet spectra near $150 \, {\rm GeV}$ results from vector-boson-mediated processes, which are included in the SMM but not in the LHC DMF model.  Section~\ref{sec:VBF} discusses the vector boson mediated (VBM) contributions in more detail.  

\begin{figure}[!t]
        \centering
        \includegraphics[width=0.49\textwidth]{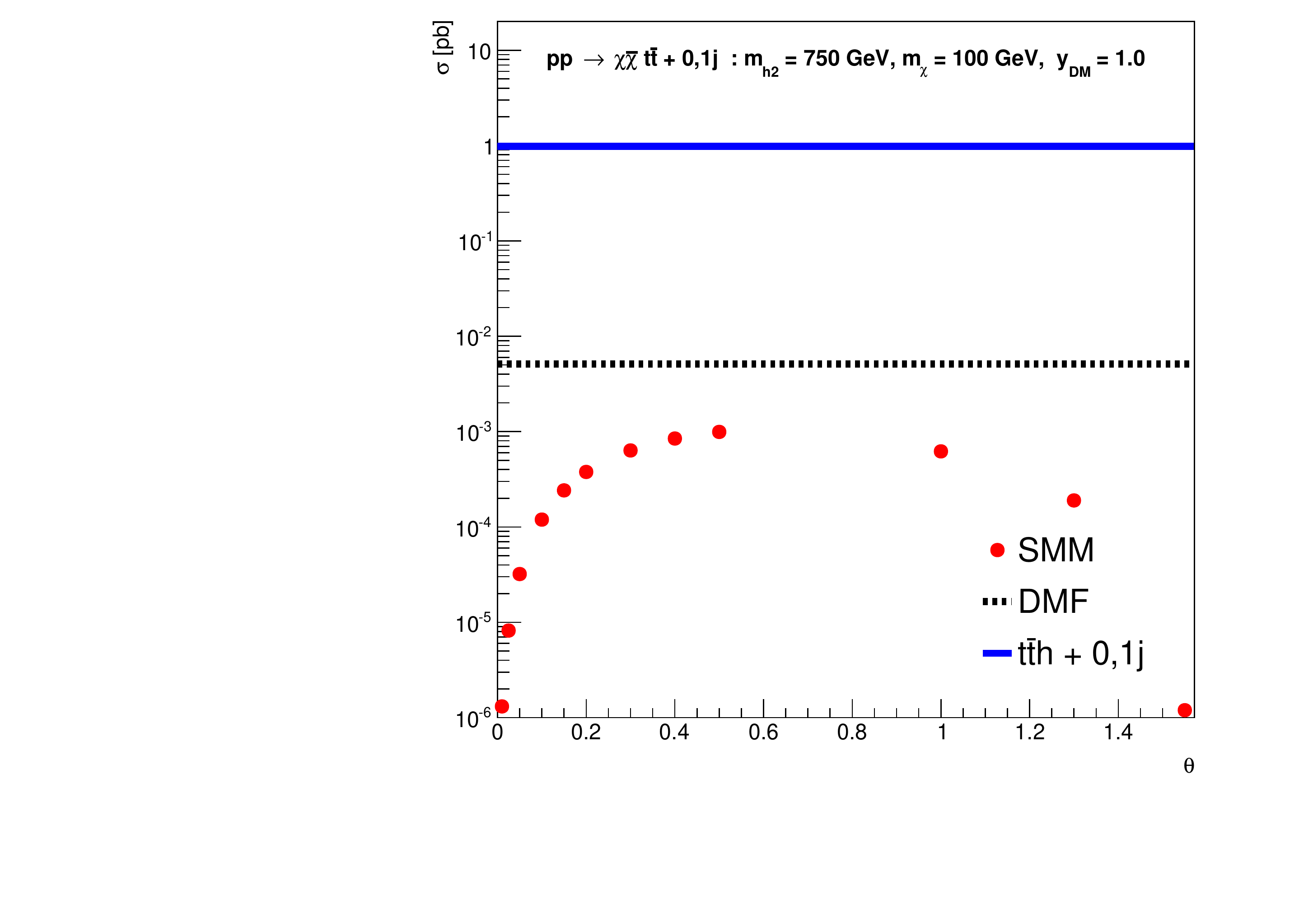}
        \caption{\it
	Scenario A cross sections:  the SMM $\ttb + \MET$ production cross section for  $\mhTwo=750\rm~GeV$, $\mchi=100\rm~GeV$, $\yDM=1.0$ as a function of mixing angle compared with the LHC DMF model and SM $\ttb + h$ cross sections.  Because the $h_{1} \rightarrow \chichi$ channel is inaccessible in this scenario, the SMM cross section remains below the LHC DMF model equivalent. 
        }   
        \label{fig:scenarioA_xsecBR}
\end{figure}

Figure~\ref{fig:scenarioA_xsecBR} shows the $t \bar t + \MET$ production cross section for the nominal case of $\yDM=1.0$.  As expected, the SMM cross section times branching fraction approaches zero as the mixing angle $\theta$ tends to $0$ or $\pi/2$.  Previous studies have shown that $\chi \bar \chi$ kinematics are independent of the $\yDM$ value for the low to moderate mediator masses explored here~\cite{Abercrombie:2015wmb}.  The prediction for $\sigma ( pp \to t \bar t + h_1/h_2) \hspace{0.25mm} {\rm Br} ( h_1/h_2 \to \chi \bar \chi)$ is smaller than both the SM Higgs boson and corresponding LHC DMF model cross sections due to mixing between $\hOne$ and $\hTwo$, and because on-shell $\chichi$ production via  $\hOne$ exchange is forbidden in scenario A.  These results also generalize to the monojet process. 

\subsection{Scenario B} 

\begin{figure}[!t]
        \centering
        \includegraphics[width=0.49\textwidth]{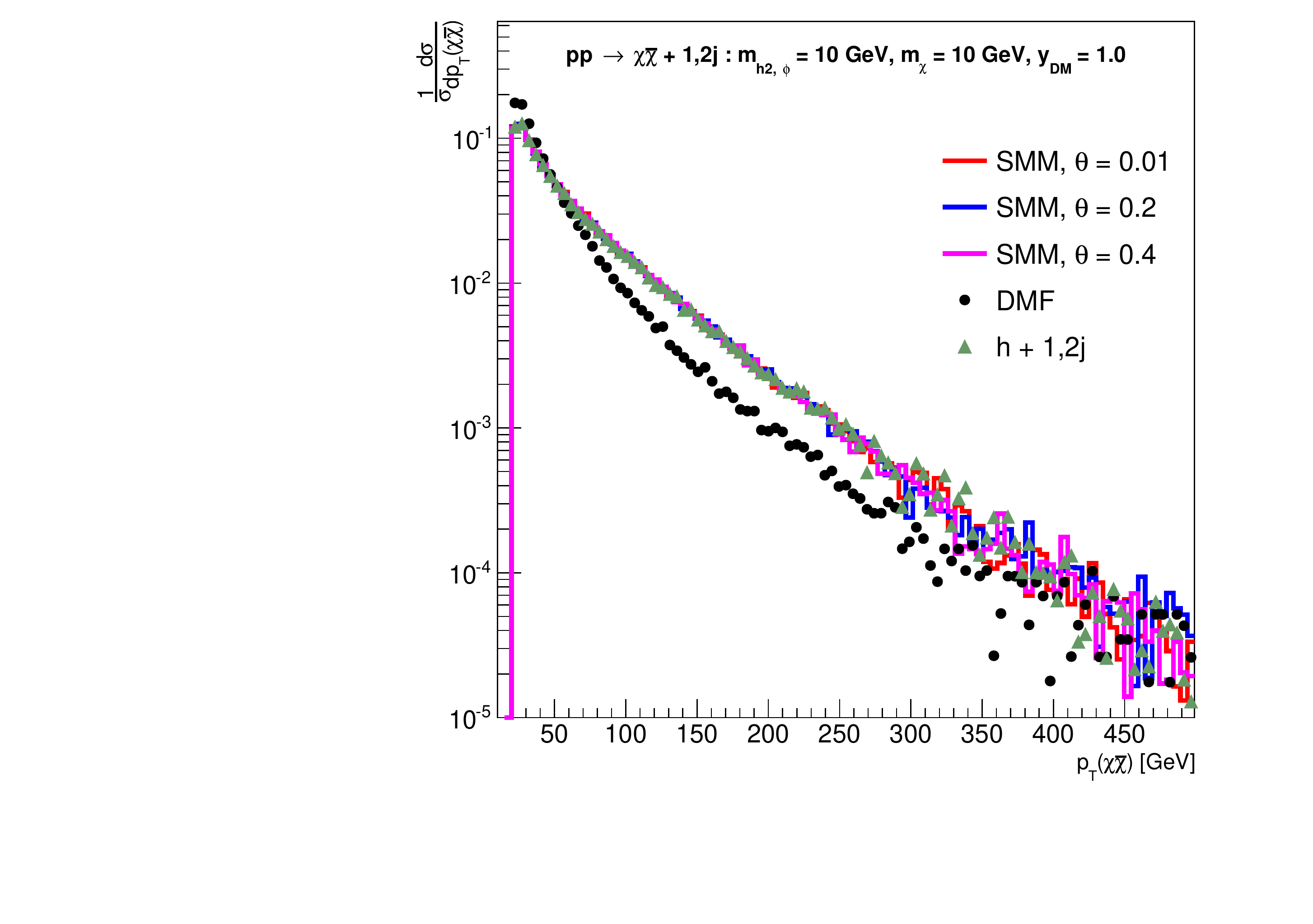}
        \includegraphics[width=0.49\textwidth]{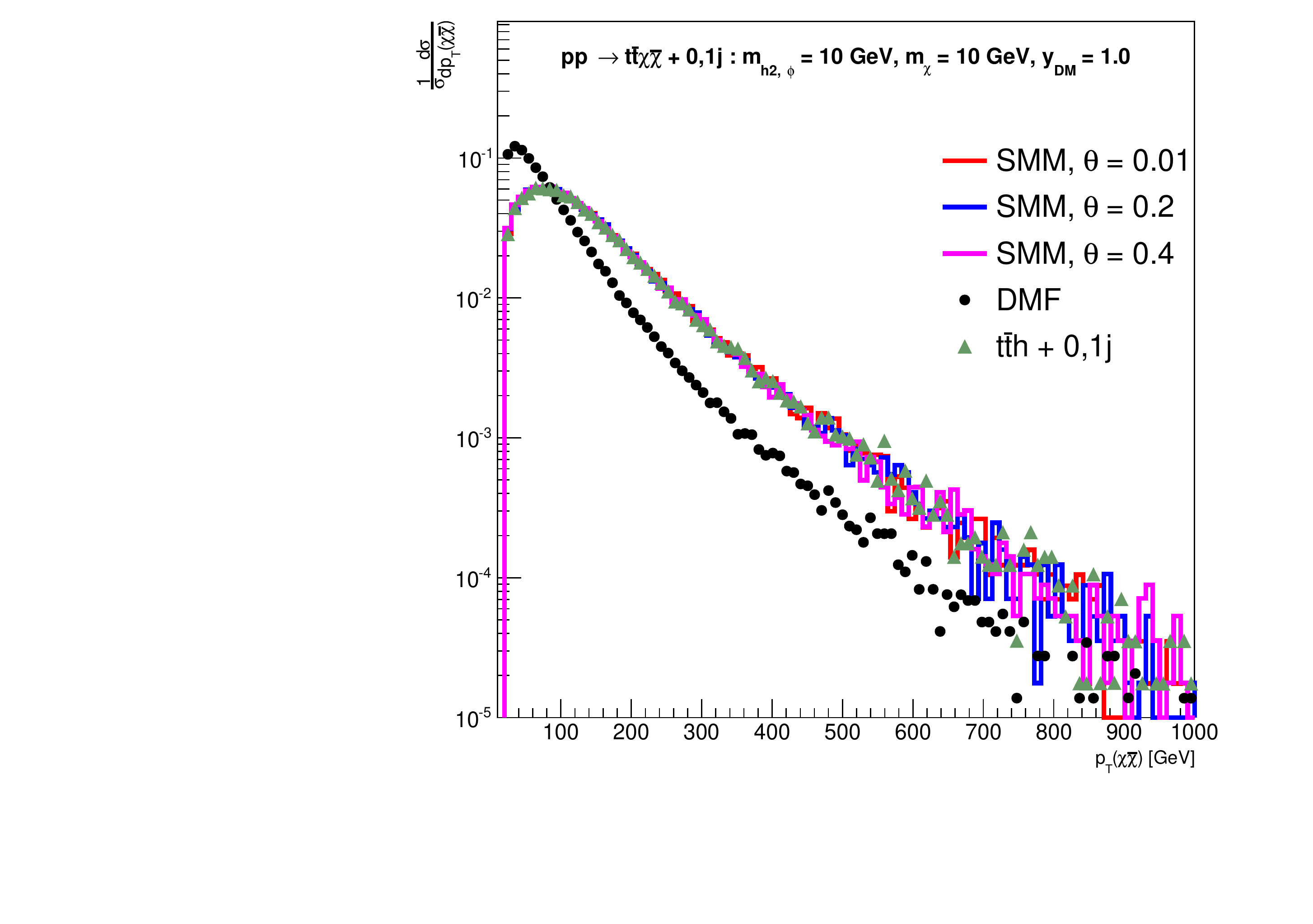}
        \caption{\it
          Scenario B kinematics: $\ptchichi$ distributions for the monojet (left) and $t \bar t + \MET$ (right) process with~$\mchi=10 \, {\rm GeV}$, $\mhTwo=10 \, {\rm GeV}$, and $\yDM=1.0$. The rest of the notations are as in Fig.~\protect\ref{fig:scenarioA_kin}.
        }
        \label{fig:scenarioB_kin}
\end{figure}

\begin{figure}[!t]
        \centering
        \includegraphics[width=0.49\textwidth]{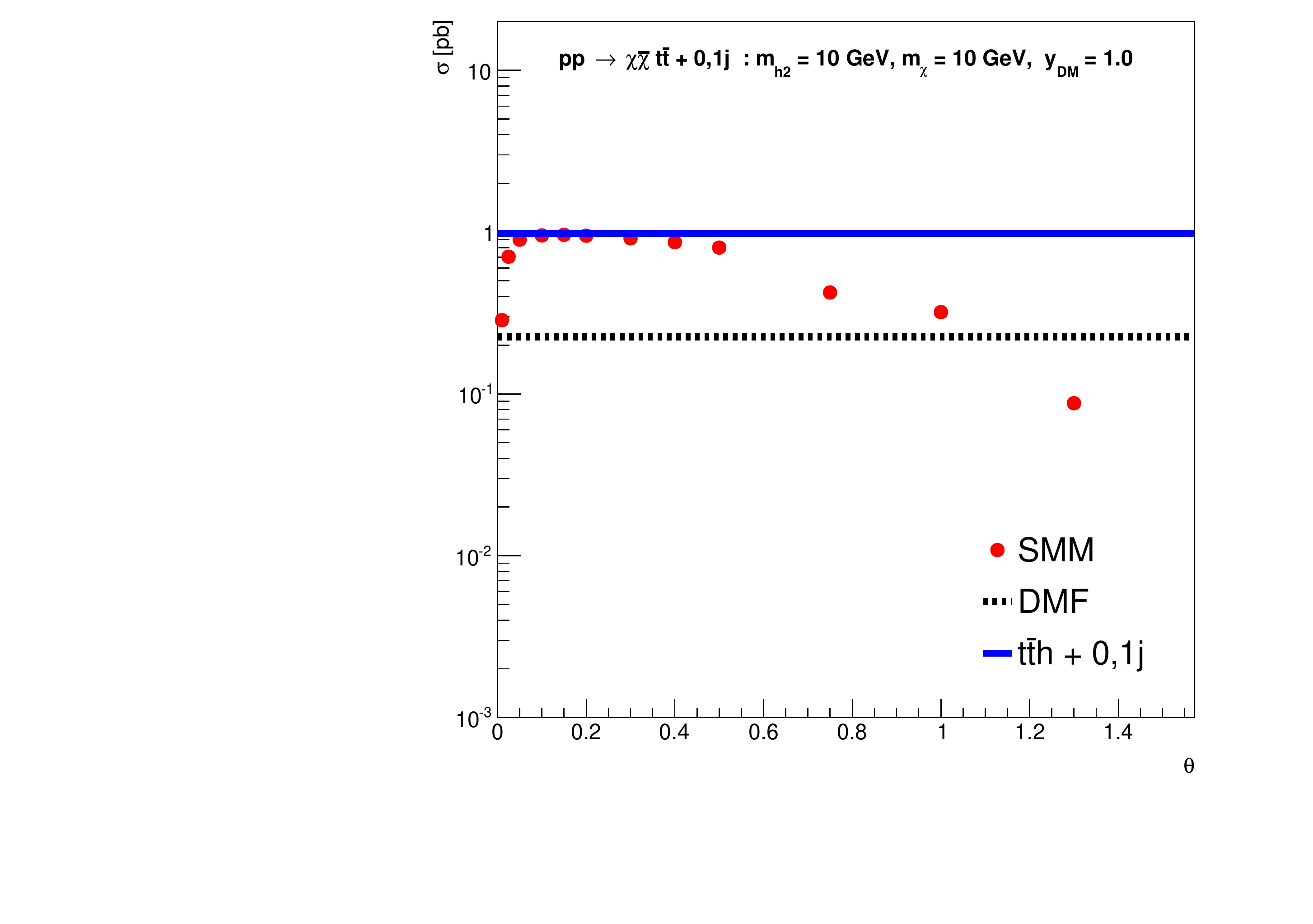}
        \caption{\it
	Scenario B cross sections: the SMM $\ttb + \MET$ production cross section for $\mchi=10 \, {\rm GeV}$, $\mhTwo=10 \, {\rm GeV}$, and $\yDM=1.0$. 
}
        \label{fig:scenarioB_xsecBR}
\end{figure}

Figures~\ref{fig:scenarioB_kin} and~\ref{fig:scenarioB_xsecBR}, respectively, compare $\ptchichi$ distributions and cross sections for scenario B.  In this scenario, SMM kinematics clearly correspond to SM Higgs boson production rather than to the LHC DMF model predictions.  Figure~\ref{fig:scenarioB_xsecBR} displays $\sigma ( pp \to t \bar t + h_1/h_2) \hspace{0.25mm} {\rm Br} ( h_1/h_2 \to \chi \bar \chi)$  for representative $\mhTwo,\mchi$ values in the nominal case of $\yDM=1.0$.  For intermediate values of the mixing angle, $\sigma ( pp \to t \bar t + h_1/h_2) \hspace{0.25mm} {\rm Br} ( h_1/h_2 \to \chi \bar \chi)$ lies between the corresponding LHC DMF model and SM Higgs boson cross sections.  The finding that $h_1$ drives the prediction in scenario~B also applies to the monojet case. 

\subsection{Scenarios C and D} 

\begin{figure}[!t]
        \centering
        \includegraphics[width=0.49\textwidth]{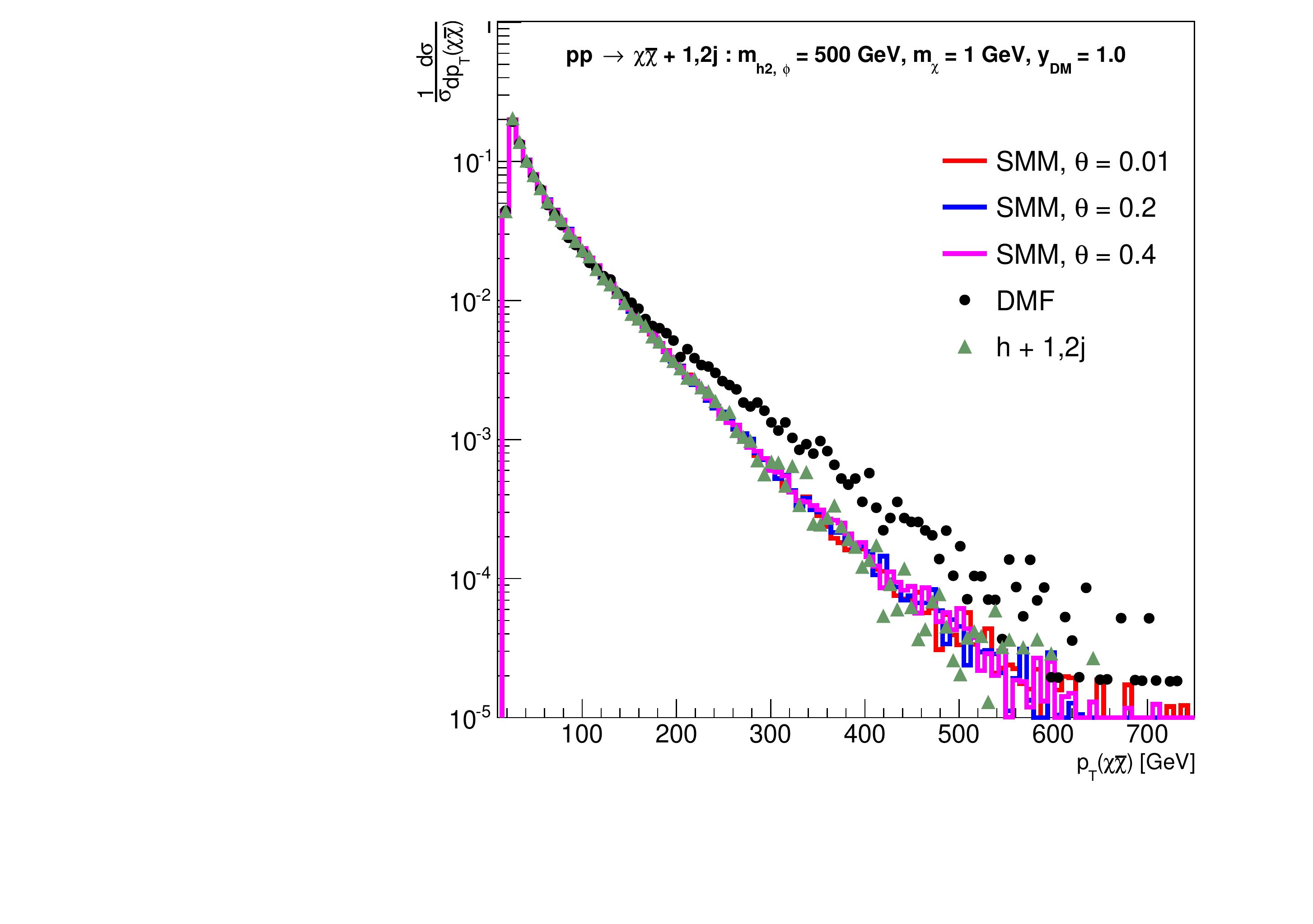}
        \includegraphics[width=0.49\textwidth]{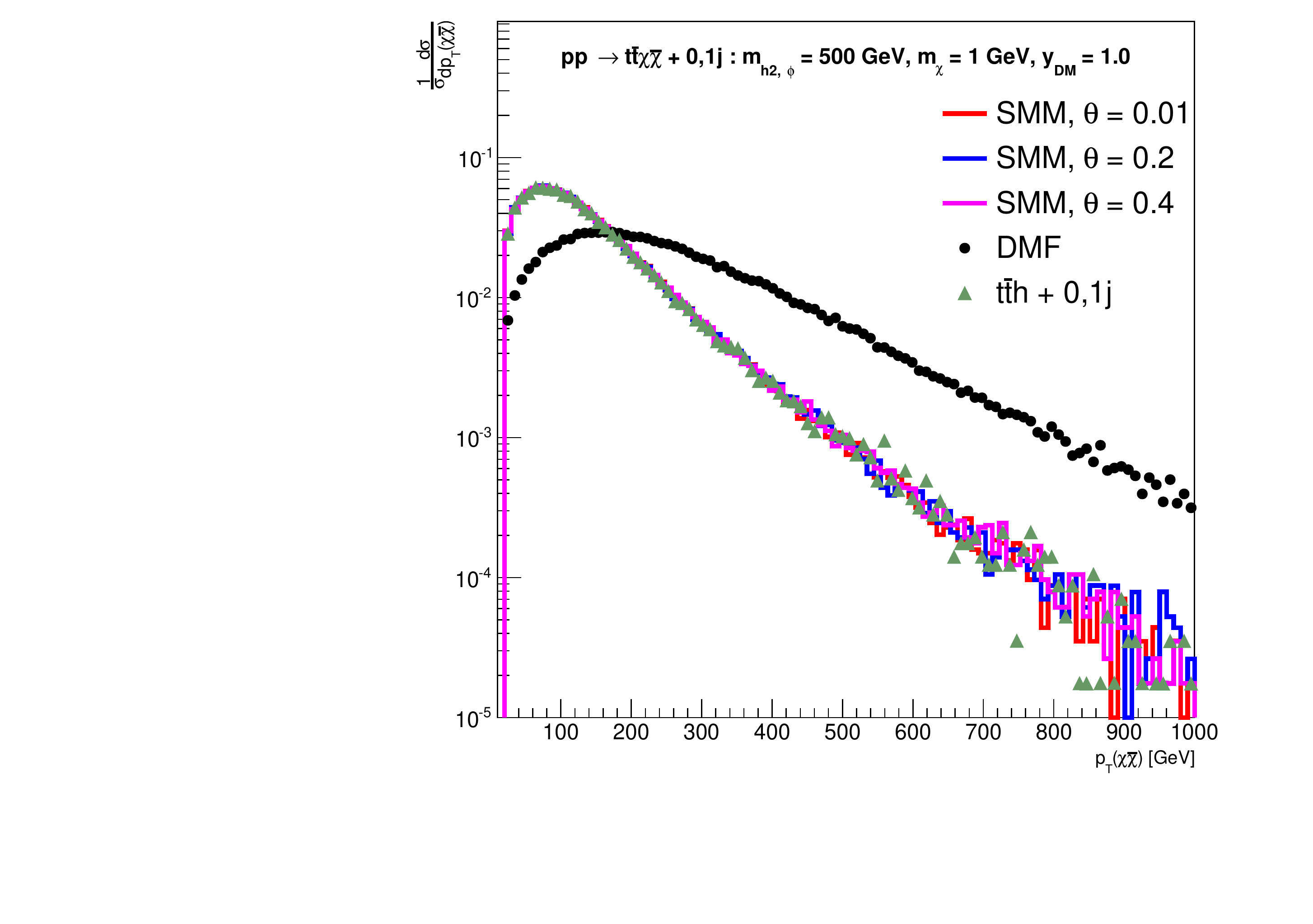}
        \caption{\it
          Scenario C kinematics: $\ptchichi$ distributions for the monojet~(left) and $\ttb + \MET$ (right) channels. The results shown correspond to $\mhTwo= 500 \, {\rm GeV}$, $\mchi= 1 \, {\rm GeV}$, and $\yDM=1.0$. The rest of the notations are as in Fig.~\protect\ref{fig:scenarioA_kin}.
        }   
        \label{fig:scenarioC_kin}
\end{figure}

\begin{figure}[!t]
        \centering
        \includegraphics[width=0.49\textwidth]{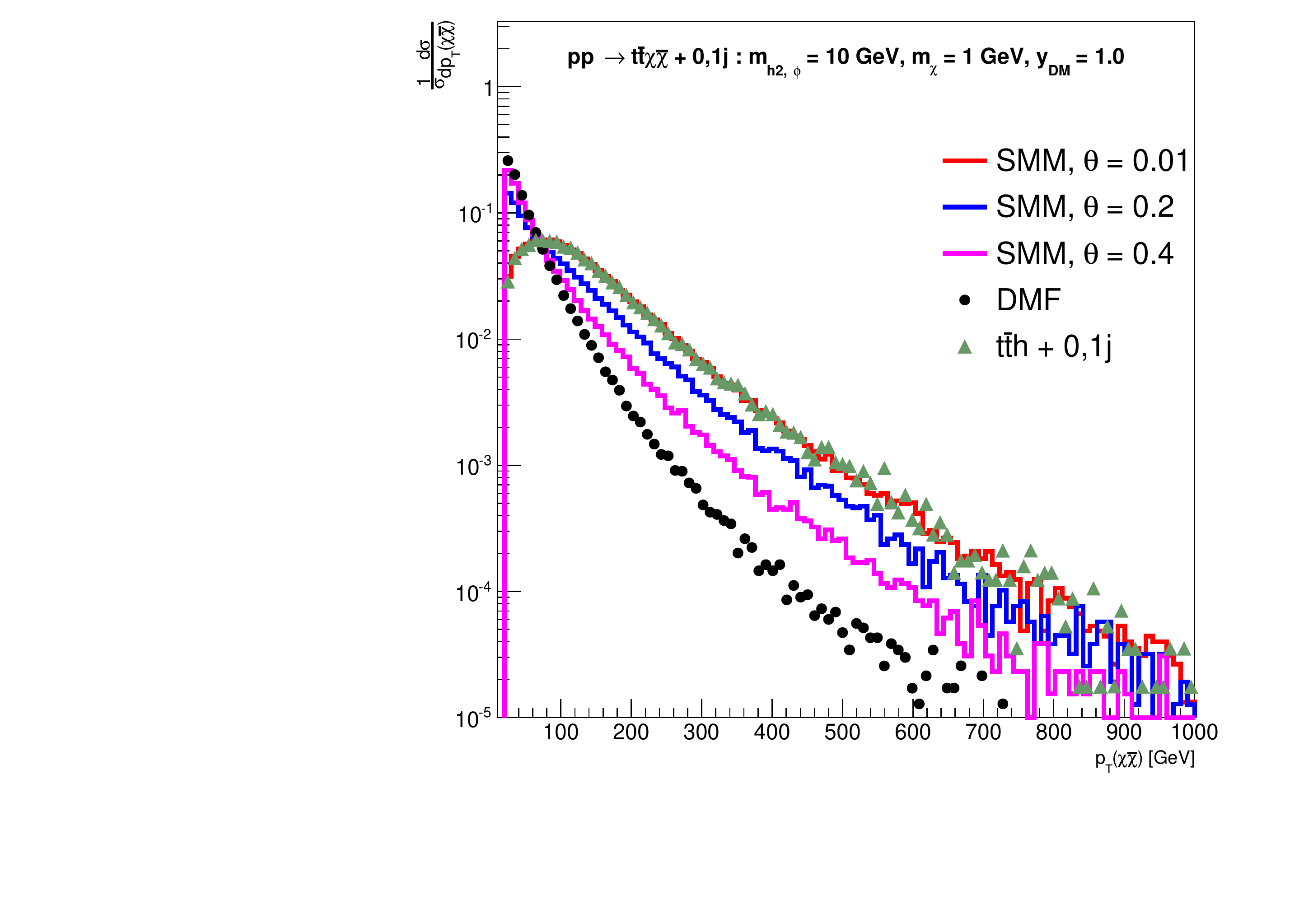}
        \includegraphics[width=0.49\textwidth]{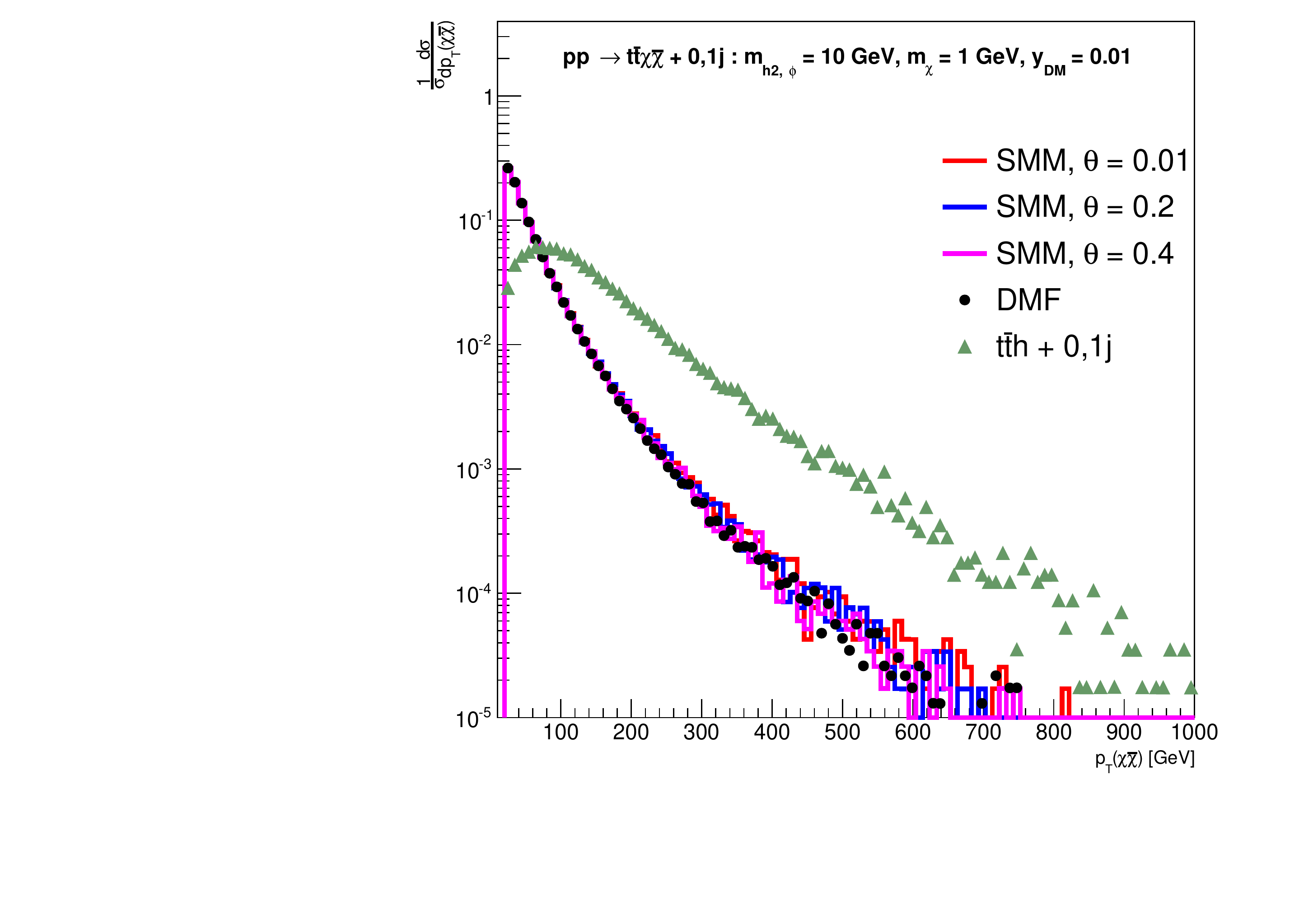}
        
        \vspace{2mm}
        
        \includegraphics[width=0.49\textwidth]{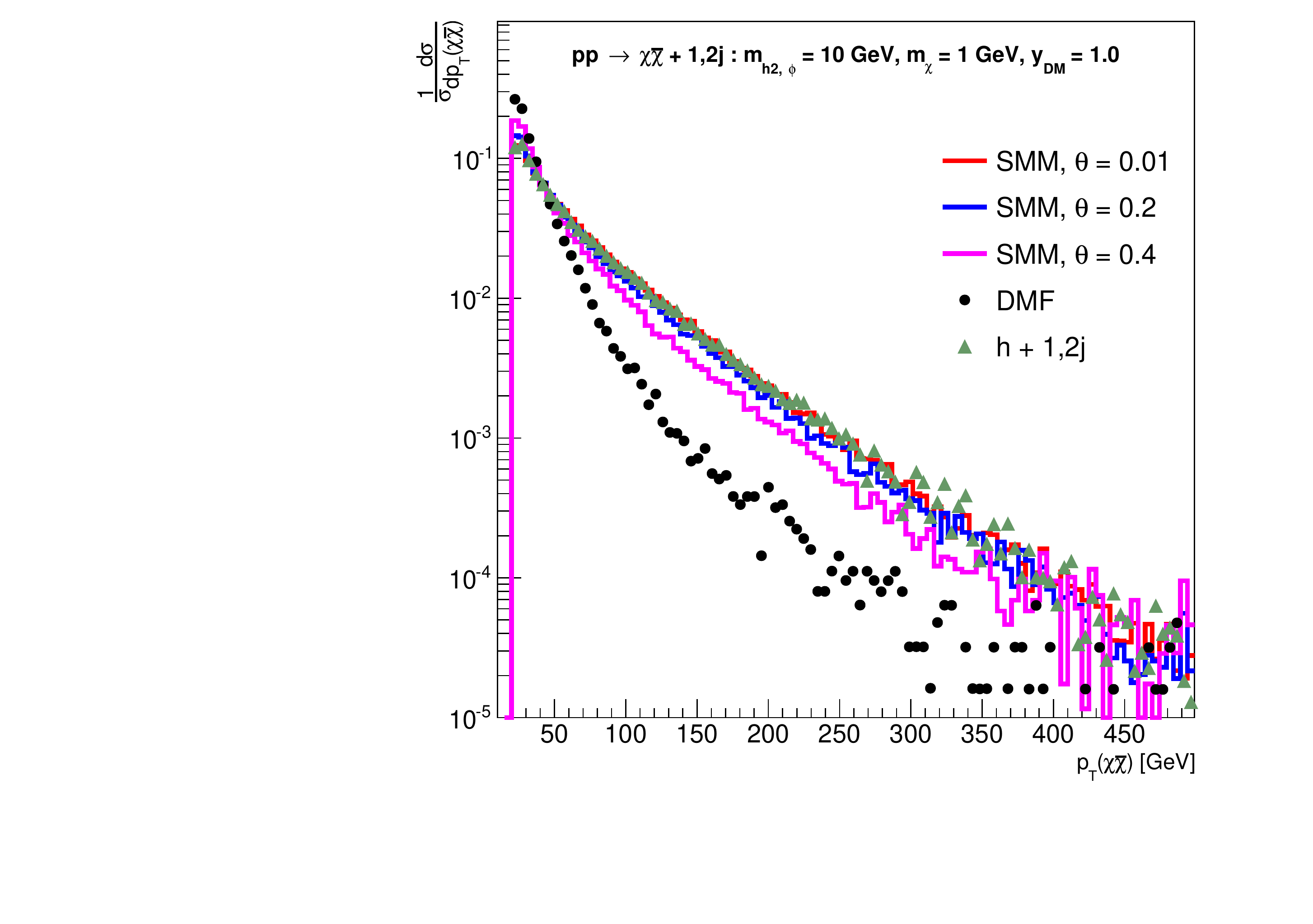}
        \includegraphics[width=0.49\textwidth]{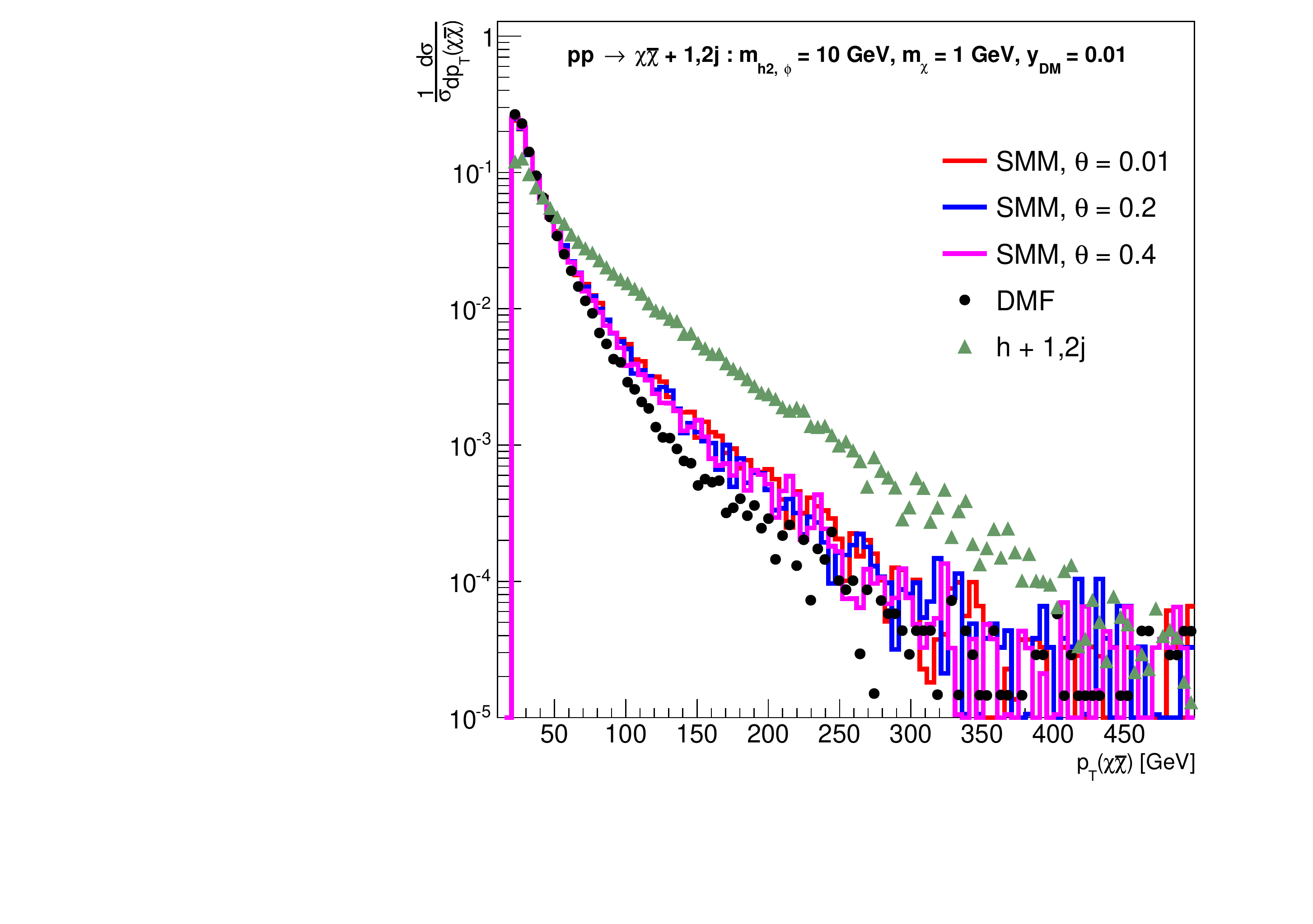}
        \caption{\it
          Scenario D kinematics:  $\ptchichi$ distributions for $\ttb + \MET$ (top row) and  monojet (bottom row) channels. The displayed results are obtained for $\mhTwo=10 \, {\rm GeV}$, and $\mchi=1 \, {\rm GeV}$.  The left and right columns correspond to $\yDM=1.0$ and $0.01$, respectively.  The rest of the notations are as in Fig.~\protect\ref{fig:scenarioA_kin}.
        }
        \label{fig:scenarioD_kin}
\end{figure}

\begin{figure}[!htp]
        \centering
        \includegraphics[width=0.49\textwidth]{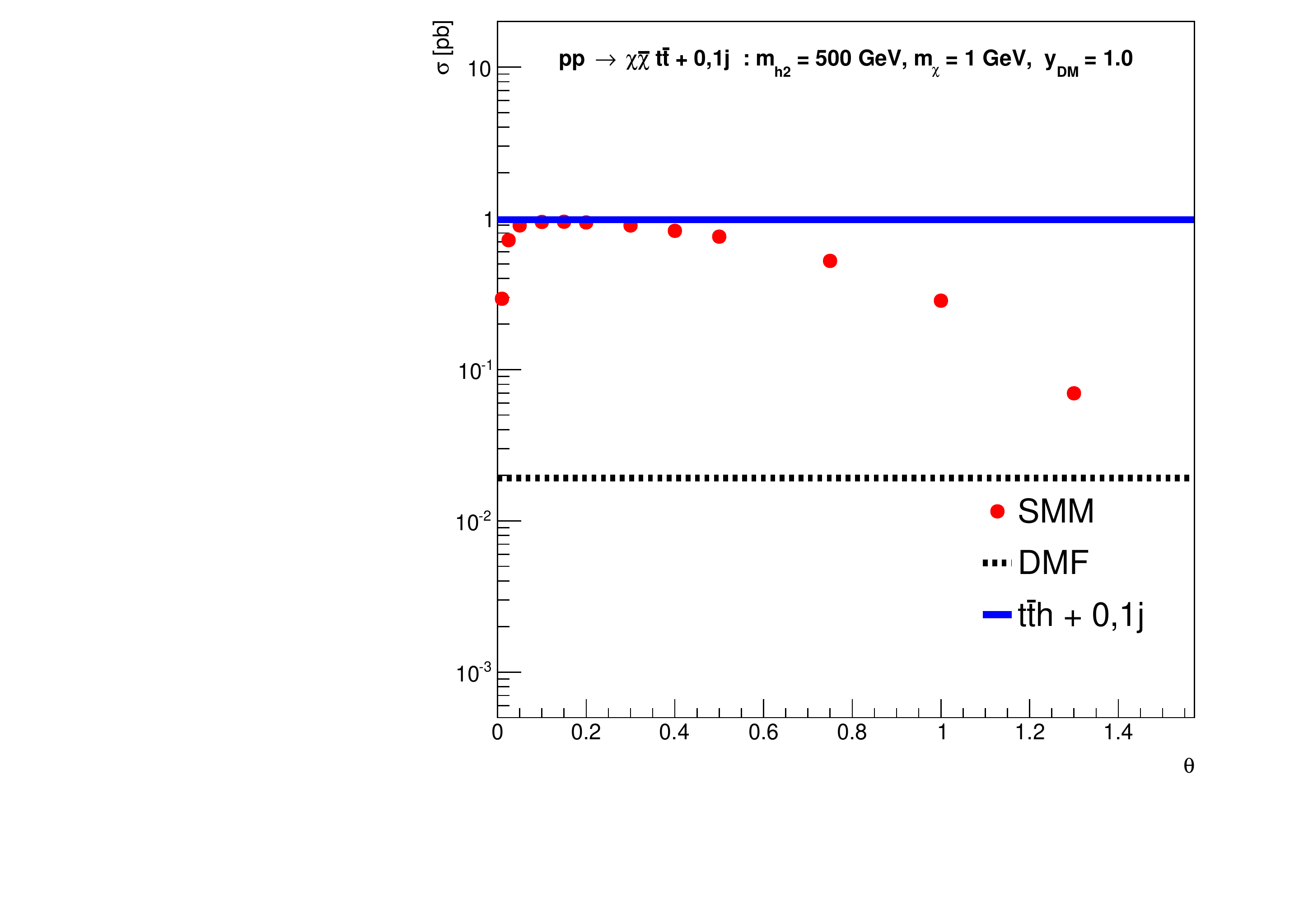}
        \includegraphics[width=0.49\textwidth]{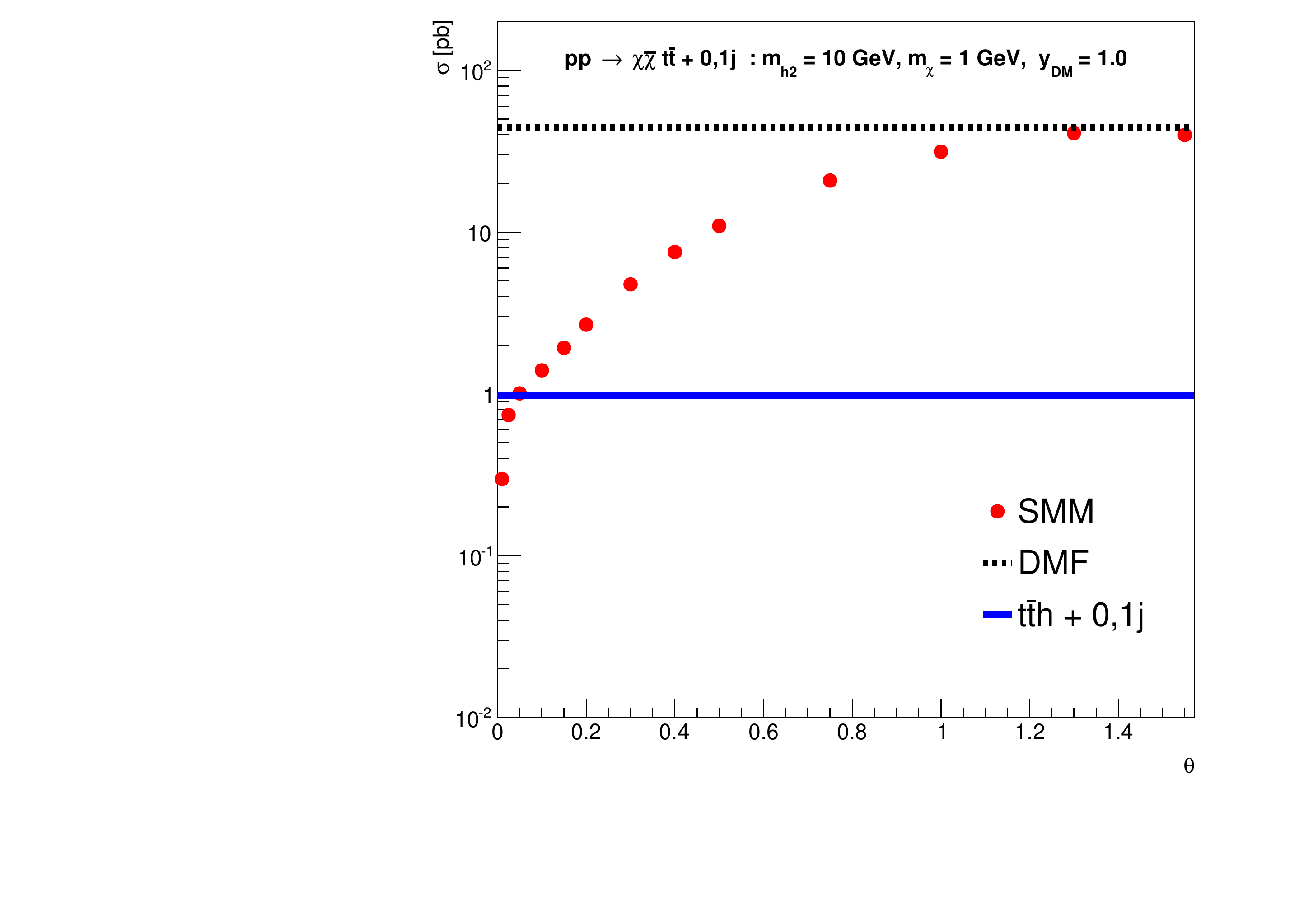}
        \caption{\it
	Scenario C and D cross sections:  $\ttb + \MET$ production cross sections compared with those for the SM $\ttb+h$ and the corresponding LHC DMF model cross sections for $\mhTwo=500 \, {\rm GeV}$~(left) and $\mhTwo = 10 \, {\rm GeV}$~(right).  All results use $\mchi=1 \, {\rm GeV}$ and $\yDM=1.0$.  
        }   
        \label{fig:scenarioCD_xsecBR}
\end{figure}

Scenarios C and D in the SMM are similar in that on-shell decays of both the $\hOne$ and $\hTwo$ mediators are possible.  In principle, both mediators can therefore contribute to the~$\MET$ production cross sections in the different channels.  Figures~\ref{fig:scenarioC_kin} and~\ref{fig:scenarioD_kin} show representative $\ptchichi$ distributions for scenarios C and D, respectively.  Representative cross sections for these scenarios are shown in Fig.~\ref{fig:scenarioCD_xsecBR}.  

From Fig.~\ref{fig:scenarioC_kin} it is evident that the SMM $\ptchichi$ distributions are generally softer than those of the LHC DMF model.  Within statistical uncertainties the kinematics of the SMM signals are essentially identical to Higgs production in the SM.  Figure~\ref{fig:scenarioCD_xsecBR} demonstrates that the $\ttb + \MET$ cross section in scenario~C is generally larger than that of the LHC DMF model, and approaches the SM~$\ttb  + h$ cross section for intermediate values of the mixing angle.

Figure~\ref{fig:scenarioD_kin}, which corresponds to scenario D, clearly shows the impact of $\hOne/\hTwo$ mixing.  Significant differences between the LHC DMF model and SMM kinematics are found for large $\yDM$.  As $\yDM$ decreases from $1.0$, kinematics approach those of the LHC DMF model.  The SMM production cross section is shown as a function of mixing angle in Fig.~\ref{fig:scenarioCD_xsecBR}.  The situation here is essentially the reverse of scenario C, with the production cross section remaining below that of the LHC DMF model.


\subsection{VBM production}\label{sec:VBF} %

\begin{figure}[!htp]
        \centering
        \includegraphics[width=0.49\textwidth]{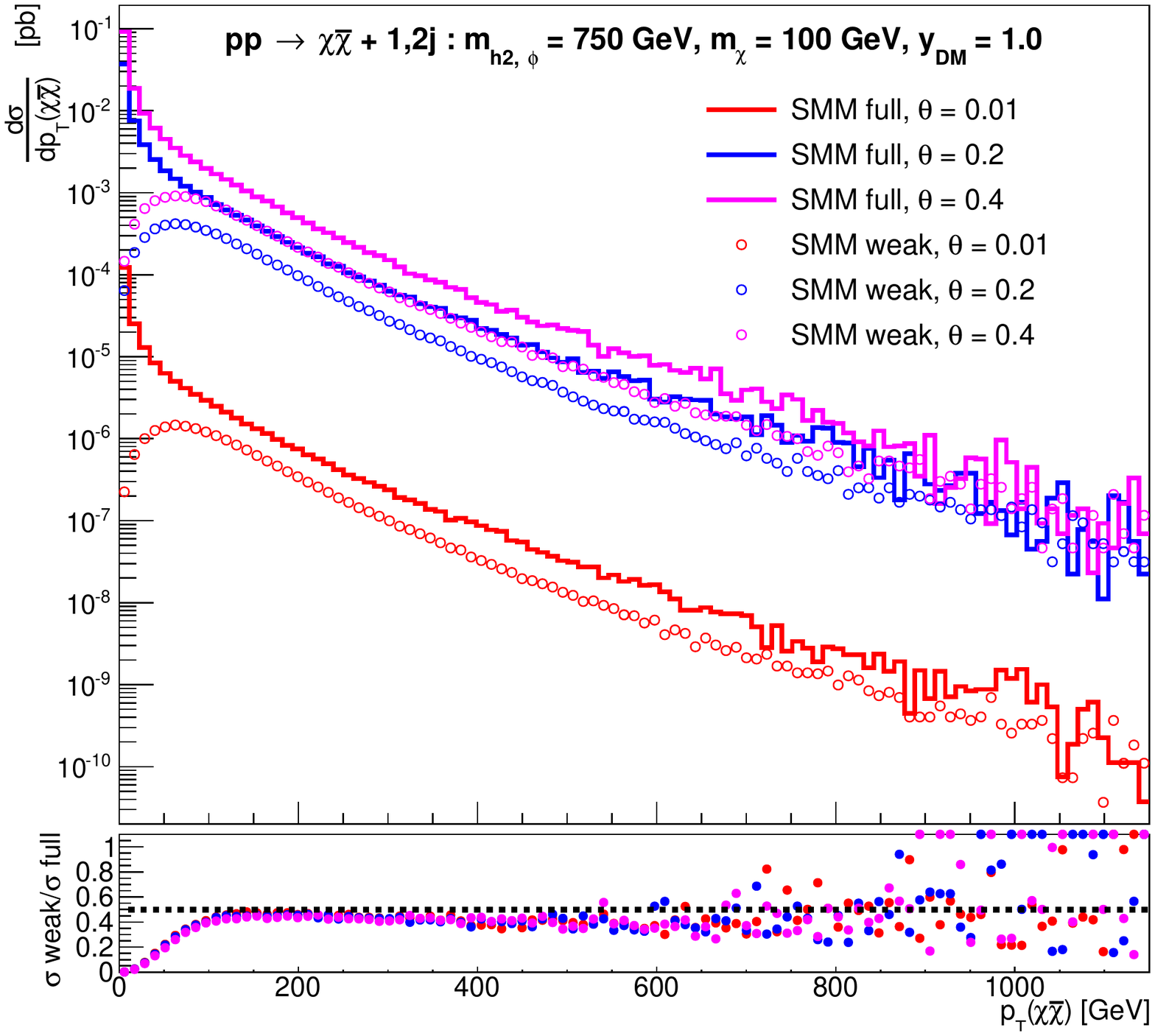}
        \includegraphics[width=0.49\textwidth]{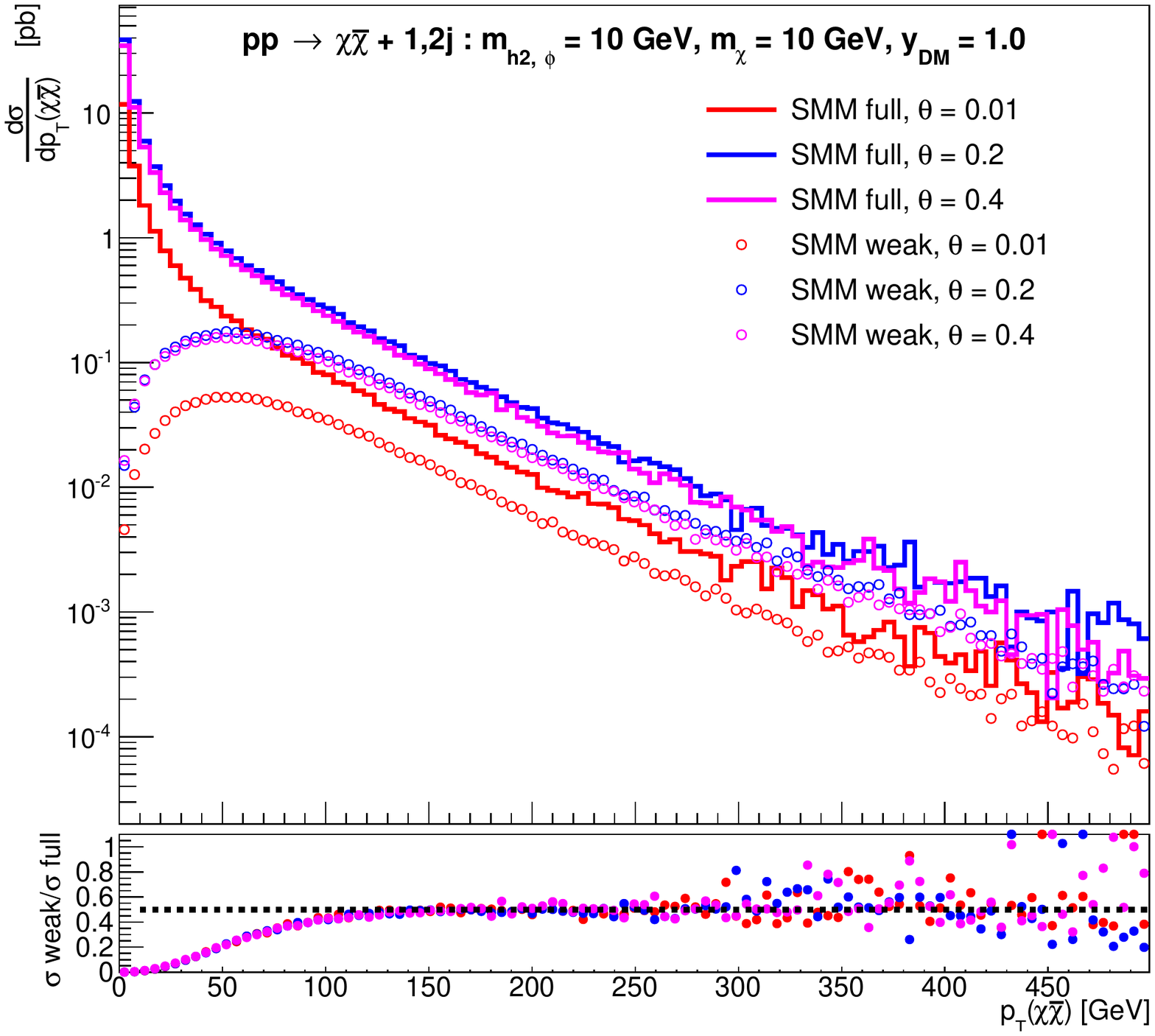}
        \caption{\it
        Comparison of total and VBM $\MET + {\rm jets}$ cross sections for scenarios~A and B: the full  differential cross sections for the SMM are shown as solid lines, while the weak contributions due to $W/Z$-associated and VBF production are indicated with open circles.  The results correspond to $\mhTwo=750 \, {\rm GeV}$,  $\mchi=100 \, {\rm GeV}$~(left) and $\mhTwo = 10 \, {\rm GeV}$,  $\mchi=10 \, {\rm GeV}$~(right).  All results use $\yDM=1.0$. 
        }   
        \label{fig:VBF}
\end{figure}

The effects of Higgs boson mixing were not considered in the LHC DMF monojet model, which includes mediator production via top quark loop diagrams only.  The SMM, on the other hand, also accounts for possible mediator production via $s$-channel or $t$-channel massive gauge boson exchange. The corresponding $W/Z$-associated and VBF-like topologies are shown in the first two panels of Fig.~\ref{fig:scalardiagrams}.  Figure~\ref{fig:VBF} compares the VBM $\MET + \rm{jets}$ cross section against the full  result for scenarios~A and~B. We observe that the VBM processes constitute an appreciable fraction of the total cross section already at  \ptchichi \ values of the order of two times the massive gauge boson masses.

\subsection{Relic density}\label{sec:relic} %
\begin{figure}[!htp]
        \centering
        \includegraphics[width=0.49\textwidth]{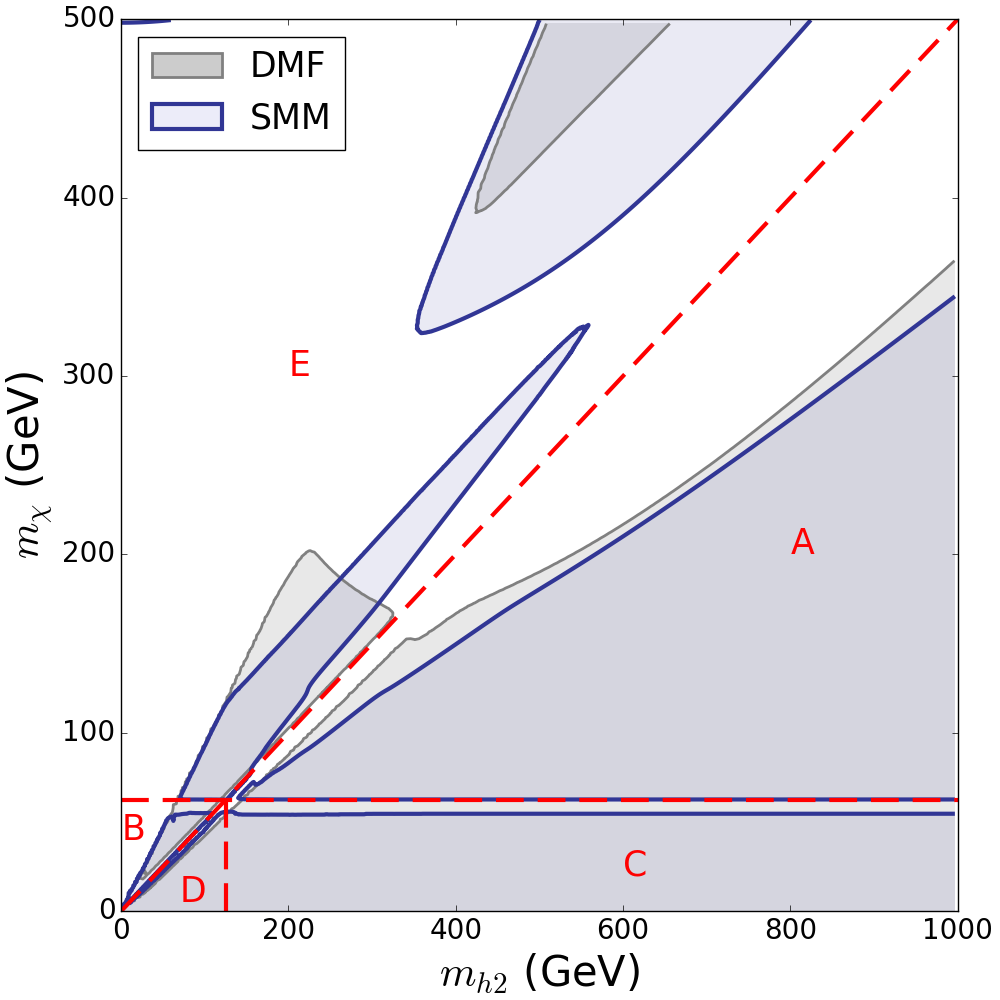}
        \includegraphics[width=0.49\textwidth]{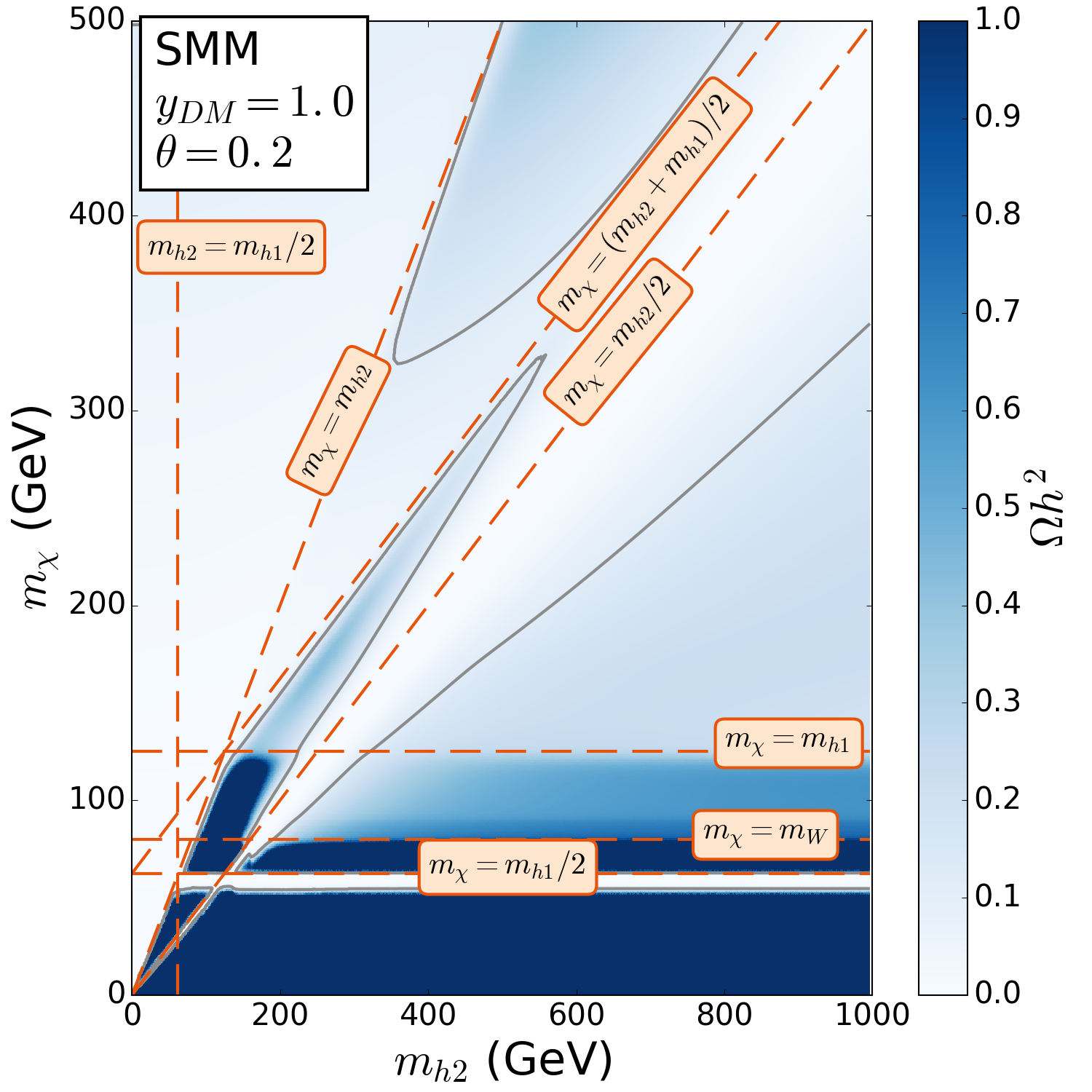}
        \caption{\it
        Left: comparison of the DM relic abundance obtained from the SMM and DMF models.  The solid lines indicate regions for which the calculated relic density matches the observation of $\Omega_{c}h^{2} = 0.12$.  The $m_{\chi}$--$\hspace{0.5mm} m_{h_2}$ mass hierarchies introduced in Section~\ref{scalar} are delineated with dashed red lines.  Right: the relic abundance for the SMM overlaid with phenomenologically relevant mass relations.  In both plots $\yDM=1.0$ and $\theta=0.2$ are used.}   
        \label{fig:relic}
\end{figure}

The DMF and SMM scalar models can be used to obtain the DM relic density in the universe with an assumption that there is only a single species of DM particle and that no mechanisms can generate/annihilate DM beyond those contained in the models.  We compute the relic density using {\sc MadDM} version 2.0.6~\cite{Backovic:2013dpa,Backovic:2015tpt}, which considers all $2 \rightarrow 2$ interactions between DM and SM particles.  The contours shown in the left panel of~Fig.~\ref{fig:relic}, which are estimated following the procedure described in~\cite{Pree:2016hwc}, correspond to the DMF and SMM model parameter spaces for which the computed relic density matches the $\Omega_{c}h^{2} = 0.12$ observation from the Planck collaboration~\cite{Ade:2015xua}.  Regions interior (exterior) to the contours are those in which the obtained relic density is over-abundant (under-abundant) with respect to observation.  Note that the color scale is truncated at 1.0; larger values of the relic density are indicated in the same shade of dark blue.  As before, the Yukawa coupling strength in the SMM and DMF models is set to 1.0.  The SMM results shown correspond to a mixing angle of $\theta = 0.2$.  The mass hierarchies of scenarios A to E are indicated in the left panel.  The right panel shows the relic abundance for the SMM model together with several relevant mass relations.

The plots show that the observed relic density can be obtained from both the DMF and SMM models over a wide range of parameter space.  Dashed lines are added to illustrate which processes contribute to enhanced annihilation along the corresponding relic density contours.  For example, the vertical line labeled $m_{\chi}=m_{h_1}/2$ corresponds to an enhancement of the $\chi \bar{\chi} \rightarrow h_{1}$ process.  Likewise, the line labeled $m_{\chi}=m_{h_2}$ ($m_{\chi}=(m_{h_2}+m_{h_1})/2$) corresponds to the enhancement of the $\chi \bar{\chi} \rightarrow h_{2} h_{2}$ ($\chi \bar{\chi} \rightarrow h_{2} h_{1}$). 

Perhaps the most obvious difference in the results obtained from the two models lies in the region near $m_{\chi} = m_{h_1}/2$.  This region is depleted in the SMM due to the resonant enhancement of DM annihilation to SM particles through the light $h_{1}$ mediator.  In region~E, the lower SMM contour departs from the line of $m_{\chi} = m_{h_2}$, which corresponds to $t$-channel $\chi\bar{\chi} \rightarrow h_{2}h_{2}$ annihilation, at a value of $m_{h_2} = m_{h_1}$.  For $m_{h_2} > m_{h_1}$, the SMM contour instead follows the line of $m_{\chi} = (m_{h_1} + m_{h_2})/2$,  corresponding to $\chi\bar{\chi} \rightarrow h_{1}h_{2}$.  The upper contour in region E also stems from $t$-channel $\chi\bar{\chi} \rightarrow h_{2}h_{2}$ annihilation.  This region is enlarged for the SMM because the coupling between the $h_{2}$ mediator and SM particles is relatively weaker (by a factor of $\sin^{2} \theta$) than the analogous coupling in the DMF model.  The relic density shown in the right panel of Fig.~\ref{fig:relic} indicates a series of steps in the DM abundance of region A at low $m_{\chi}$ that are not apparent for the DMF model.  These steps coincide with $m_{\chi} = m_{h_1},m_{W}$, and are due to the additional Higgs and VBM interactions present in the SMM.     

~\\

\noindent In summary, we have studied a simple extension of the LHC DMF scalar mediator model that incorporates mixing between the new scalar DM mediator field and the SM Higgs boson.  We have shown that in several cases mixing leads to kinematics and cross sections that significantly differ from those of the LHC DMF model.  We have also shown that the addition of $W/Z$-associated and VBF production processes leads to changed kinematics and cross sections relative to the those obtained with loop-only topologies.  The scalar mixing model also gives rise to several distinct features in the DM relic density distribution that do not appear in the LHC DMF analog.  Overall, our results reveal that simplified scalar models with Higgs boson mixing typically display a much richer phenomenology than the simple LHC DMF model.  The SMM introduced in~(\ref{eq:smh1h2}) represents the simplest extension of the LHC DMF model that includes Higgs boson mixing, and thus allows for a consistent comparison and combination of individual~$\MET$ channels such as monojet, mono-$V$, and~$\ttb + \MET$.

\newcommand{\GeV}{~GeV}
\newcommand{\METP}{$\slashed{E}_T$~}
\newcommand{\fb}{{\rm{fb}}}

\section{Connecting an LHC discovery of a mediator particle with DM signals}
\label{750}

In this Section we discuss the possibility of connecting a new physics signal in a channel visible at the LHC with DM using a simplified model with effective couplings. For definiteness we focus on a particular example consistent with the 750 GeV diphoton excess observed by the ATLAS and CMS experiments with the 2015 data sample. Although the 750 GeV excess did not survive in the 2016 data of ATLAS and CMS (see, e.g.,~\cite{Khachatryan:2016yec} ) this example still provides interesting insights on a potential strategy that could be followed in case of a signal. Our pragmatic Ansatz links such a potential collider signal to other experimental signatures, which can be used to verify/falsify a specific signal hypothesis and to study its underlying nature. With this approach it is possible to define discovery scenarios combining different signatures, which can be vital for guiding the experimental search programme in case of a discovery.

\subsection{Simplified DM model}
\label{750model}
In order to see how a direct-channel resonance such as the diphoton excess seen in 2015 by ATLAS and CMS~\cite{Aaboud:2016tru,Khachatryan:2016hje} could be linked to a stable DM candidate, one can augment the SM by a scalar $S$ or pseudoscalar $P$ particle with the mass of $750 \, {\rm GeV}$ and a Dirac fermion $\chi$.\footnote{The formalism we present is also directly applicable  to a Majorana DM particle.} The relavant interactions can be written in the scalar case as 
\begin{equation} \label{eq:LS}
{\cal L}_S  = g_{\chi} \hspace{0.25mm} S \hspace{0.25mm} \bar \chi \chi + \frac{\alpha_s}{4 \pi} \hspace{0.25mm} \frac{c_G}{\Lambda} \hspace{0.25mm} S \hspace{0.25mm} G_{\mu \nu}^a G^{a\, \mu \nu} + \frac{\alpha}{4 \pi s_w^2}\hspace{0.25mm}  \frac{c_W}{\Lambda} \hspace{0.25mm} S \hspace{0.25mm} W_{\mu \nu}^i W^{i \,  \mu \nu} + \frac{\alpha}{4 \pi c_w^2}\hspace{0.25mm}  \frac{c_B}{\Lambda} \hspace{0.25mm} S \hspace{0.25mm} B_{\mu \nu} B^{ \mu \nu}   \,.
\end{equation}
Here $G^a_{\mu \nu}$, $W_{\mu \nu}^i$, and~$B_{\mu \nu}$ are the $SU(3)_C$, $SU(2)_L$, and $U(1)_Y$ field strength tensors, $\alpha_s$ and~$\alpha$ denote the strong and electromagnetic coupling constants, and $s_w$ and $c_w$ are the sine and the cosine of the weak mixing angle. The scale that suppresses the higher-dimensional interactions that couple the mediator $S$ to gauge fields is denoted by $\Lambda$, while the Wilson coefficients $c_V$ with $V=G,\;W,\;B$ describe how strongly $S$ couples to $V_{\mu \nu} V^{\mu \nu}$. The corresponding Lagrangian~${\cal L}_P$ for the pseudoscalar case is obtained from (\ref{eq:LS}) by replacing $\bar \chi \chi$ by $\bar \chi \hspace{0.25mm} i \gamma_5 \hspace{0.25mm} \chi$, $V_{\mu \nu} V^{\mu \nu}$ by $V_{\mu \nu} \tilde V^{\mu \nu}$ with $\tilde V_{\mu \nu} = \frac{1}{2} \, \epsilon_{\mu \nu \alpha \beta} \hspace{0.25mm} V^{\alpha \beta}$ and, finally, $g_\chi$ by $\tilde g_\chi$ and $c_V$ by $\tilde c_V$. The DM phenomenology of such simplified models ${\cal L}_{S,P}$ has recently been studied in~\cite{Backovic:2015fnp,Barducci:2015gtd,Mambrini:2015wyu,D'Eramo:2016mgv,Backovic:2016gsf}.

The production cross section and the rates for the decays of the resonance $S$ ($P$)  into SM and DM  particles can all be expressed in terms of the Wilson coefficients $c_V$ ($\tilde c_V$), the scale $\Lambda$, the DM coupling $g_\chi$ ($\tilde g_\chi$) and the DM mass $m_\chi$. Assuming that the production of a new spin-0 state is dominated by gluon-gluon fusion and considering
for definiteness the excess reported in~\cite{Aaboud:2016tru,Khachatryan:2016hje} and  one obtains \cite{Bauer:2016lbe}
\begin{equation} \label{eq:xsecS}
\sigma_{8 \, {\rm TeV}} \left ( p p \to S \right ) \simeq 46.7 \, {\rm fb} \left ( \frac{c_G \, {\rm TeV}}{\Lambda} \right )^2 \,, \quad \sigma_{13 \, {\rm TeV}}  \left ( p p \to S \right ) \simeq 208 \, {\rm fb} \left ( \frac{c_G \, {\rm TeV}}{\Lambda} \right )^2  \,, 
\end{equation}
for the total cross section at $\sqrt{s} = 8$ and 13~TeV, respectively. These results  hold to first approximation also for a pseudoscalar~$P$ after obvious replacements. 

The partial decay rates of such a scalar resonance into pairs of vector bosons and DM particles
can be written as~\footnote{The expressions for the partial widths to $\gamma Z$, $ZZ$ and $WW$ are only approximations that hold in the limit of vanishing $W$ and $Z$ boson masses, as is appropriate for any heavy spin-0 state. They reproduce the full results (see~e.g.~\cite{Bauer:2016lbe}), which will be used in the numerical analysis, to better than 10\%.}:
\begin{equation} \label{eq:Gammas}
\begin{split}
\Gamma \left (S \to gg \right ) & = \frac{\alpha_s^2 \hspace{0.25mm} M_S^3}{8 \pi^3} \hspace{0.5mm} K_G \left ( \frac{c_G}{\Lambda}  \right )^2 \simeq 1.93 \cdot 10^{4} \, {\rm GeV}^3 \left ( \frac{c_G}{\Lambda}  \right )^2 \,, \\[2mm]
\Gamma \left (S \to \gamma \gamma \right ) & = \frac{\alpha^2 \hspace{0.25mm}  M_S^3}{64 \pi^3} \left ( \frac{c_W}{\Lambda}  +  \frac{c_B}{\Lambda}  \right )^2 \simeq 11.3 \, {\rm GeV}^3 \left ( \frac{c_W}{\Lambda}  +  \frac{c_B}{\Lambda}  \right )^2 \,, \\[2mm]
\Gamma \left (S \to \gamma Z \right ) & \simeq \frac{\alpha^2 \hspace{0.25mm} M_S^3}{32 \pi^3} \left (  \frac{c_w}{s_w} \frac{c_W}{\Lambda}  -  \frac{s_w}{c_w} \frac{c_B}{\Lambda}   \right )^2 \simeq 26 \, {\rm GeV}^3 \left (  1.82 \, \frac{c_W}{\Lambda}  - 0.55 \, \frac{c_B}{\Lambda}   \right )^2 \,, \\[2mm]
\Gamma \left (S \to Z Z \right ) & \simeq \frac{\alpha^2   \hspace{0.25mm} M_S^3}{64 \pi^3} \left (  \frac{c_w^2}{s_w^2} \frac{c_W}{\Lambda}  + \frac{s_w^2}{c_w^2} \frac{c_B}{\Lambda}   \right )^2 \simeq 13 \, {\rm GeV}^3 \left (  3.32 \, \frac{c_W}{\Lambda}  + 0.30 \, \frac{c_B}{\Lambda}   \right )^2  \,, \\[2mm]
\Gamma \left (S \to W W \right ) & \simeq \frac{\alpha^2  \hspace{0.25mm} M_S^3}{32 \pi^3 \hspace{0.25mm} s_w^4} \left (  \frac{c_W}{\Lambda}     \right )^2 \simeq 485 \, {\rm GeV}^3 \left (  \frac{c_W}{\Lambda}  \right )^2  \,, \\[2mm]
\Gamma \left (S \to \chi \bar \chi \right ) & =  \frac{g_\chi^2 \hspace{0.25mm} M_S}{8 \pi} \left ( 1 - \frac{4 m_\chi^2}{M_S^2} \right )^{3/2} \simeq 29.8 \, {\rm GeV} \, g_\chi^2 \,.
\end{split}
\end{equation}
These numbers are simple to rescale for any other possible new spin-0 state, assumin that couplings of the new resonance to SM quarks are negligible. For definiteness in our numerical results we use $K_G = 1.348$~\cite{Bauer:2016lbe}, $\alpha_s= 0.092$, $\alpha = 1/137.04$ for the diphoton decay and $\alpha = 1/127.94$ otherwise, and $s_w^2 = 0.2313$. In the case of the invisible decay width, we set the DM particle mass to zero, as would be a good approximation for any heavy spin-0 state decaying into light DM particles. After replacing $c_V$ by $\tilde c_V$, the above results for the partial decay widths of $S$ to gauge bosons also apply to the case of a pseudoscalar, while to obtain the invisible decay rate of $P$ one has to change the exponent $3/2$ appearing in $\Gamma \left (S \to \chi \bar \chi \right )$ with $1/2$. 

Looking at the expressions in (\ref{eq:Gammas}), one observes that new physics scenarios that lead to~$c_W \ll c_B$  are generically less constrained than models that predict $c_W \gg c_B$, because in the former case the decays to $\gamma Z$ and $ZZ$ are suppressed by a  factor $\left (s_w/c_w \right ) ^4 \simeq 0.1$ and    $\left ( s_w/c_w \right ) ^8  \simeq 0.01$, respectively, and decays to $WW$ are absent in the limit of~$c_W$ going to zero. In the following we will focus on the model realisations with~$c_W = 0$ and $c_B \neq 0$ (or $\tilde c_W = 0$ and $\tilde c_B \neq 0$, thereby
avoiding constraints on the simplified model~(\ref{eq:LS}) arising from resonance searches in the $\gamma Z$, $ZZ$, and $WW$ channels~\footnote{We do not examine scenarios with $c_W \simeq (s_w/c_w)^2 \hspace{0.25mm} c_B$ or $c_W \simeq -(s_w/c_w)^4 \hspace{0.25mm} c_B$, which would evade constraints  from $\gamma Z$ or $Z Z$ resonance searches by tuning $c_W$ and $c_B$.}.

In the narrow-width approximation, the signal strength for the process $pp \to XY$ with $XY = \{ gg, \,\gamma\gamma, \,\gamma Z, \,ZZ,$ $WW\}$  factorises into the product of the total production cross section and the relevant branching fraction
\begin{equation} \label{eq:muSM}
\mu_{\sqrt{s}} \left (pp \to XY  \right ) = \sigma_{\sqrt{s}} \left (pp \to S \right ) \, {\rm Br} \left ( S \to X Y \right ) \,,
\end{equation}
and a similar factorisation also applies in the pseudoscalar case. 

\subsection{Monojet signatures}
\label{sec:monojet}

Since the couplings of the mediators to gluon pairs are implemented by means of effective operators $\big($see (\ref{eq:LS})$\big)$,  the factorisation of the signal strength (\ref{eq:muSM}) is expected also to apply to the case of a monojet signature for any spin-0 state in the general class considered here. This means, in particular, that varying the coupling~$c_G$~($\tilde c_G$) should only result in an overall rescaling of the total $pp \to \MET + j$ cross section, but should leave the shape of all kinematic distributions unchanged. To validate the extent to which the kinematic distributions can be affected by the detector effects relevant for modern searches for DM  at the LHC \cite{Aad:2015zva,Aaboud:2016tnv,Khachatryan:2014rra,CMS-PAS-EXO-12-055,CMS-PAS-EXO-16-013}, we simulate the~$\MET + 0,1,2$ jet  spectra resulting from the model (\ref{eq:LS}) using {\tt MadGraph5\_aMC@NLO}~\cite{Alwall:2014hca} with  the MLM merging scheme~\cite{MLM}, {\tt FastJet}~\cite{Cacciari:2011ma}, and {\tt PYTHIA~8}~\cite{Sjostrand:2007gs}.  The modelling of the experimental resolution for $\MET$ and the recoiling system is done by using resolutions typical of the ATLAS and CMS detectors. 

\begin{figure}[!t]
\begin{center}
\includegraphics[width=0.65\textwidth]{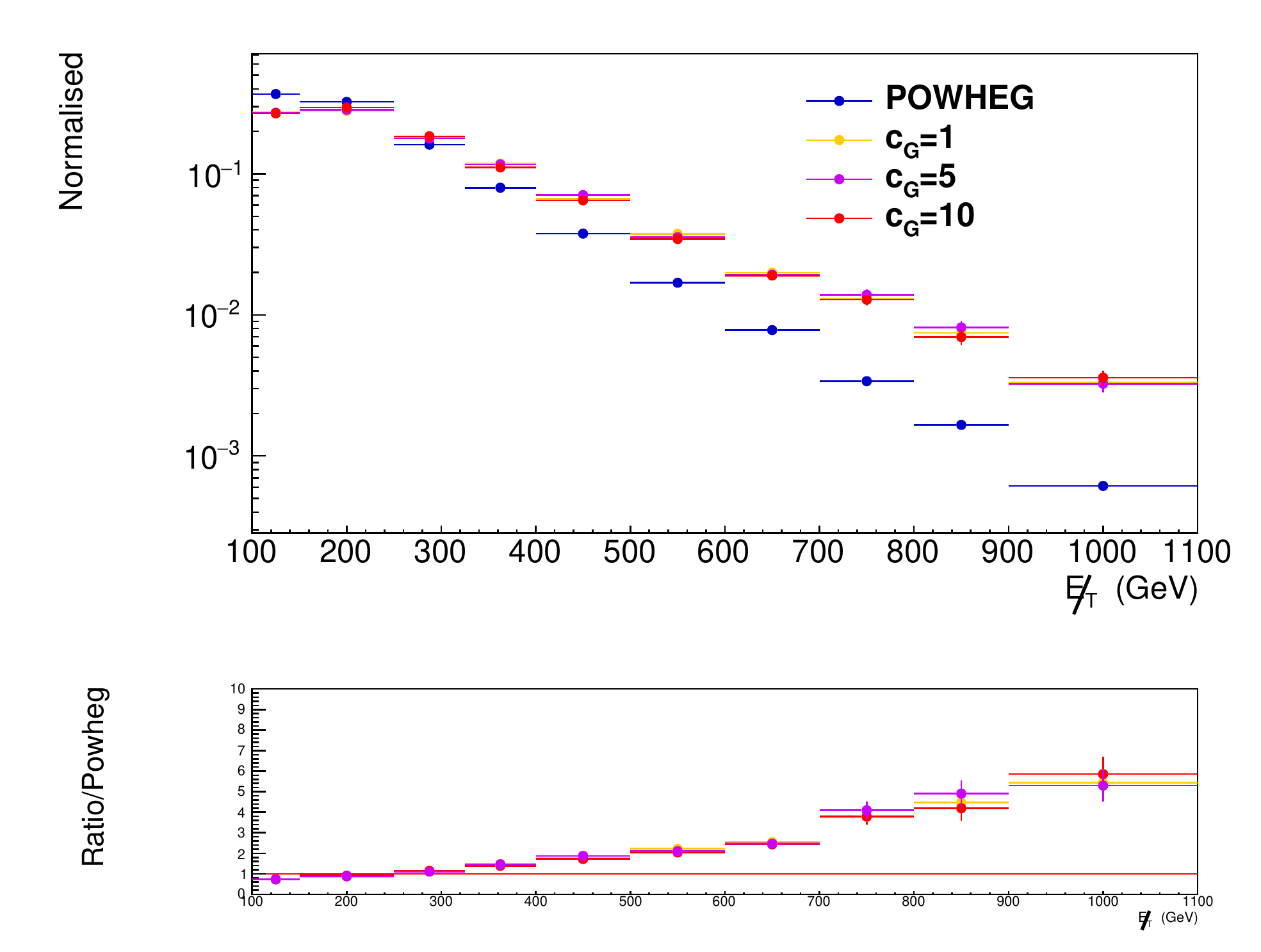} 
\caption{\it Normalised detector-level $\MET$ distributions for different values of $c_G$. For comparison, we show also the normalised $\MET$ distribution predicted in the scalar-mediated DM model recommended by the LHC DMF~\protect\cite{Abercrombie:2015wmb}. The latter predictions have been obtained by the ${\tt POWHEG}$ implementation~\protect\cite{Haisch:2013ata}.}
\label{fig:kine}
\end{center}
\end{figure} 

Fig.~\ref{fig:kine} shows the normalised $\MET$ distributions predicted in the model (\ref{eq:LS}) for different values of the coupling $c_G$. Within the detector resolution  no significant variations in the~$\MET$ shapes are observed, which  implies that the signal strength  $\mu_{\sqrt{s}} \left ( pp \to \MET + j \right )$ for a given monojet signal region is proportional to the total production cross section of the mediator times its invisible branching fraction. For instance, for the recent CMS monojet search~\cite{CMS-PAS-EXO-16-013}, one finds 
\begin{equation}  \label{eq:mumono}
\mu_{13 \, {\rm TeV}} \left (pp \to \MET + j  \right ) \simeq 3.2 \cdot 10^{-2} \; \sigma_{13 \, {\rm TeV}} (pp \to S) \, {\rm Br} \left ( S \to \chi \bar \chi \right ) \,.
\end{equation}
The same expression also holds in the pseudoscalar case, since the $\MET$ spectrum is  insensitive to the mediator type. 

It is also illustrative to compare the normalised $\MET$ shapes resulting from ${\cal L}_S$ with the spectra predicted in the LHC DMF model. Such a comparison is presented in Fig.~\ref{fig:kine} as well.  One observes that  the effective interactions present in (\ref{eq:LS})  lead to a significantly harder spectrum than the top quark loop diagrams (see the third panel in Fig.~\ref{fig:scalardiagrams}) that produce the $\MET + j$ events in the LHC DMF model.  Numerically, the observed suppression amounts to a factor of around 7  for $\MET \simeq 1 \, {\rm TeV}$. This  is an expected feature, because high-energy jet radiation is able to resolve the structure of the top quark loops~\cite{Haisch:2012kf,Haisch:2013fla,Buckley:2014fba,Harris:2014hga,Haisch:2015ioa}, while the production mechanism cannot be resolved in the model~(\ref{eq:LS}) where the coupling of $S$ (or $P$) to gluons  is implemented through  a dimension-five operator. 

\subsection{LHC constraints}
\label{sec:LHCconstraints}

\begin{table}[!t]
\begin{center}
\caption{\label{tab:mu} \it The 95\% confidence level  (CL)  upper bounds on the relevant signal strengths arising from different  LHC Run 1 and 2 searches.}
\vspace{5mm}
\begin{tabular}{ccccc}
\hline $gg$ (8 TeV) & $\gamma Z$ (8 TeV) & $Z Z$ (8 TeV) & $W W$ (8 TeV) & $\MET + j$ (13 TeV) \\ \hline 
 $<2.5 \, {\rm pb}$ \cite{Khachatryan:2016ecr} & $<4 \, {\rm fb}$ \cite{Aad:2014fha} & $<12 \, {\rm fb}$ \cite{Aad:2015kna} & $<40 \, {\rm fb}$ \cite{Aad:2015agg} & $<14 \, {\rm fb}$ \cite{CMS-PAS-EXO-16-013} \\ \hline
\end{tabular}
\vspace{0mm}
\end{center}
\end{table}

\begin{figure}[t!]
\begin{center}
\includegraphics[width=0.975\textwidth]{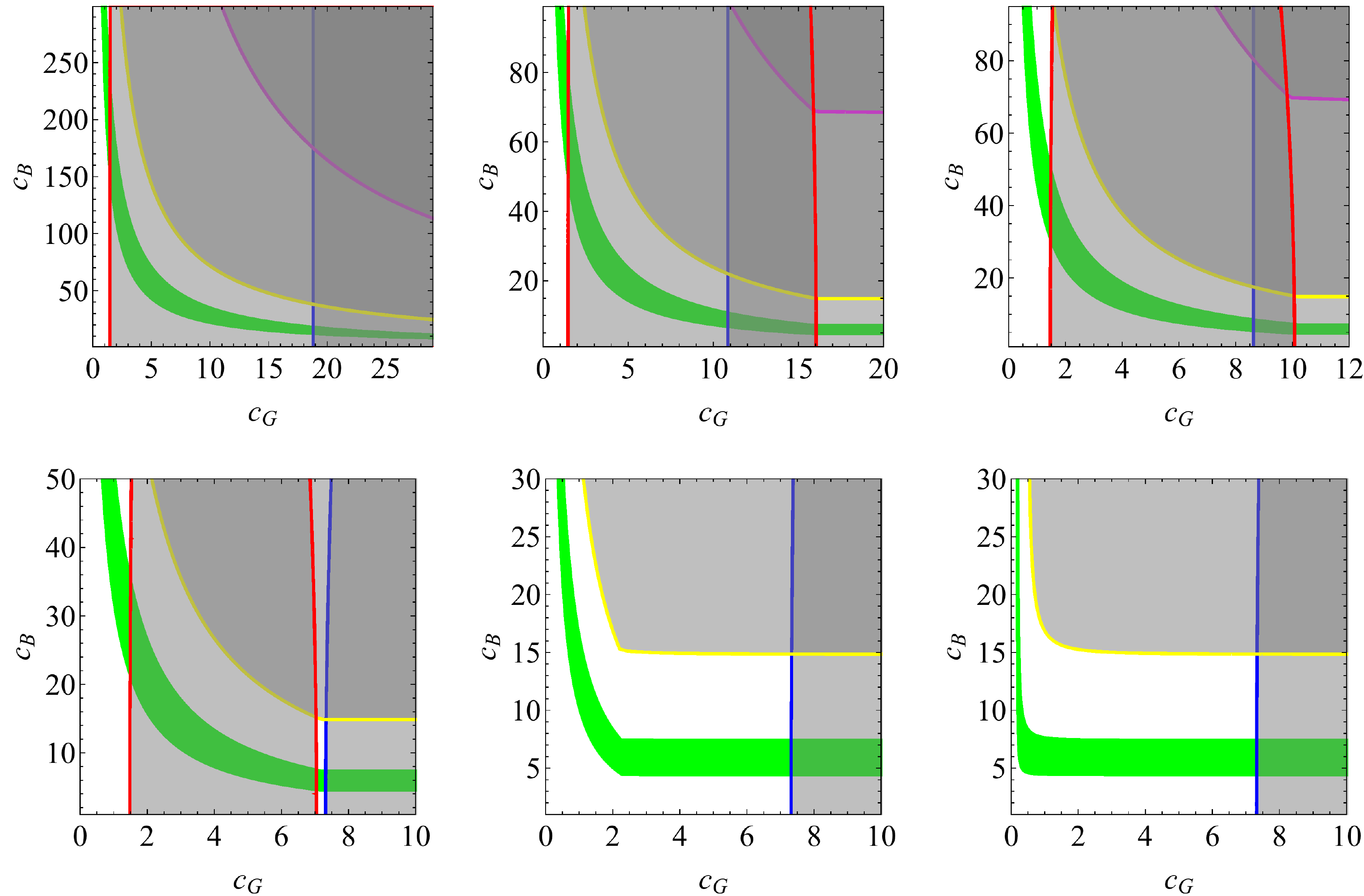}
\caption{\it \label{fig:pseudoB} Fit to the LHC diphoton excess in the $c_G \hspace{0.25mm}$--$\hspace{0.25mm}  c_B$ plane for $\Gamma_S = 45 \, {\rm GeV}$, $5 \, {\rm GeV}$, $2 \, {\rm GeV}$, $1 \, {\rm GeV}$, $0.1 \, {\rm GeV}$ and $g_\chi = 0$ from upper left to lower right. The 95\% CL regions favoured by the reported diphoton excess are shaded  green, while bounds from Run 1 and 2 data are shown as contour lines coloured blue for dijets, yellow for $\gamma Z$, magenta for $ZZ$, and red for monojets. There are no bounds from the $WW$ final states. The grey shaded areas are excluded at the 95\% CL. }
\end{center}
\end{figure}

For the purposes of our subsequent illustration of the interplay between collider and astrophysical constraints,
we first explore for which parameters the simplified model (\ref{eq:LS}) could have explained the putative diphoton excess 
reported in the 2015 data~\cite{Buttazzo:2015txu}:
\begin{equation} \label{eq:mugammagamma}
\mu_{13 \, {\rm TeV}} \left (pp \to \gamma \gamma \right ) = (4.6 \pm 1.2) \, {\rm fb} \,,
\end{equation}
while at the same time respecting existing bounds  from dijet, diboson, and monojet searches. The bounds are collected in Tab.~\ref{tab:mu}. Notice that dijet production arises in the context of (\ref{eq:LS}) to first order only from the process $pp \to S/P \to gg$. In order to suppress contributions to the diboson channels, we study the scalar scenario with $c_W = 0$ and~$c_B \neq 0$. After setting~$M_S = 750 \, {\rm GeV}$ and $\Lambda = 1 \, {\rm TeV}$, the full phenomenology in the simplified model can thus be characterised by the four parameters $c_G$, $c_B$, $g_\chi$,  and $m_\chi$. In fact, one can trade the two parameters $g_\chi$  and $m_\chi$ for the total decay width $\Gamma_S$ by correctly adjusting the DM coupling~$g_\chi$ for any choice of $c_G$, $c_B$, and $m_\chi$. If this is done, one can derive the constraints in the $c_G \hspace{0.25mm}$--$\hspace{0.25mm} c_B$ plane for different values of $\Gamma_S$. The outcome of this exercise is depicted in the six panels of Fig.~\ref{fig:pseudoB}. We note that very similar plots would be obtained for a pseudoscalar scenario with $\tilde c_W = 0$ and~$\tilde c_B \neq 0$.

For the width $\Gamma_S = 45 \, {\rm GeV}$ preferred by the ATLAS 2015 data (upper left panel), one observes that monojet searches severely constrain the region in the $c_G \hspace{0.25mm}$--$\hspace{0.25mm}  c_B$ plane in which the diphoton excess can be explained.\footnote{From the discussion in Section~\ref{sec:monojet} it should be clear that the strength of the monojet constraints is partly due to that fact that the production of $S,P$ proceeds via higher-dimensional operators. Milder bounds are expected to apply to any weakly coupled model with an ultraviolet completion of (\ref{eq:LS}) at a low scale~$\Lambda$.}  This observation has also been made in~\cite{Backovic:2015fnp,Barducci:2015gtd,Mambrini:2015wyu,D'Eramo:2016mgv,Backovic:2016gsf}. In fact, the allowed values of $c_G \simeq 1$ and $c_B \simeq 200$ translate into the following effective digluon and diphoton couplings, 
\begin{equation} \label{eq:Cga}
C_g = \frac{\alpha_s}{4\pi} \, \frac{c_G}{\Lambda} \simeq \frac{0.007}{{\rm TeV}}  \,, \qquad C_\gamma = \frac{\alpha}{4\pi} \, \frac{c_B}{\Lambda} \simeq  \frac{0.12}{{\rm TeV}}  \,.
\end{equation}
The effective spin-0 mediator coupling to gluons  is hence of  similar size to the effective SM Higgs digluon coupling, while the $S$ field interacts with photons  20 times more strongly than the Higgs. For smaller total widths of $\Gamma_S = 5 \, {\rm GeV}$ and $2 \, {\rm GeV}$  (upper middle and right panel) the constraints on the Wilson coefficient $c_B$ become weaker by a factor of 3 and 5, while the minimal allowed value of $c_G$  remains basically the same. Reducing the total width   to $\Gamma_S = 1 \, {\rm GeV}$ (lower left panel), one sees that the regions disfavoured by the monojet and dijet searches do not overlap anymore, allowing for an explanation of the diphoton excess with $c_G \simeq  c_B \simeq 7$. For  $\Gamma_S = 0.1 \, {\rm GeV}$~(lower middle panel) monojet searches do  not provide a direct constraint on $c_G$ and $c_B$ any more, but invisible decays $S \to \chi \bar \chi$ indirectly still play a role compared to the case of $g_\chi =0$~(lower right panel), since significantly larger values of $c_B$ are needed for  $g_\chi \neq 0$ to fit the diphoton excess if~$c_G \lesssim 2$. 

\subsection{Direct detection}
\label{sec:DD}

Direct detection experiments can also be used to constrain the generic scalar model ${\cal L}_S$, since it leads to a spin-independent~(SI) DM-nucleon scattering cross section, but not the pseudoscalar scenario~${\cal L}_P$, because it predicts spin-dependent and momentum-suppressed rates.  After the mediator $S$ has been integrated out, the interactions~(\ref{eq:LS}) induce couplings between DM and gluons, photons, as well as EW gauge bosons.  If the gluon coupling is non-vanishing at the scale~$M_S$, the couplings to photons and EW gauge bosons can be shown to provide a subleading contribution to direct detection rates~\cite{Frandsen:2012db,Crivellin:2014gpa,D'Eramo:2016mgv}. To keep the discussion simple, we ignore such effects and include only QCD corrections. The SI DM-nucleon scattering cross section takes the form
\begin{equation} \label{eq:SI}
(\sigma_{\rm SI}^N)_S \simeq \frac{g_\chi^2 \hspace{0.5mm} \mu_{N\chi}^2 \hspace{0.25mm}   m_N^2 \hspace{0.25mm}   f^2(c_G)}{\pi} \,,
\end{equation}
where $\mu_{N\chi} = m_N \hspace{0.25mm} m_\chi/(m_N + m_\chi)$ is the DM-nucleon reduced mass with $m_N \simeq 0.939 \, {\rm GeV}$ the nucleon mass. The mediator-nucleon coupling $f (c_G)$ reads (see, e.g.~\cite{D'Eramo:2016mgv,Frandsen:2012db,Vecchi:2013iza,Hill:2014yxa})
\begin{equation} \label{eq:fcg}
f (c_G) \simeq \frac{c_G}{4 \pi \hspace{0.1mm} \Lambda \hspace{0.1mm} M_S^2} \left  ( 6 \alpha_s f_q^N + \frac{8 \pi}{9} f_{\rm TG}^N \right  ) \simeq  \frac{2 \hspace{0.25mm} c_G  \hspace{0.1mm}  f_{\rm TG}^N }{9 \Lambda \hspace{0.1mm} M_S^2} \,.
\end{equation}
Notice that to first approximation the coupling $f(c_G)$ only depends on the gluonic component of the nucleon,~i.e.~$f^{N}_{\rm TG}=1-\sum_{q=u,d,s }f_{q}^{N} \simeq 0.894$ \cite{Junnarkar:2013ac,Hoferichter:2015dsa}, while contributions from the light-quark form factors~$f_{q}^{N}$ are suppressed by an additional power of $\alpha_s$. 

\begin{figure}[!t]
\begin{center}
\includegraphics[height=0.3\textwidth]{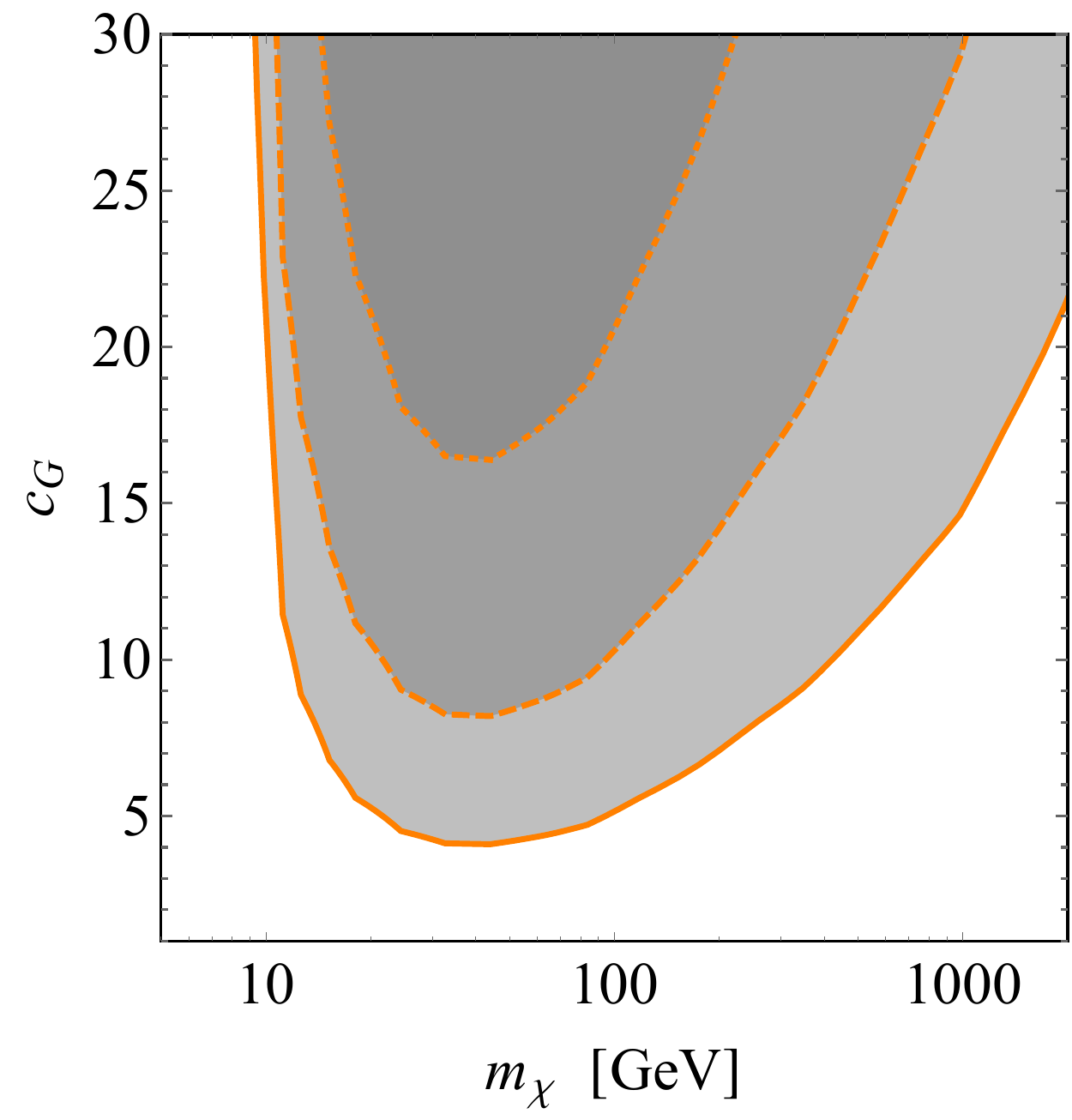}   \quad 
\includegraphics[height=0.3\textwidth]{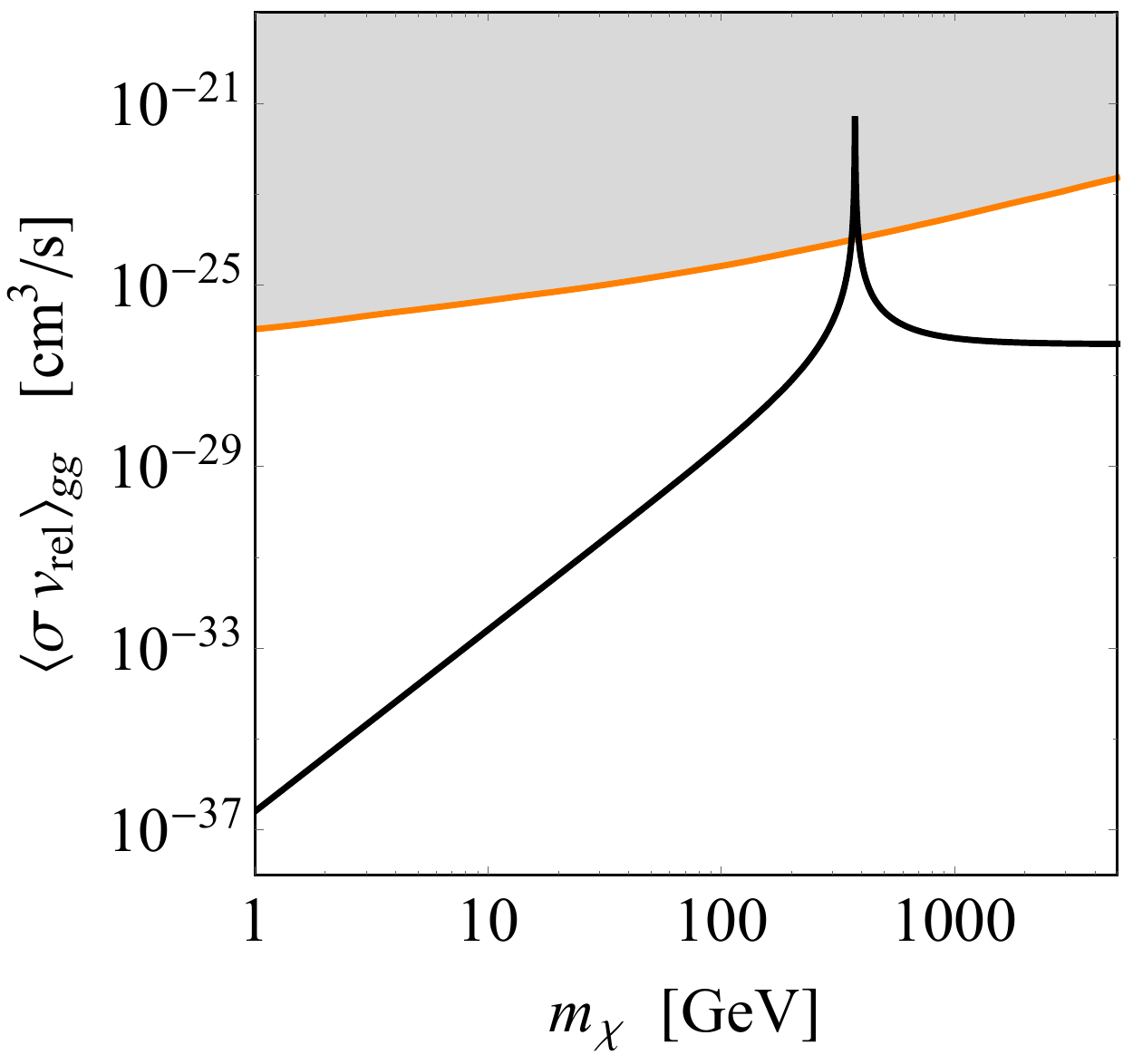}   \quad 
\includegraphics[height=0.3\textwidth]{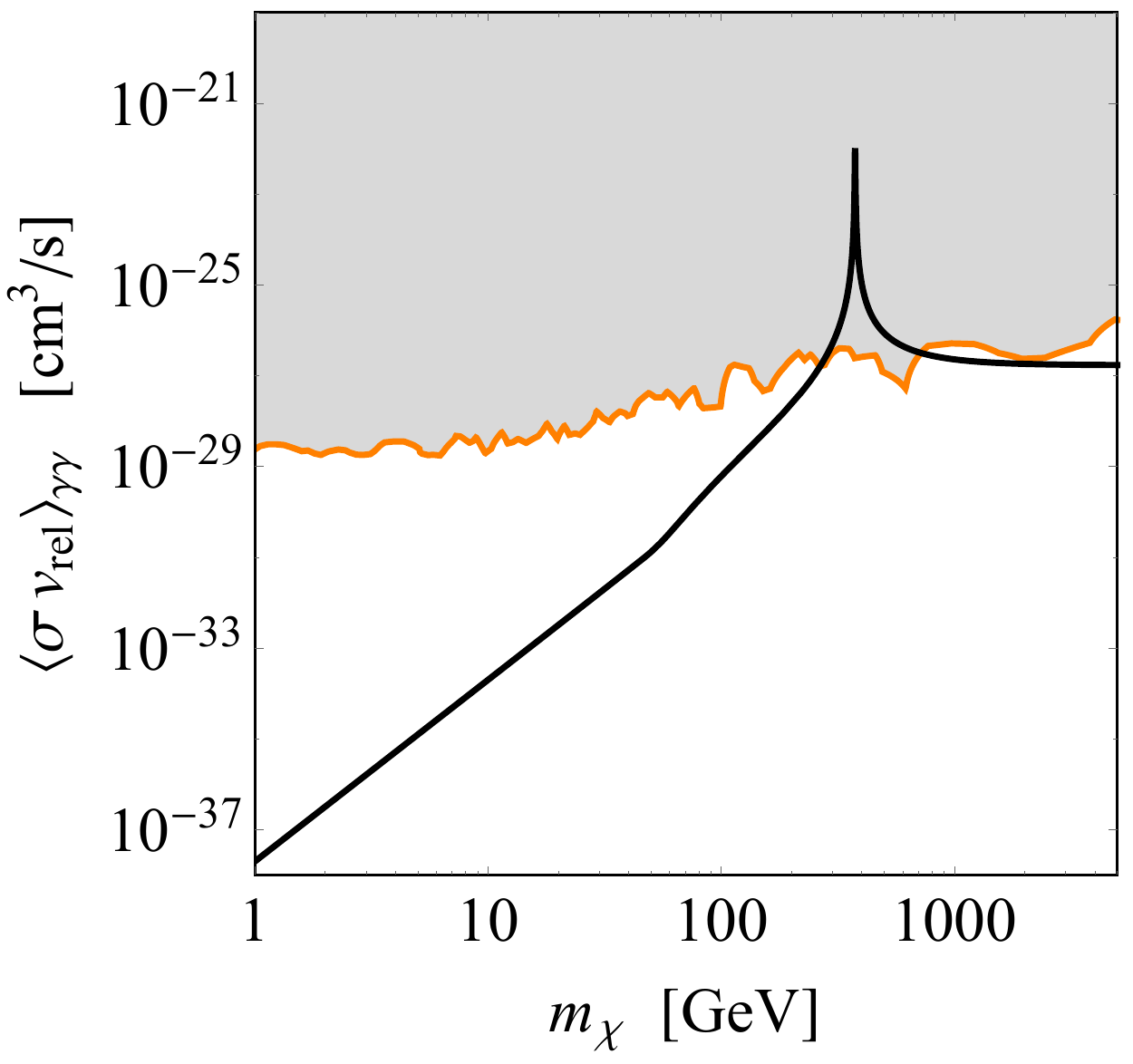}   
\caption{\it Left: Constraints in the $m_\chi$--$\hspace{0.5mm} c_G$ plane arising from the LUX bound~\cite{Akerib:2015rjg}  on $\sigma_{\rm SI}^N$ for  $g_\chi = 0.5$ (dotted orange), $g_\chi = 1$ (dashed orange), and $g_\chi = 2$~(solid orange). The grey shaded regions are excluded at a 90\% CL. Middle: DM annihilation rate into $gg$ (black line) for $\tilde g_\chi = 1$, $\tilde c_G = 5$, $\tilde c_W = \tilde c_B = 0$ and $\Lambda = 1 \, {\rm TeV}$. The orange curve indicates the corresponding 95\% CL bound from~\cite{Ackermann:2015zua} and the region shaded  grey is excluded.  Right: DM annihilation rate into $\gamma \gamma$ (black line) for $\tilde g_\chi = 1$, $\tilde c_B = 50$, $\tilde c_G = \tilde c_W = 0$ and $\Lambda = 1 \, {\rm TeV}$. For comparison the 95\% CL bound from \cite{Ackermann:2015lka} is indicated employing the same colour scheme as in the middle panel. }
\label{fig:DDID}
\end{center}
\end{figure}

Combining~(\ref{eq:SI}) and (\ref{eq:fcg}), one finds that for $M_S = 750 \, {\rm GeV}$ prototype axample the size of the SI cross section is
\begin{equation}
(\sigma_{\rm SI}^N)_S \simeq 1.4 \cdot 10^{-47} \, {\rm cm}^2  \, g_\chi^2 \, c_G^2 \, \left ( \frac{1 \, {\rm TeV}}{\Lambda} \right )^2 \left ( \frac{\mu_{N\chi}}{1 \, {\rm GeV}} \right )^2 \,.
\end{equation}
Using this expression one can derive the region in the~$m_\chi$--$\hspace{0.5mm} c_G$ plane that is disfavoured by direct detection experiments. The left panel in Fig.~\ref{fig:DDID} shows the resulting constraints for three different values of $g_\chi$, employing the recent LUX bound~\cite{Akerib:2015rjg}\footnote{In this study we use the LUX result as an example to illustrate the impact from direct detection experiments. As outlined in detail in the the LHC DM WG recommendation~\cite{Boveia:2016mrp}, also other experiments constrain this parameter space. The PandaX-II experiment~\cite{Tan:2016zwf} possess similar sensitivity than LUX and together they provide the strongest limits for DM-neutron scattering cross sections. For DM-proton scattering cross sections the strongest limits are from the PICO collaboration~\cite{Amole:2016pye,Amole:2015pla}, while for DM particles lighter than $\mathcal{O}(10 \, \mathrm{GeV})$, solid-state cryogenic detectors as used by the SuperCDMS~\cite{Agnese:2015nto} and CRESST-II~\cite{Angloher:2015ewa} collaborations are more constraining than xenon experiments as their energy threshold is lower. The IceCube~\cite{Aartsen:2016exj} and Super-Kamiokande~\cite{Choi:2015ara} neutrino observatories are also able to provide constrains.}  
One sees that depending on whether $g_\chi $ is $0.5, 1$, or $2$, current direct detection experiments can exclude $c_G$ values larger than  around~16, 8, or 4 for DM masses around $40 \, {\rm GeV}$. For smaller and larger DM masses the LUX constraints soften and exclude only values of $c_G$ that are typically in conflict with dijet bounds (see Fig.~\ref{fig:pseudoB}).  Below we will combine the above direct detection constraint with the LHC bounds for a scalar benchmark model. 

\subsection{Indirect detection}

In contrast to direct detection, indirect detection is only relevant for the case of a pseudoscalar mediator, since DM annihilations mediated by scalar exchange are $p$-wave suppressed. Constraints on the couplings of the pseudoscalar mediator arise from $\gamma$ ray line searches~\cite{Abramowski:2013ax,Ackermann:2013uma,Ackermann:2015lka} as well as continuum limits  from observations of dwarf spheroidal galaxies~\cite{Abramowski:2014tra,Ackermann:2015zua}. 

The velocity-averaged DM annihilation rates relevant for the following discussion are given in terms of the couplings $\tilde c_G$, $\tilde c_W$, and $\tilde c_B$ by  
\begin{equation}
\begin{split}
& \hspace{5mm} \langle \sigma ( \chi \bar \chi \to gg ) \hspace{0.25mm} v_{\rm rel} \rangle \simeq \frac{\alpha_s^2 \hspace{0.25mm} g_\chi^2 \hspace{0.5mm} \tilde c_G^2}{\pi^3 \hspace{0.25mm} \Lambda^2}  \, \frac{m_\chi^4}{\left (4 m_\chi^2 - M_P^2 \right )^2 + \Gamma_P^2 M_P^2} \,, \\[2mm]
&  \langle  \sigma ( \chi \bar \chi \to \gamma \gamma ) \hspace{0.25mm} v_{\rm rel} \rangle \simeq \frac{\alpha^2 \hspace{0.25mm} g_\chi^2 \left (\tilde c_W + \tilde c_B \right)^2}{8 \pi^3 \hspace{0.25mm} \Lambda^2}  \, \frac{m_\chi^4}{\left (4 m_\chi^2 - M_P^2 \right )^2 + \Gamma_P^2 M_P^2} \,, \\[2mm]
& \hspace{-5mm} \langle \sigma ( \chi \bar \chi \to \gamma Z ) \hspace{0.25mm} v_{\rm rel} \rangle \simeq \frac{\alpha^2 \hspace{0.25mm} g_\chi^2 \left ( \displaystyle \frac{c_w}{s_w} \, \tilde c_W  - \displaystyle \frac{s_w}{c_w} \, \tilde c_B \right)^2}{4 \pi^3 \hspace{0.25mm} \Lambda^2} \,  \frac{m_\chi^4 \left ( 1 -  \displaystyle \frac{M_Z^2}{4 m_\chi^2} \right )^3}{\left (4 m_\chi^2 - M_P^2 \right )^2 + \Gamma_P^2 M_P^2} \,.
\end{split}
\end{equation}
Notice that the given expressions are all  independent of $v_{\rm rel} \simeq 1.3 \cdot 10^{-3} \, c$, since the annihilation rates all proceed via $s$-wave. To constrain the parameter space of the pseudoscalar model, we compare the limits on $\chi \bar \chi \to u \bar u$ from~\cite{Ackermann:2015zua} with the predicted annihilation cross section into gluons, while we use \cite{Ackermann:2015lka} with an Einasto R16 DM profile when comparing with annihilation into the combination $\gamma \gamma + \gamma Z/2$.  We rescale all indirect limits by a factor of 2 to take into account that they are obtained for Majorana DM while we are considering Dirac DM.

Our results for the  DM annihilation rates into digluons and diphotons are shown in the middle and on the right of Fig.~\ref{fig:DDID}. The parameters that we have employed to obtain the plots are specified in the figure caption. From both panels it is evident that the existing indirect detection limits exclude only realisations of the pseudoscalar model if $m_\chi = {\cal O}( M_P/2) \simeq 375 \, {\rm GeV}$,~i.e.,~DM can annihilate resonantly  into SM final states via~$P$ exchange. Notice that in this mass region also the DM relic density constraints are most easily fulfilled~\cite{Backovic:2015fnp,Barducci:2015gtd,Mambrini:2015wyu,D'Eramo:2016mgv,Backovic:2016gsf}, since due to the resonance enhancement DM can annihilate efficiently into SM states in the early universe. In order to give an example, in the case of the scalar model the parameter choices $m_\chi = 323 \, {\rm GeV}$, $g_\chi = 2.7$, $c_G = 1.9$, and $c_B = 132$~\cite{Backovic:2015fnp} allow for instance  to reproduce the observed relic abundance $\Omega h^2 \simeq 0.12$, if  standard thermal freezeout is assumed. For this parameters the predicted diphoton cross section is consistent with~(\ref{eq:mugammagamma}) and the total decay width of the scalar amounts to $\Gamma_S = 29 \, {\rm GeV}$.\footnote{Note that the quoted  parameters are viable if $8 \, {\rm TeV}$  LHC data is considered,  as done in \cite{Backovic:2015fnp},  but they are incompatible with the latest $13 \, {\rm TeV}$ CMS $\MET + j$ results~\cite{CMS-PAS-EXO-16-013}, as they lead to a signal strength of~$20 \, {\rm fb}$.} 

\subsection{Benchmark scenarios}

To compare the constraints from collider DM searches with other experiments, the best-fit diphoton bounds, the indirect search bounds, and the direct detection searches for the prototype 750~GeV case can be plotted in the $g_{\chi}\hspace{0.25mm}$--$\hspace{0.5mm}c_{G}$ plane with a fixed DM mass. The choice of this plane constitutes all allowed free couplings since the diphoton cross section measurement constrains the photon coupling in terms of the other DM and gluon couplings. 

The final collider bounds on this prototype scenario are determined from the observed \METP distribution~\cite{CMS-PAS-EXO-16-013} through a CL$_s$ fit~\cite{Junk,Read,Cowan:2010js,CMS-NOTE-2011-005,cls} exploiting the full shape. The quoted change in likelihood corresponding to the 95\% CL is presented as the exclusion bound. 

\begin{figure}[!t]
\begin{center}
\includegraphics[width=0.45\textwidth]{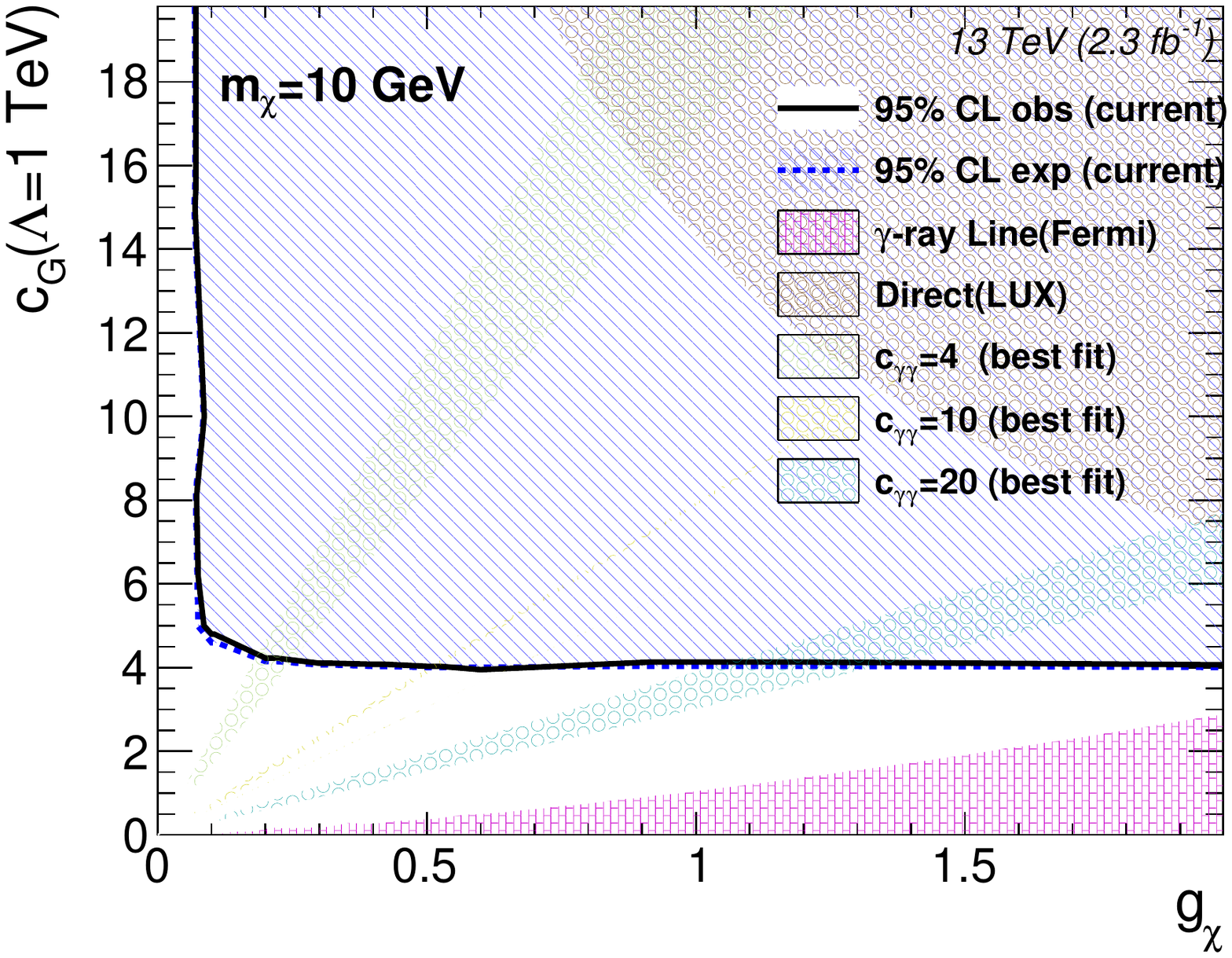}  
\includegraphics[width=0.45\textwidth]{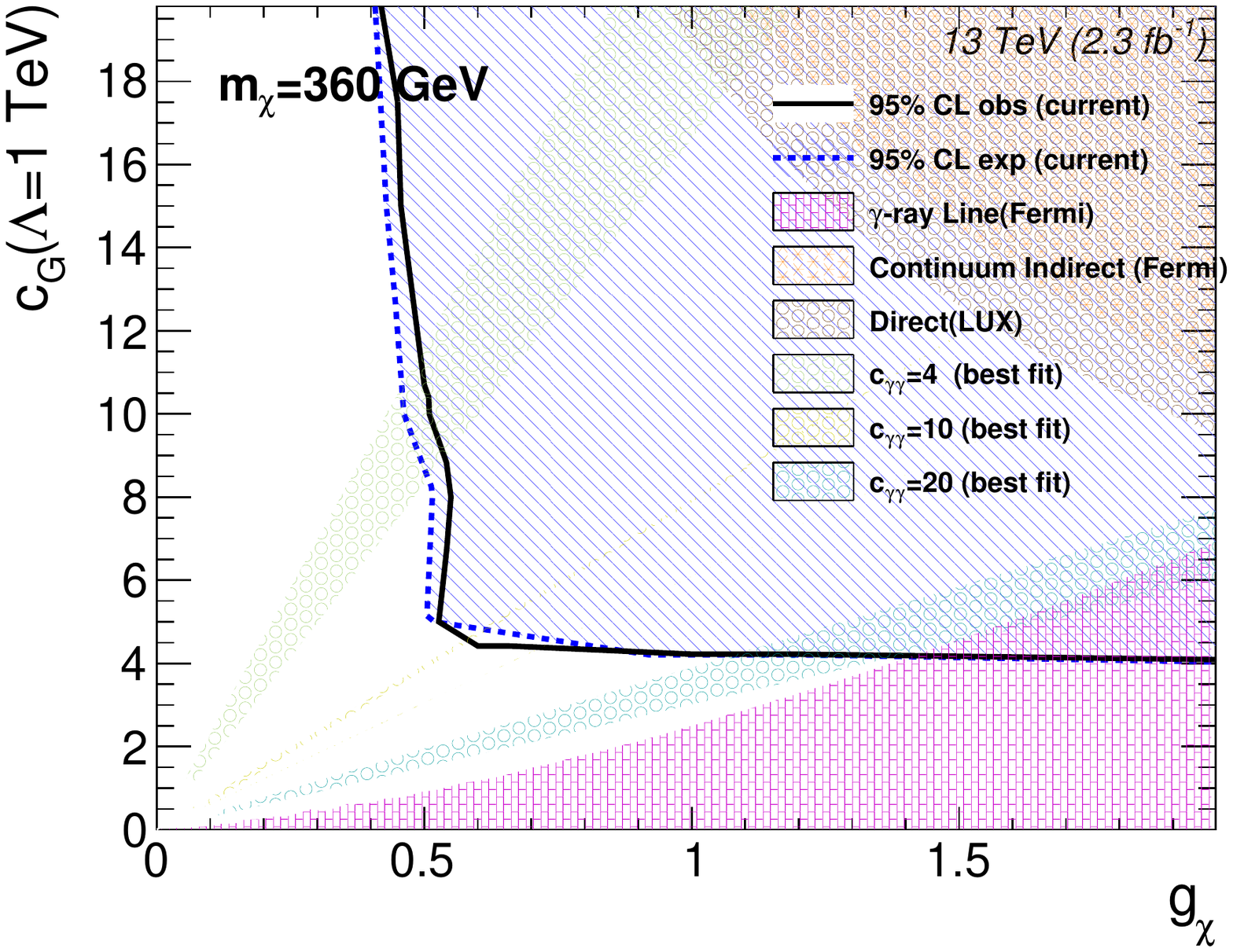}  \\
\includegraphics[width=0.45\textwidth]{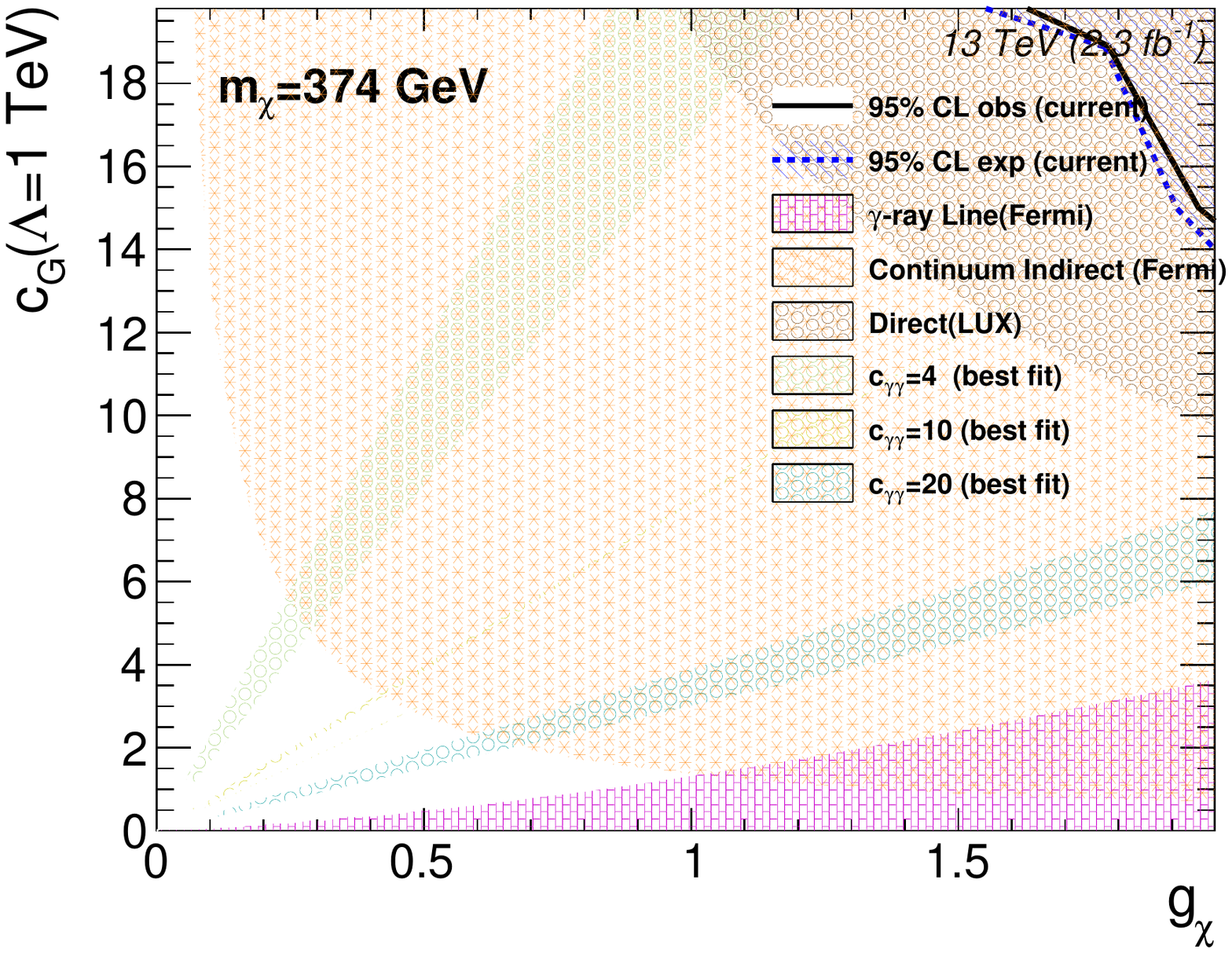}  
\caption{\it Comparison of the bounds from direct detection~\cite{Akerib:2015rjg}, $\gamma$ ray line searches~\cite{Ackermann:2013uma,Abramowski:2013ax}, continuum indirect detection~\cite{Ackermann:2015zua,Abramowski:2014tra}, and constraints from collider monojet searches with the best fit measurements of the putative diphoton excess assuming a fixed diphoton coupling. Three plots are shown for three different DM masses, to show the modification in the sensitivity of the searches in the resonant and non-resonant scenario. Finally, bounds for direct detection can only be applied when the $750 \, {\rm GeV}$ mediator is assumed to be scalar, whereas bounds from the photon line and indirect searches are only valid when the mediator is a pseudoscalar. Bounds from collider searches are valid in both cases.}
\label{fig:all750Bounds}
\end{center}
\end{figure}

The combined bounds are shown in Fig.~\ref{fig:all750Bounds}. The collider searches are essentially independent of the coupling structure, so the bound for the mediator  holds if the mediator is either a scalar or a pseudoscalar. For direct detection, the shown bounds only apply in the case that the mediator is a scalar, while  the continuum and $\gamma$ ray line searches limits constrain only the pseudoscalar mediator.  For large DM couplings where the total width is dominated by the DM contribution $\Gamma(S \rightarrow\bar{\chi}\chi)$ additional modifications of the production cross section occur when the DM particle mass encroaches the region of resonant annihilation $m_\chi = {\cal O}( M_{S}/2) \simeq 375 \, {\rm GeV}$. These modifications can be taken into account by considering three benchmark DM masses on-shell production $m_{\chi}=1 \, {\rm GeV}$, resonant production $m_{\chi}=374 \, {\rm GeV}$, and near resonant production $m_{\chi}=360 \, {\rm GeV}$. All three scenarios are shown in Fig.~\ref{fig:all750Bounds}. 

For DM searches both on-shell and near-resonant production, the collider searches drive the constraining power. Constraints from the collider are strongest for large values of $c_{G}$ or $g_{\chi}$. For pseudoscalar mediators, bounds from the $\gamma$ ray line searches exclude the region of large diphoton coupling. The continuum indirect detection bounds exceed the collider bounds and provide the strongest constraints in the region of resonant production where $m_{\chi}=374\, {\rm GeV}$.  For scalar mediators, direct detection provides the strongest current bound when  $m_{\chi}=374\, {\rm GeV}$. For the other scenarios, collider constraints dominate. 

Although the detailed numbers in the above analysis apply to the specific 750~GeV diphoton excess reported by ATLAS and CMS,
the manner in which the LHC, direct and indirect constraints interplay is more general, as is the approach described above
for modelling the potential discovery of a mediator particle and linking it to DM physics.

\section{Other simplified models of interest}
\label{otherSMS}
In this Section we highlight other interesting options for simplified models that could be studied in more detail in the future. In Subsection~\ref{t-channel} we outline interesting features of $t$-channel simplified models, while in Subsection~\ref{Spin-2} we discuss some interesting models with spin-2 mediators. We conclude this part with a discussion about simplified models with pseudo-Dirac DM in Subsection~\ref{pseudo-Dirac}. 

\subsection{A few representative $t$-channel simplified models}     
\label{t-channel}

So far, we have focused on simplified models with the $s$-channel exchange of the mediatior. However, it is now relevant to build and explore simplified models with the mediator exchanged in the $t$ channel. While for the $s$-channel mediator exchange the EFT description breaks down spectacularly in the case of a light mediator that can be produced on-shell, also in the $t$-channel case there are sizable regions of the parameter space probed by the LHC where the EFT approach breaks down~\cite{Busoni:2014haa}, and therefore a simplified model description is important in this case as well.

Since we wish to produce the DM via colored particles at the LHC, but the DM particle itself cannot be colored, the mediator exchanged in the $t$ channel needs to carry non-trivial color. Thus, it is possible to search for the mediator via its direct QCD pair production. This is the \textit{leitmotiv} of all $t$-channel models.

By restricting ourselves to scalar or fermion DM and to tree-level mediation only, there are four possible $t$-channel simplified models, see, e.g., the recent review~\cite{DeSimone:2016fbz} for more details. 

Perhaps the most relevant one is the case of a ``squark-like'' mediator (the model $0t\frac12$, following the nomenclature of~\cite{DeSimone:2016fbz}). 

The reason why we believe this simplified model is particularly interesting is twofold:  
on the one hand, in this model the mediator has the quantum numbers of squarks in SUSY, and therefore the analyses for squark searches  can be efficiently readapted, and,
on the other hand, the other $t$-channel models either share with it very similar collider phenomenology, or involve suppressed higher-dimensional interactions.
For these reasons we recommend the experimental collaborations to start from this model and to consider it as a benchmark for $t$-channel simplified models.

Let us now discuss the $0t\frac12$ model in more detail. The DM is a Dirac or Majorana fermion $\chi$ and the interactions with the quarks are  mediated by a  set of colored scalar particles $\eta^{(i)}$. For simplicity, we will only consider the case where the DM is a total singlet under the SM symmetries, in particular DM carries no flavor or EW quantum numbers (see, e.g.,~\cite{1109.3516, 1308.0584, 1404.1373} for scenarios with ``flavored'' DM). This implies that the scalar mediator carries not only color charge, but also EW and flavor charges, and hence it resembles the squarks of the minimal supersymmetric extension of the SM.

The most general Lagrangian describing the renormalizable interactions between the SM quark doublets ($Q_L^{(i)}=(u_L, d_L)^{(i)}$) and singlets ($u_R^{(i)}$ and $d_R^{(i)}$) of flavour $i=1,2,3$, a fermion singlet DM $\chi$, and the colored mediators $\eta^{(i)}$ is given by
\begin{equation}
\mathcal{L}=\sum_{i=1,2,3} g_i \,\chi
\left(\bar Q_L^{(i)} \eta^{(i)}_L+\bar u_R^{(i)}\eta^{(i)}_{u,R}
+\bar d_R^{(i)}\eta^{(i)}_{d,R}
  \right)+\textrm{ h.c.},
  \label{tchannelLagr}
\end{equation}  
where the mediators $\eta_L^{(i)}, \eta_{u,R}^{(i)}, \eta_{d,R}^{(i)}$ transform 
under the SM gauge group $SU(3)_c\otimes SU(2)_L\otimes U(1)_Y$ according to the representations $(3,2,-1/6), (3,1,2/3)$, and $(3,1,-1/3)$, respectively.

The minimal width of the generic mediator $\eta^{(i)}$ of mass $M_i$ decaying to a generic quark $q_i$ and the DM particle is simply
\begin{equation}
\Gamma(\eta_i\to \bar q_i \chi)=\frac{g_i^2}{16\pi}\frac{M_i^2-m_{q_i}^2-m_\chi^2}{M_i^3}\sqrt{(M_i^2-m_\chi^2-m_{q_i}^2)^2-4m_\chi^2m_{q_i}^2}
\simeq \frac{g_i^2 M_i}{16\pi}\left[1-\frac{m_\chi^2}{M_i^2}\right]^2,
\end{equation}
where the last expression holds for $M_i,m_\chi \gg m_{q_i}$.

A simplification of the Lagrangian (\ref{tchannelLagr}) arises by assuming Minimal Flavour Violation (MFV), which implies that the $\eta_i$'s have equal masses $M_1=M_2=M_3\equiv M$ and couplings $g_1=g_2=g_3\equiv g$, and therefore the model has only three parameters: $\{m_\chi, M,g\}$, with the restriction $m_\chi<M$ to ensure the stability of the DM.

The MFV hypothesis implies that couplings to third-generation quarks should be nonzero. However, from the point of view of flavor constraints, in some particular situations it may also be safe to violate MFV and restrict the Lagrangian  (\ref{tchannelLagr}) to the first two generations $i=1,2$. In any case, we recommend to stick to MFV and to include the couplings to heavy flavors, as they also induce interesting collider phenomenology, allowing the possibility of exploiting also the searches with b jets in the final state. 

For the parameter scan, we recommend to fix the value of the flavor-universal coupling $g$ (e.g., $g=1$) while performing a scan over $m_\chi, M$, 
with $m_\chi<M$. The parameter space points need to satisfy the narrow-width condition for the mediator $\Gamma/M<1$ and allow for a sufficient number of events to pass the experimental selections
(see, e.g., Table 2.8 of the LHC DMF report \cite{Abercrombie:2015wmb}).

Considering a subset of the general model described by (\ref{tchannelLagr}) may represent a convenient starting point to perform the experimental analyses. 
Several choices for the mediators in the general Lagrangian of Eq.~(\ref{tchannelLagr}) have been studied in the literature:
all mediator species  $\eta_L^{(i)}, \eta_{u,R}^{(i)}, \eta_{d,R}^{(i)}$ (for $i=1,2$)  \cite{Papucci:2014iwa,Abdallah:2014hon, Ko:2016zxg},
only $\eta_{L}^{(i)}$ \cite{Bell:2012rg, Chang:2013oia,DiFranzo:2013vra,Busoni:2014haa, Garny:2014waa}, 
only $\eta_{u,R}^{(i)}$ \cite{DiFranzo:2013vra},
only $\eta_{d,R}^{(i)}$ \cite{DiFranzo:2013vra,Papucci:2014iwa,Abdallah:2014hon},
or combinations \cite{Bai:2013iqa, An:2013xka}.
For instance, one can choose
to couple the mediators only to left-handed  quarks 
$Q_L^{(i)}$ $(i=1,2,3)$. Of couse, other choices can (and should) be explored in a similar fashion.
In this simpler setup, there are three scalar colored mediators $\eta_L^{(i)}$, with the quantum numbers of the left-handed squarks.
The Lagrangian, with the MFV assumption,  is then simply given by
a subset of the interactions of the general Lagrangian (\ref{tchannelLagr})
\begin{equation}
\mathcal{L}= \sum_{i=1,2,3} 
g \,\eta_L^{(i)}\,\bar Q_L^{(i)}\, \chi+\textrm{ h.c.}\,,
\end{equation}
with $\chi$ being either a Dirac or a Majorana fermion singlet.

\begin{figure}[t!]
\centering
\includegraphics[width=0.3\textwidth]{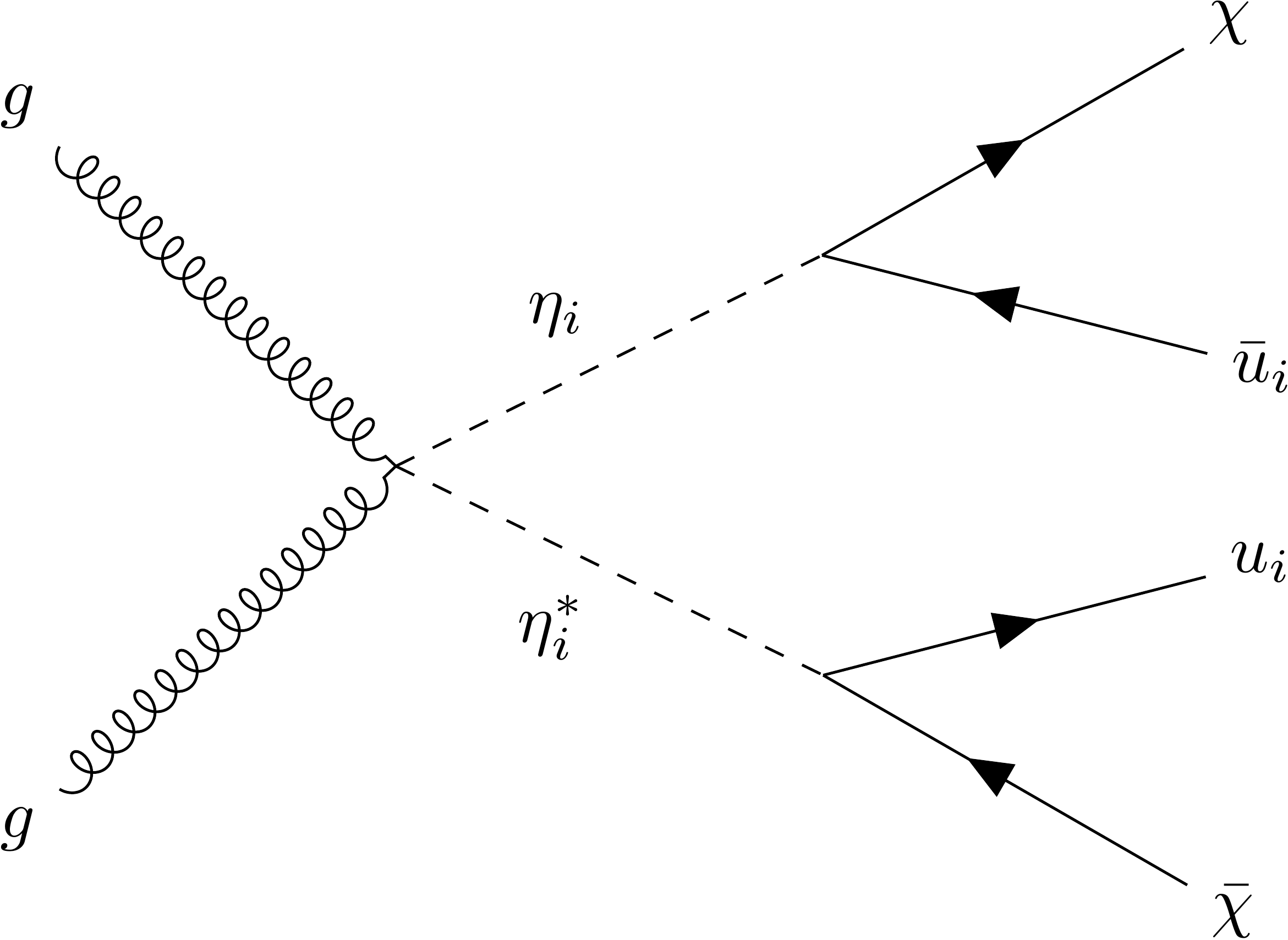}\quad
\includegraphics[width=0.3\textwidth]{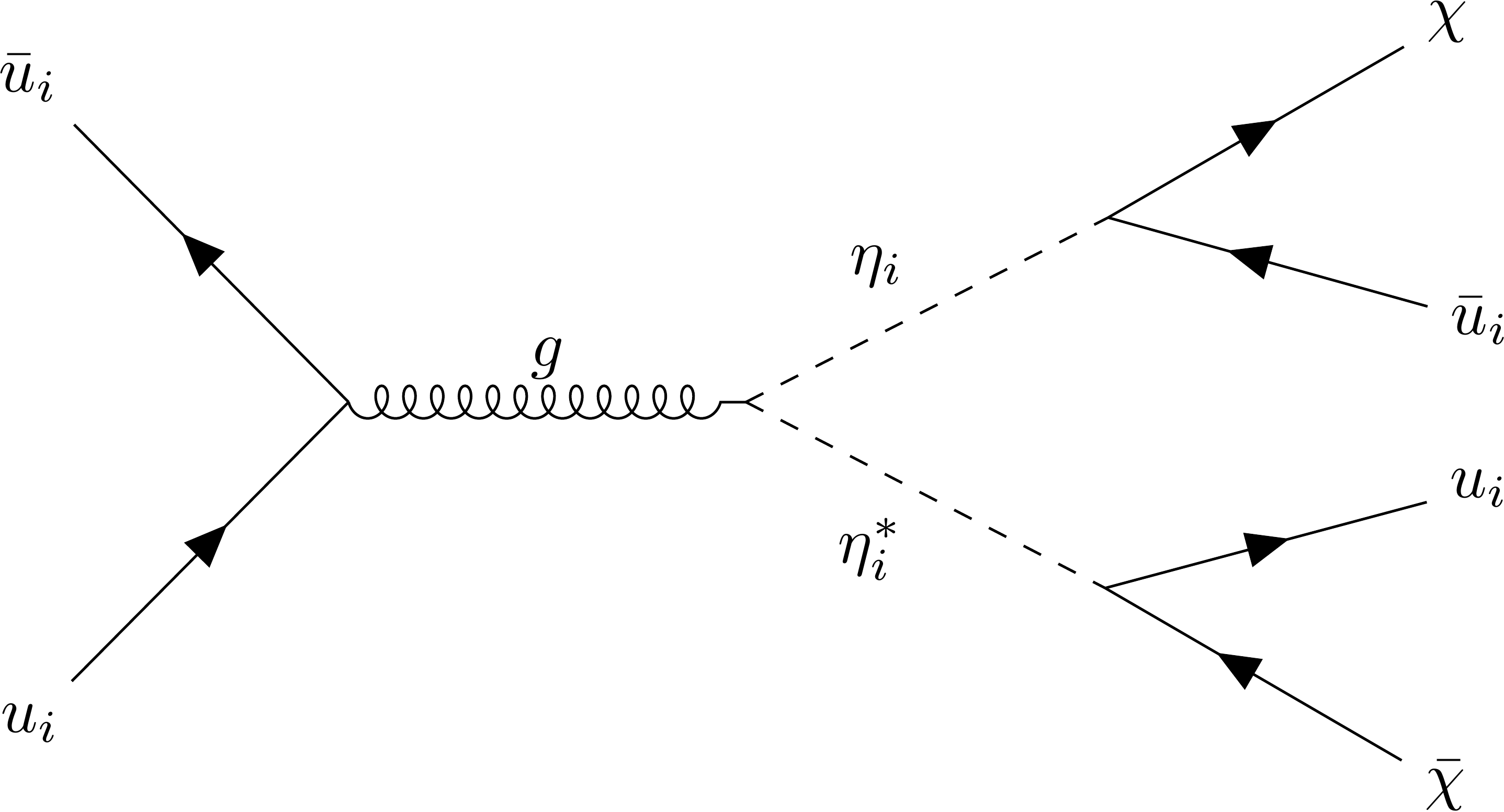}\quad
\includegraphics[width=0.2\textwidth]{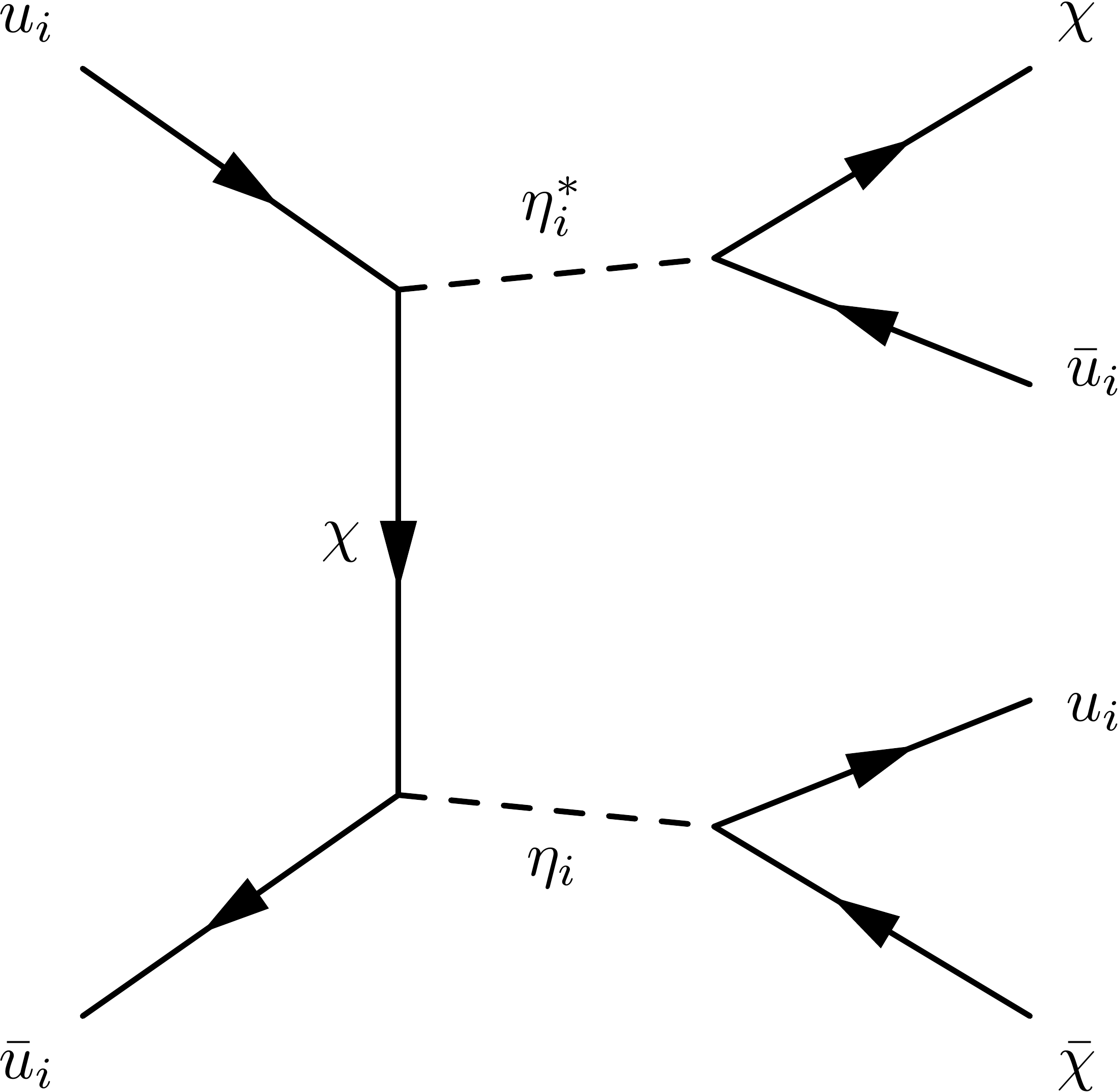}
\caption{\it
Some representative diagrams for mediator pair production in the $t$-channel $0t\frac{1}{2}$ model, which contribute to  2 jets+$\slashed{E}_T$ events.}
\label{fig:0t12}
\end{figure}

The collider phenomenology of this model is mainly driven by the 1 jet + $\slashed{E}_T$ and 2 jets + $\slashed{E}_T$ signals. The former is mostly due to the usual initial-state radiation of a parton from the processes of DM pair production  with $t$-channel exchange of the $\eta_L^{(i)}$  (radiation of a gluon from the $\eta_L^{(i)}$ is also possible but suppressed, although this process and the analogue ones with EW radiation play a relevant role in indirect detection, see, e.g.~\cite{Bell:2010ei, Bell:2011eu, Bell:2011if, Ciafaloni:2011gv, 
DeSimone:2013gj}).
The latter process (2 jets+$\slashed{E}_T$) is instead a rather distinctive feature of $t$-channel models, because it is mostly arises from pair production of the mediator, followed by the decay $\eta_i\to u_i\chi$, see diagrams in Fig.~\ref{fig:0t12}.

Mediator pair production is typically dominated by QCD interactions, being initiated by two gluons or $\bar u_i u_i$. However, since the mediator has EW charges, also Drell--Yan pair production is possible and, more importantly, it is possible to produce a pair of $\eta^{(i)}$ from $\bar u_i u_i$ (or even via the leading channel $u_i u_i$, for Majorana DM) through the $t$-channel exchange of a DM particle (right diagram of Fig.~\ref{fig:0t12}). 
This process is controlled by the Yukawa coupling $g$ of the simplified model.
An interesting feature to keep in mind is that $g$ is a free parameter, whereas in SUSY the coupling of squarks and neutralinos is set to be a combination of gauge couplings. Therefore, depending on the value of $g$  compared to the strong gauge coupling, the relative importance of the diagrams for the $\eta$ pair production (QCD with respect to  DM exchange) can be varied. Another difference with respect to the SUSY case is that the DM can be a Dirac fermion, unlike the neutralino.

Powerful analyses can be carried out by exploiting the combination of the searches for the monojet signal with mediator pair production
(see, e.g.~\cite{An:2013xka, Bai:2013iqa, DiFranzo:2013vra, Papucci:2014iwa} for early work), and by a proper reformulation of the squark searches  with the Yukawa coupling $g$ as free parameter.

As far as the comparison with other experiments is concerned, the limits from direct detection are rather strong for Dirac DM, as it leads to spin-independent DM-nucleus scattering, while in the Majorana DM case only spin-dependent scattering is allowed, for which current constraints are much weaker. 
Combining LHC and direct detection results with the requirement of a correct relic abundance excludes the $0t\frac12$ model with Dirac DM with masses below TeV, while a Majorana DM candidate is still viable for a DM particle mass above $\sim$100 GeV \cite{An:2013xka}. However, it should be noted that the relic density constraint may be evaded by either alternative (non-thermal) production mechanisms in the early universe,
or by assuming that only a fraction of the present energy density of DM has originated from freeze-out, or by additional DM candidates
with respect to the one produced at LHC.

\vspace{1em}

Another representative $t$-channel model is the model with a ``vector-like quark'' mediator (the model $\frac12 t 0$, following the nomenclature of~\cite{DeSimone:2016fbz}), where the DM particle is a scalar singlet $\phi$ and the mediator $\psi$ is a vector-like colored fermion.
By choosing to couple the mediator and the DM particle to right-handed quarks, and assuming MFV, the Lagrangian is
\begin{equation}
\mathcal{L}=\frac12\left[(\partial_\mu\phi)^2-m_\phi^2\phi^2\right]
+\bar\psi(i\slashed{D}-M)\psi+ (y\,\phi\,\bar\psi\, q_R+\textrm{h.c.}).
\end{equation}
The case of the mediator coupling to left-handed quarks can be worked out similarly.
The mediator is a color triplet and electrically charged, so
 it is pair produced mainly via QCD interactions (the processes are depicted in~\cite{Giacchino:2015hvk}, Fig.~6).

For phenomenological studies of this simplified model, including an analysis of the LHC constraints, see~\cite{Vasquez:2009kq, Ibarra:2014qma, 
Giacchino:2014moa, Giacchino:2015hvk}.
The reach for this model is improved by
combining the DM searches with the collider searches for vector-like quarks, to be interpreted as searches for the mediator.

We expect the collider phenomenology of this model to be similar, although not identical, to the one of the $0t\frac12$ (squark-like mediator) model discussed at the beginning of this Subsection, as some processes for mediator pair production are different. As for the combinations with other searches, it should be noted that the direct and indirect detection phenomenology of this model is very different from the case with a squark-like mediator (see, e.g.,~\cite{DeSimone:2016fbz}). 
Combining all the available limits from LHC, direct, and indirect detection, as well as relic density constraints, one ends up with a rather constrained scenario, but still some parameter space is available for $m_\phi\gtrsim 200$ GeV and 
$m_\psi/m_\phi\lesssim 2$ \cite{Giacchino:2015hvk}, which is worth exploring 
with analyses of the upcoming data.

\subsection{Spin-2 models}
\label{Spin-2}

The behaviour of DM depends on the way it interacts with the SM and, in particular, on the quantum numbers of both DM and the mediator. An interesting possibility for a new kind of spin-two mediator has been proposed in Ref.~\cite{Lee:2013bua} in the so-called Gravity-Mediated Dark Matter (GMDM) model.  In this scenario, the origin of DM is linked to a new sector of strong interactions (gravity or its gauge dual) and mediation to the SM is via spin-two and spin-zero states. In the gravitational interpretation of the scenario, these mediator states can be identified as the lightest Kaluza--Klein graviton and radion of the compact extra dimension. In the gauge dual scenario, these states are glueballs linked to the spontaneous breaking of symmetries. Fortunately, the behaviour of the GMDM spin-2 mediator $h_{\mu\nu}$ is determined by the low-energy symmetries of the theory~\cite{Fok:2012zk}, and is largely independent of the interpretation of the spin-two state. In particular, DM and SM fields would interact with the mediator via dimension-five operators involving the same type of couplings as from the stress-tensor $T^{\mu\nu}$, namely
\begin{equation}
\mathcal{L}_{int} = \frac{-c_i}{\Lambda} \, h_{\mu\nu} \, T^{(i) \mu\nu} \ ,
\end{equation}
where the index $i$ denotes any kind of field, DM or SM, and $\Lambda$ is the scale suppressing the interactions. In~\cite{Lee:2013bua,Lee:2014caa}, it was found that an unsuppressed s-wave annihilation is possible for scalar, vector and fermionic DM.  

This distinctive scenario for DM is largely unexplored and may deserve a closer look. The DM phenomenology via a spin-2 mediator necessarily involves higher-dimensional operators,  which emulates the suppression of contact interactions, yet in a resonant regime where the mediator is on-shell. In particular, one would need to explore further the implications in direct detection and collider phenomenology of these models. For example, current searches for spin-2 Kaluza--Klein gravitons could be made in the context of GMDM and linked to the DM searches at the LHC, as illustrated recently in the interpretation of GMDM in the context of the 750 GeV diphoton excess~\cite{Han:2015cty}.

\subsection{Pseudo-Dirac DM}
\label{pseudo-Dirac}
The starting point is to consider a generic new four-component Dirac fermion $\Psi$ that is a singlet under the SM gauge group. We consider the most general Lagrangian  for $\Psi$ with both Dirac ($M_D$) and Majorana ($m_{L,R}$) masses 
\cite{DeSimone:2010tf}:
\begin{equation}
\mathcal{L}_0=\bar\Psi(i\slashed{\partial}-M_D)\Psi
-\frac{m_L}{2}(\bar\Psi^cP_L\Psi+\textrm{h.c.})
-\frac{m_R}{2}(\bar\Psi^cP_R\Psi+\textrm{h.c.}),
\end{equation}
where $P_{R,L}=(1\pm \gamma^5)/2$.
We focus on the ``pseudo-Dirac'' limit of the mass matrix, 
where $M_D\gg m_L, m_R$. 
The two mass eigenstates, denoted by $\chi_{1,2}$, with masses $m_{1,2}=
M_D\mp (m_L+m_R)/2$, will be linear combinations of $\Psi,\Psi^c$.
It is then possible to construct the Majorana fields $\Psi_1^{\rm{M}}, \Psi_2^{\rm{M}}$ out of the mass eigenstates: $\Psi_1^{\rm{M}}\equiv \chi_1+\chi_1^c$ and $\Psi_2^{\rm{M}}\equiv \chi_2+\chi_2^c$.

The spectrum of this model consists of the lightest state $\Psi_1^{\rm{M}}$ with mass $M_1=M_D-(m_L+m_R)/2$, identified with a Majorana DM particle, and a slightly heavier companion state  $\Psi_2^{\rm{M}}$, with mass $M_2=M_1+\Delta M=M_1+(m_L+m_R)$. The model described by the free Lagrangian $\mathcal{L}_0$ is simply defined by the two mass parameters $M_1, \Delta M$.

The situation with pseudo-Dirac fermions may also be realized in a SUSY framework, see e.g.~Refs.~\cite{Hall:1990hq, Nelson:2002ca, Belanger:2009wf, Chun:2009zx, Hsieh:2007wq}.

At this point one needs to consider the interactions of $\Psi$ with the SM fields $f$. 
The choice made  in~\cite{DeSimone:2010tf} was to consider effective (non-renormalizable) interactions. At dimension six, one can write:
\begin{equation}
\mathcal{L}_{\rm{int}}=\frac{1}{\Lambda^2}
\left[\bar\Psi\gamma^\mu(c_L P_L+c_R P_R)\Psi\right]
\times
\left[\bar f\gamma_\mu(c_L^{(f)} P_L+c_R^{(f)} P_R)f\right],
\label{pDDMLint}
\end{equation}
where $c_{R,L}, c_{R,L}^{(f)}$ are generic operator coefficients.
Other dimension-6 operators are possible, e.g. those leading to anapole moments of Majorana Dark Matter fermions with SM gauge bosons \cite{Fitzpatrick:2010br, Ho:2012bg, Gao:2013vfa}. The analysis of such operators in the context of pseudo-Dirac Dark Matter deserves further investigation.

The Lagrangian (\ref{pDDMLint}) can be rewritten in terms of the Majorana
mass eigenstates $\Psi_{1,2}^{\rm M}$, leading to terms for the interactions
of two $\Psi_1^{\rm M}$ particles
\begin{equation}
\mathcal{L}_{\rm{int},{11}}=\frac{1}{\Lambda^2}
\frac{c_R-c_L}{4}
\left[\bar\Psi_1^{\rm M}\gamma^\mu\gamma^5\Psi_1^{\rm M}\right]
\times
\frac12\left[(c_L^{(f)}+c_R^{(f)})\bar f\gamma_\mu f
+(c_R^{(f)}-c_L^{(f)})\bar f\gamma_\mu \gamma^5 f\right] \, ,
\label{pDDMLint11}
\end{equation}
and to terms for the interaction of $\Psi_1^{\rm M}$ with  $\Psi_2^{\rm M}$
\begin{equation}
\mathcal{L}_{\rm{int},{12}}=\frac{i}{\Lambda^2}
\frac{c_R+c_L}{2}
\left[\bar\Psi_1^{\rm M}\gamma^\mu\Psi_2^{\rm M}\right]
\frac12\left[(c_L^{(f)}+c_R^{(f)})\bar f\gamma_\mu f
+(c_R^{(f)}-c_L^{(f)})\bar f\gamma_\mu \gamma^5 f\right].
\label{pDDMLint12}
\end{equation}

The pseudo-Dirac DM scenario, despite its minimality, has several interesting features, as we now describe briefly (see~\cite{DeSimone:2010tf} for more details).
\begin{itemize}
\item With a large enough splitting $\Delta M\gtrsim \mathcal{O}(10 - 100 \textrm{ keV})$, DM-nucleon scattering involves only the elastic scattering of a Majorana DM particle $\Psi_1^{\rm M}$, and it is driven by the 
interactions in (\ref{pDDMLint11}), which are spin-dependent. 
In this way the stringent constraints on spin-independent scattering are evaded.

\item 
The relic density is driven by the coannihilation channel of the two nearly-degenerate states $\Psi_1^{\rm M}\Psi_2^{\rm M}\to \bar f f$ described by the terms in Eq.~(\ref{pDDMLint12}), since the Majorana annihilations of $\Psi_1^{\rm M}\Psi_1^{\rm M}$ and $\Psi_2^{\rm M}\Psi_2^{\rm M}$ are either suppressed by $m_f$ or by the relative velocity ($p$-wave).

\item The same interaction terms in Eq.~(\ref{pDDMLint12}) responsible for the relic abundance calculation are also driving the decay of $\Psi_2^{\rm M}\to \Psi_1^{\rm M}f\bar f$. 
For mass splittings of the order of GeV, the decay lengths can naturally be of the order of a measurable displaced vertex.
By considering $f$ as a lepton, the edge of the dilepton invariant mass distribution is directly related to the mass splitting $\Delta M$. 
So  by just measuring the decay length and the dilepton edge one can determine the overall DM mass scale and the mass splitting.

\item As a consequence of the last two points, it is possible to relate the decay length
to the DM relic abundance and the mass parameters of the model, in such a way that 
one can make a prediction for the DM mass, to be tested against other independent measurements.
\end{itemize}
Pseudo-Dirac DM is particularly interesting in the context of collider physics, since it leads to a rather rich phenomenology for DM searches beyond the usual missing transverse energy signature, such as the displaced vertex signatures which have not been fully explored yet in this context.
(See also the discussion of SUSY DM signatures in the next Section.)

It was remarked during the workshop that it would be interesting to adapt the analysis of~\cite{DeSimone:2010tf} in the framework of simplified models
(Note: simplified models for coannihilation scenarios have been worked out in~\cite{Baker:2015qna}).
This effort will require replacing the effective interactions in Eq.~(\ref{pDDMLint}) with the inclusion of a mediator, which is currently under way.

\newcommand{\lsim}
{\;\raisebox{-.3em}{$\stackrel{\displaystyle <}{\sim}$}\;}
\newcommand{\gsim}
{\;\raisebox{-.3em}{$\stackrel{\displaystyle >}{\sim}$}\;}
\newcommand{\gmt}{\ensuremath{(g-2)_\mu}}
\newcommand{\br}{{\rm BR}}
\newcommand{\bsg}{BR($b \to s \gamma$)}
\newcommand{\cls}{\ensuremath{{\rm CL}_s}}
\newcommand{\btn}{BR($B_u \to \tau \nu_\tau$)}
\newcommand{\kmn}{BR($K \to \mu \nu_\mu$)}
\newcommand{\bmm}{\ensuremath{\br(B_s \to \mu^+\mu^-)}}
\newcommand{\bsmm}{\ensuremath{\br(B_s \to \mu^+\mu^-)}}
\newcommand{\bdmm}{\ensuremath{\br(B_d \to \mu^+\mu^-)}}
\newcommand{\bsdmm}{\ensuremath{\br(B_{s, d} \to \mu^+\mu^-)}}
\newcommand{\bqmm}{\ensuremath{\br(B_{q} \to \mu^+\mu^-)}}
\newcommand{\rmm}{\ensuremath{R_{\mu \mu}}}
\newcommand{\ssi}{\ensuremath{\sigma^{\rm SI}_p}}
\newcommand{\ssin}{\ensuremath{\sigma^{\rm SI}_n}}
\newcommand{\ssd}{\ensuremath{\sigma^{\rm SD}_p}}
\newcommand{\Och}{\ensuremath{\Omega_\chi h^2}}
\newcommand{\sweff}{\sin^2\theta_{\mathrm{eff}}}
\newcommand{\MW}{\ensuremath{M_W}}
\newcommand{\MZ}{\ensuremath{M_Z}}
\newcommand{\Mh}{\ensuremath{M_h}}
\newcommand{\MH}{\ensuremath{M_H}}
\newcommand{\MA}{\ensuremath{M_A}}
\newcommand{\MHSM}{M_H^{\rm SM}}
\newcommand{\mt}{m_t}
\newcommand{\nutau}{\ensuremath{\nu_{\tau}}}
\newcommand{\msusy}{M_{\rm SUSY}}
\newcommand{\mgl}{\ensuremath{m_{\tilde g}}}
\newcommand{\msq}{\ensuremath{m_{\tilde q}}}
\newcommand{\msqone}{\ensuremath{m_{\tilde q_1}}}
\newcommand{\msqtwo}{\ensuremath{m_{\tilde q_2}}}
\newcommand{\sto}[1]{\ensuremath{\tilde t_{#1}}}
\newcommand{\sbot}[1]{\ensuremath{\tilde b_{#1}}}
\newcommand{\stau}[1]{\ensuremath{\tilde \tau_{#1}}}
\newcommand{\stopone}{\ensuremath{\tilde t_{1}}}
\newcommand{\mstop}[1]{\ensuremath{m_{\tilde t_{#1}}}}
\newcommand{\msbot}[1]{\ensuremath{m_{\tilde b_{#1}}}}
\newcommand{\msqt}{\ensuremath{m_{\tilde q_3}}}
\newcommand{\msl}{\ensuremath{m_{\tilde l}}}
\newcommand{\msqr}{\ensuremath{m_{\tilde q_R}}}
\newcommand{\cha}[1]{\tilde \chi^\pm_{#1}}
\newcommand{\champ}[1]{\tilde \chi^\mp_{#1}}
\newcommand{\mcha}[1]{\ensuremath{m_{\tilde \chi^\pm_{#1}}}}
\newcommand{\neu}[1]{\tilde \chi^0_{#1}}
\newcommand{\mneu}[1]{\ensuremath{m_{\tilde \chi^0_{#1}}}}
\newcommand{\mst}[1]{m_{\tilde t_{#1}}}
\newcommand{\msb}[1]{m_{\tilde b_{#1}}}
\newcommand{\mstau}[1]{\ensuremath{m_{\tilde \tau_{#1}}}}
\newcommand{\msmu}[1]{\ensuremath{m_{\tilde \mu_{#1}}}}
\newcommand{\msel}[1]{\ensuremath{m_{\tilde e_{#1}}}}
\newcommand{\mslep}{\ensuremath{m_{\tilde \ell}}}
\newcommand{\mste}{m_{\tilde t_1}}
\newcommand{\mstz}{m_{\tilde t_2}}
\newcommand{\msbe}{m_{\tilde b_1}}
\newcommand{\staue}{\tilde \tau_1}
\newcommand{\mstaue}{m_{\staue}}
\newcommand{\astaue}{\overline{\tilde \tau_1}}
\newcommand{\sel}[1]{\tilde e_{#1}}
\newcommand{\asel}[1]{\overline{\tilde e_{#1}}}
\newcommand{\smu}[1]{\tilde \mu_{#1}}
\newcommand{\asmu}[1]{\overline{\tilde \mu_{#1}}}
\newcommand{\slep}[1]{\tilde \ell_{#1}}
\newcommand{\tb}{\ensuremath{\tan\beta}}
\newcommand{\ecm}{\sqrt{s}}
\newcommand{\tev}{\ensuremath{\,\, \mathrm{TeV}}}
\newcommand{\mev}{\ensuremath{\,\, \mathrm{MeV}}}
\newcommand{\SM}{``SM''}
\newcommand{\infb}{\ensuremath{{\rm fb}^{-1}}}
\def\order#1{\ensuremath{{\cal O}(#1)}}
\def\refeq#1{\mbox{Eq.~(\ref{#1})}}
\def\refeqs#1{\mbox{Eqs.~(\ref{#1})}}
\def\reffi#1{\mbox{Fig.~\ref{#1}}}
\def\reffis#1{\mbox{Figs.~\ref{#1}}}
\def\refta#1{\mbox{Table~\ref{#1}}}
\def\reftas#1{\mbox{Tabs.~\ref{#1}}}
\def\refse#1{\mbox{Sect.~\ref{#1}}}
\def\refses#1{\mbox{Sects.~\ref{#1}}}
\def\refapp#1{\mbox{App.~\ref{#1}}}
\def\citere#1{\mbox{Ref.~\cite{#1}}}
\def\citeres#1{\mbox{Refs.~\cite{#1}}}
\newcommand{\lhccol}{\ensuremath{{\rm LHC8}_{\rm col}}}
\newcommand{\lhcewk}{\ensuremath{{\rm LHC8}_{\rm EWK}}}
\newcommand{\lhcstop}{\ensuremath{{\rm LHC8}_{\rm stop}}}
\newcommand{\atom}{{\tt Atom}}
\newcommand{\scorpion}{{\tt Scorpion}}
\newcommand{\chisqlhccol}{{\ensuremath{\chi^2(\lhccol)}}}
\newcommand{\chisqlhcewk}{{\ensuremath{\chi^2(\lhcewk)}}}
\newcommand{\chisqlhcstop}{{\ensuremath{\chi^2(\lhcstop)}}}
\newcommand{\chisqatom}{{\ensuremath{\chi^2(\atom)}}}
\newcommand{\chisqscorpion}{{\ensuremath{\chi^2(\scorpion)}}}
\newcommand{\chisqatomscorpion}{{\ensuremath{\chi^2(\atom{\rm~and~}\scorpion)}}}

\newcommand{\ETslash}{\ensuremath{/ \hspace{-.7em} E_T}}
\newcommand{\pTslash}{/ \hspace{-.7em} p_T}
\newcommand{\delphes}{{\it {\tt DELPHES~3}}}
\newcommand{\atlasfive}{ATLAS 5/fb jets + $\ETslash$}
\newcommand{\lhco}{LH8$_{\rm 1/fb}$}
\newcommand{\lhcf}{LHC$_{\rm 5/fb}$}
\newcommand{\atlaso}{ATLAS$_{\rm 1/fb}$}
\newcommand{\atlasf}{ATLAS$_{\rm 5/fb}$}
\newcommand{\atlastwenty}{ATLAS 20/fb jets + $\ETslash$}

\section{What can we learn about simplified DM models from SUSY?}
\label{SUSY}

{\it In this Section we review general features of DM in complete SUSY models, and propose how these could inspire and guide the development of improved SDMMs.} 


In the absence of clear theoretical guidance, much experimental and phenomenological
effort has gone into probing models with universal soft SUSY breaking at the GUT scale, 
such as the CMSSM in which universality is postulated for the gauginos 
and all scalars, and models with non-universal Higgs masses (NUHM1,2). These models
are already significantly constrained by the LHC Run 1 data (with $p$-values $\sim 0.1$), if one
attempts to explain the $g_\mu - 2$ anomaly~\cite{NUHM2}. On the other hand, if one treats $n$ soft
supersymmetry-breaking masses as independent phenomenological inputs at the EW
scale, as in pMSSMn models, retaining only the degree of universality motivated by the upper limits 
on flavour-changing neutral currents, the LHC constraints are less restrictive ($p \sim 0.3$)
and the $g_\mu - 2$ anomaly can still be accommodated~\cite{pMSSM10}.

\subsection{The DM mechanisms in SUSY}

Generically, assuming standard Big Bang cosmology and requiring that the relic density of lightest supersymmetric particles (LSPs)
respect the upper limit imposed by the Planck satellite and other measurements imposes
an upper limit on the range of possible soft SUSY breaking masses in universal models.
Within this range, many different mechanisms for bringing the DM density into the
allowed cosmological range may come into play, not only the conventional annihilation and
freeze-out mechanism. For example, there may be enhanced, rapid annihilation through
direct channel resonances such as $Z$, $h$, $H/A$, $X(750)$. Also, coannihilation with some other,
almost-degenerate SUSY particle species such as the lighter stau ($\tilde\tau_1$), top squark ($\tilde t_1$), wino, or sneutrino, may be
important.

\begin{figure*}[htb!]
\vspace{0.5cm}
\begin{center}
\resizebox{7.5cm}{!}{\includegraphics{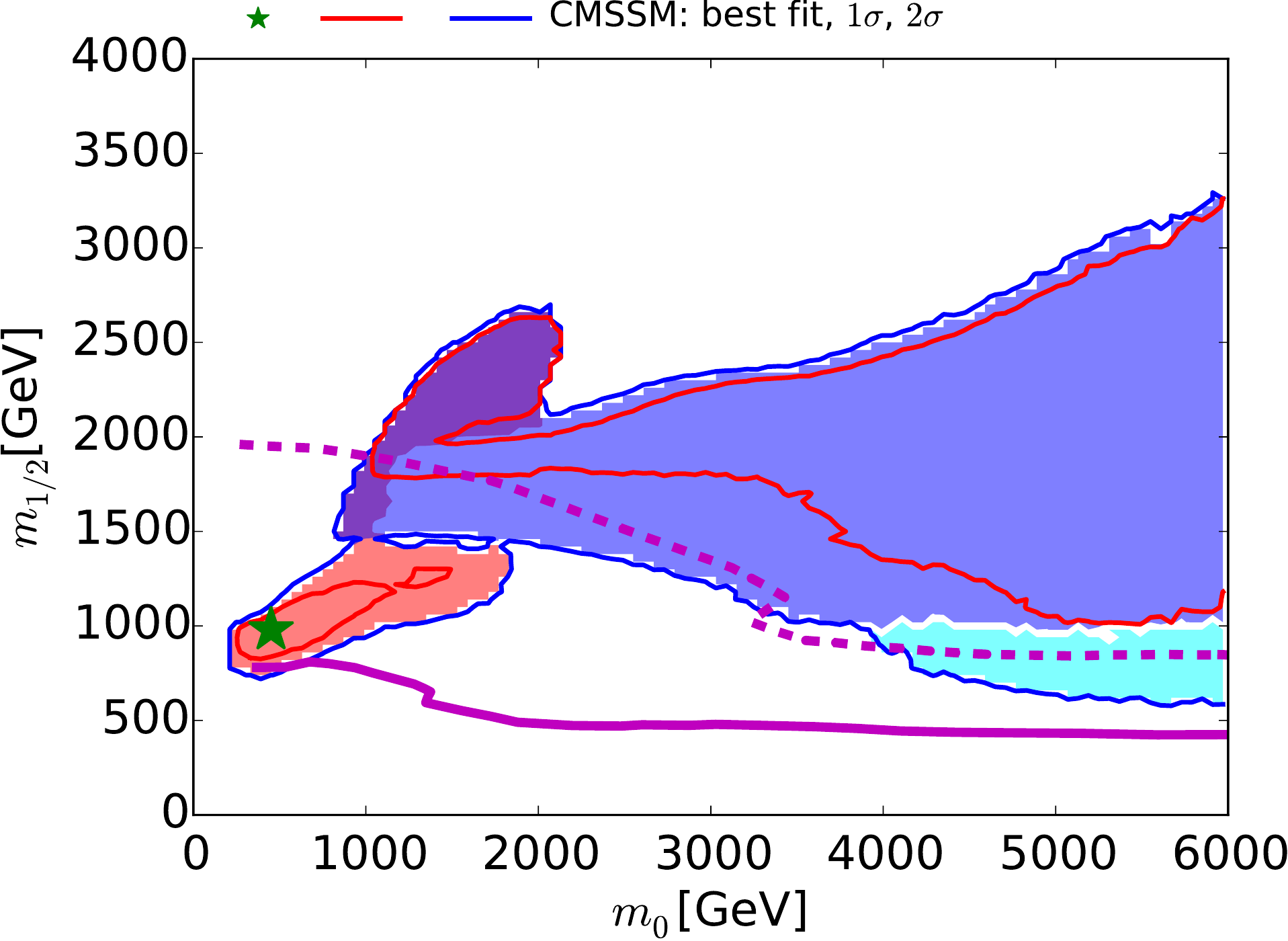}}
\resizebox{7.5cm}{!}{\includegraphics{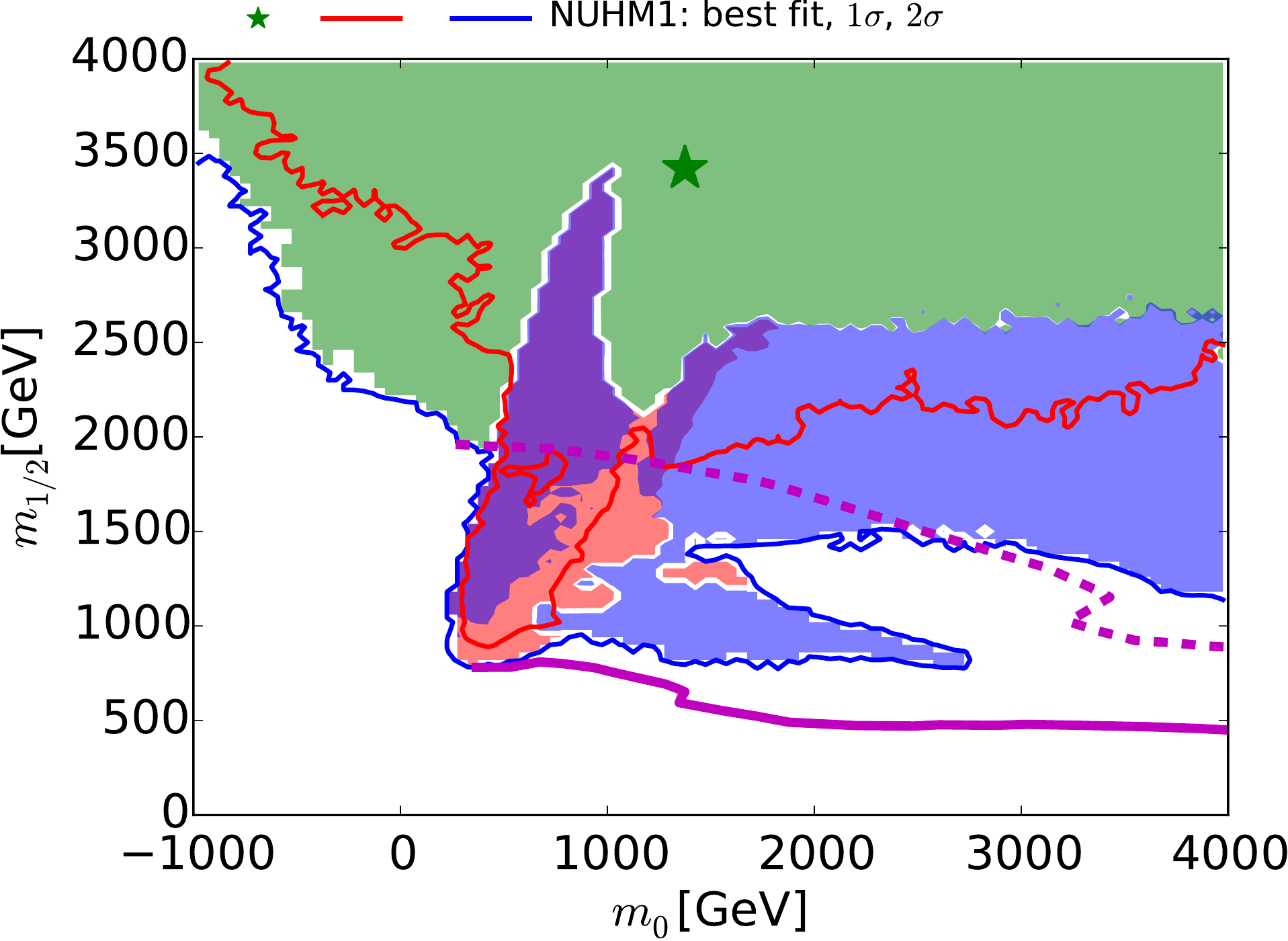}}\\[1em]
\resizebox{7.5cm}{!}{\includegraphics{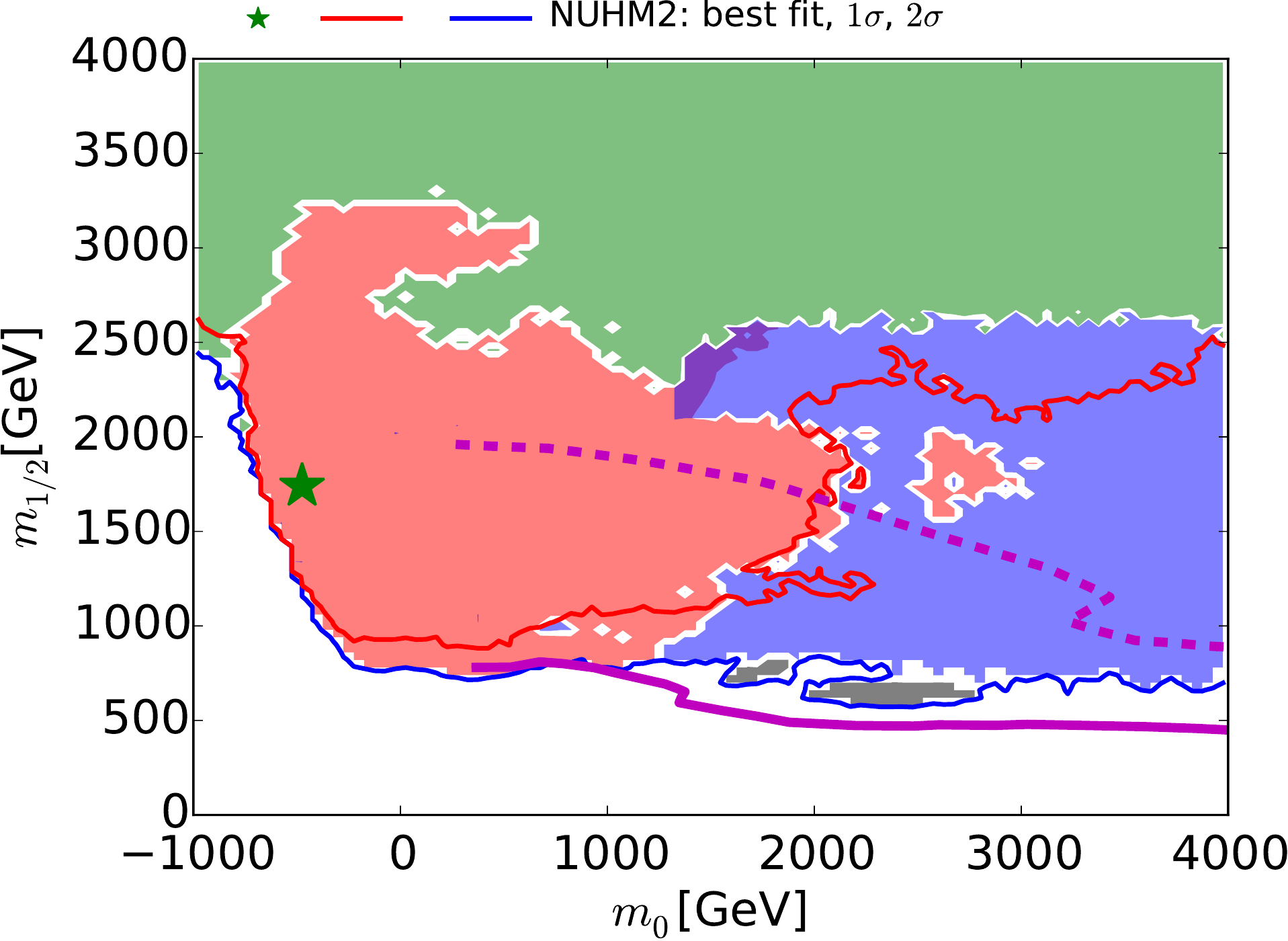}}
\resizebox{7.5cm}{!}{\includegraphics{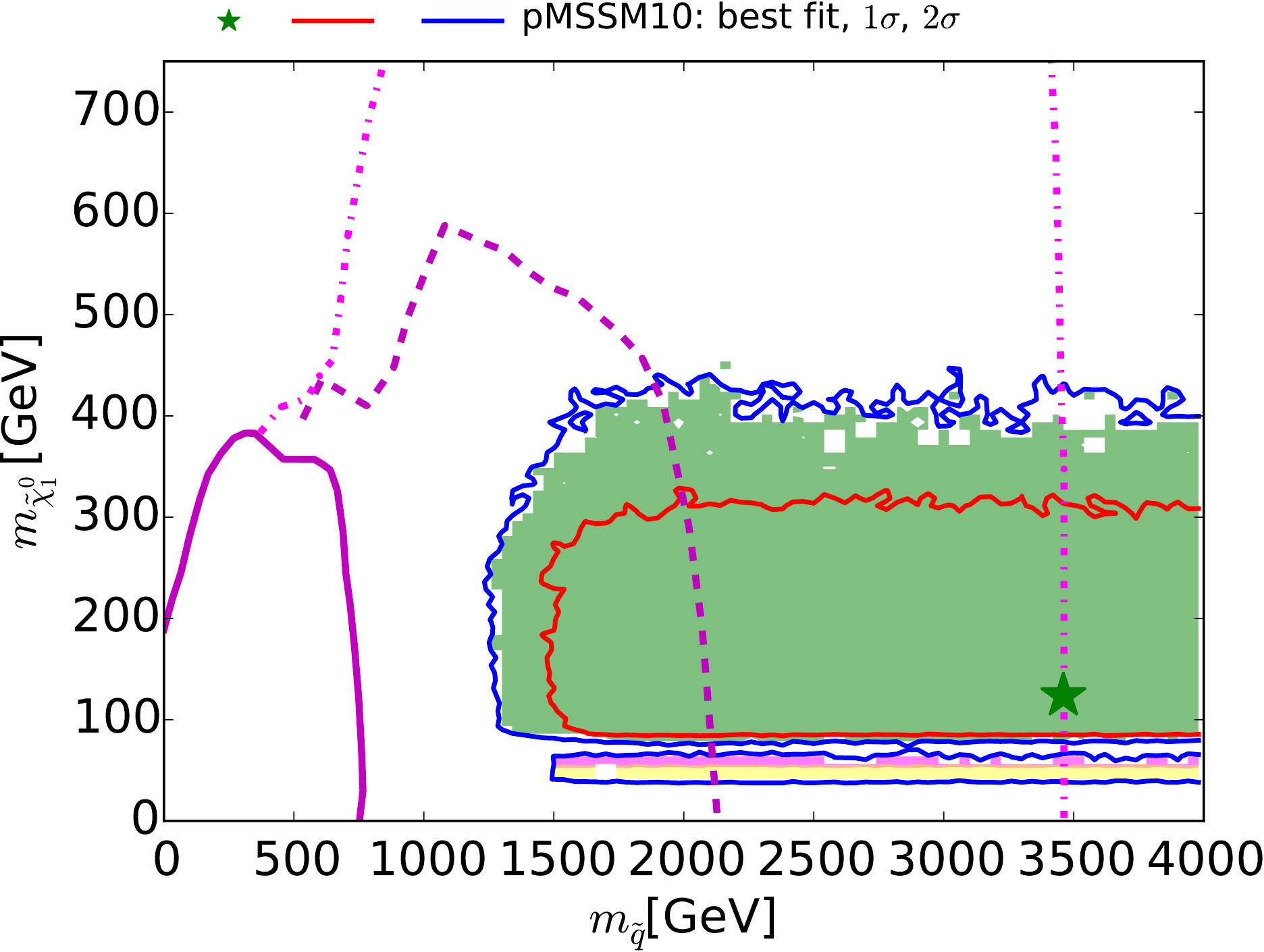}} \\
\resizebox{15cm}{!}{\includegraphics{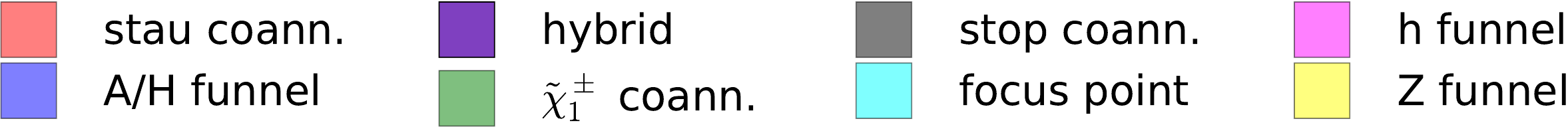}}
\end{center}
\caption{\it The $(m_0, m_{1/2})$ planes in the CMSSM (upper left), the NUHM1 (upper right), and the NUHM2 (lower left),
and the $(\msq, \mneu1)$ plane in the pMSSM10~\protect\cite{MCDM}. Regions in which different mechanisms bring the DM
density into the allowed range are shaded as described in the legend and discussed in the text. The red and blue
contours are the $\Delta \chi^2 = 2.30$ and 5.99 contours found in global fits
to these models, corresponding approximately to the 68 and 95\% CL
contours, with the green stars indicating the best fit points, and the solid purple contours
show the current LHC 95\% exclusions from ~$\ETslash$ searches.
In the CMSSM, NUHM1, and NUHM2 cases, the dashed purple contours show the prospective 5$\sigma$
discovery reaches for ~$\ETslash$ searches at the LHC with 3000 fb$^{-1}$ of data at $\sqrt{s} = 14$~TeV. 
In the pMSSM10 case, the dashed purple contour shows the 95\% CL exclusion sensitivity
of the LHC with 3000 fb$^{-1}$, assuming $\mgl \gg \msq$, and the dash-dotted lines 
bound the corresponding sensitivity region assuming $\mgl = 4.5 \tev$.
}
\label{fig:SUSY1}
\end{figure*}

Figure~\ref{fig:SUSY1} illustrates the most important DM mechanisms in the CMSSM
(upper left panel), NUHM1 (upper right panel), NUHM2 (lower left panel), and pMSSM10
(lower right panel), colour-coded as indicated in the legend~\cite{MCDM}. We see immediately
the importance of including coannihilation with staus (pink), stops (grey), and charginos
(green), as well as the need to take into account enhanced annihilations through direct channel
resonances such as the $h$ (pink), heavy SUSY Higgs bosons (dark blue), and the $Z$ boson
(yellow), often in combination as indicated by the hybrid regions (purple).

{\it It is, therefore, desirable to extend the simplified model approach to include at least some of these possibilities in order to achieve a more realistic description of relevant DM mechanisms in SDMMs.} 
(See Ref.~\cite{Baker:2015qna} for a discussion of simplified models for coannihilation.)

\subsection{Collider signatures}

\begin{figure*}[htb!]
\begin{center}
\resizebox{7.5cm}{!}{\includegraphics{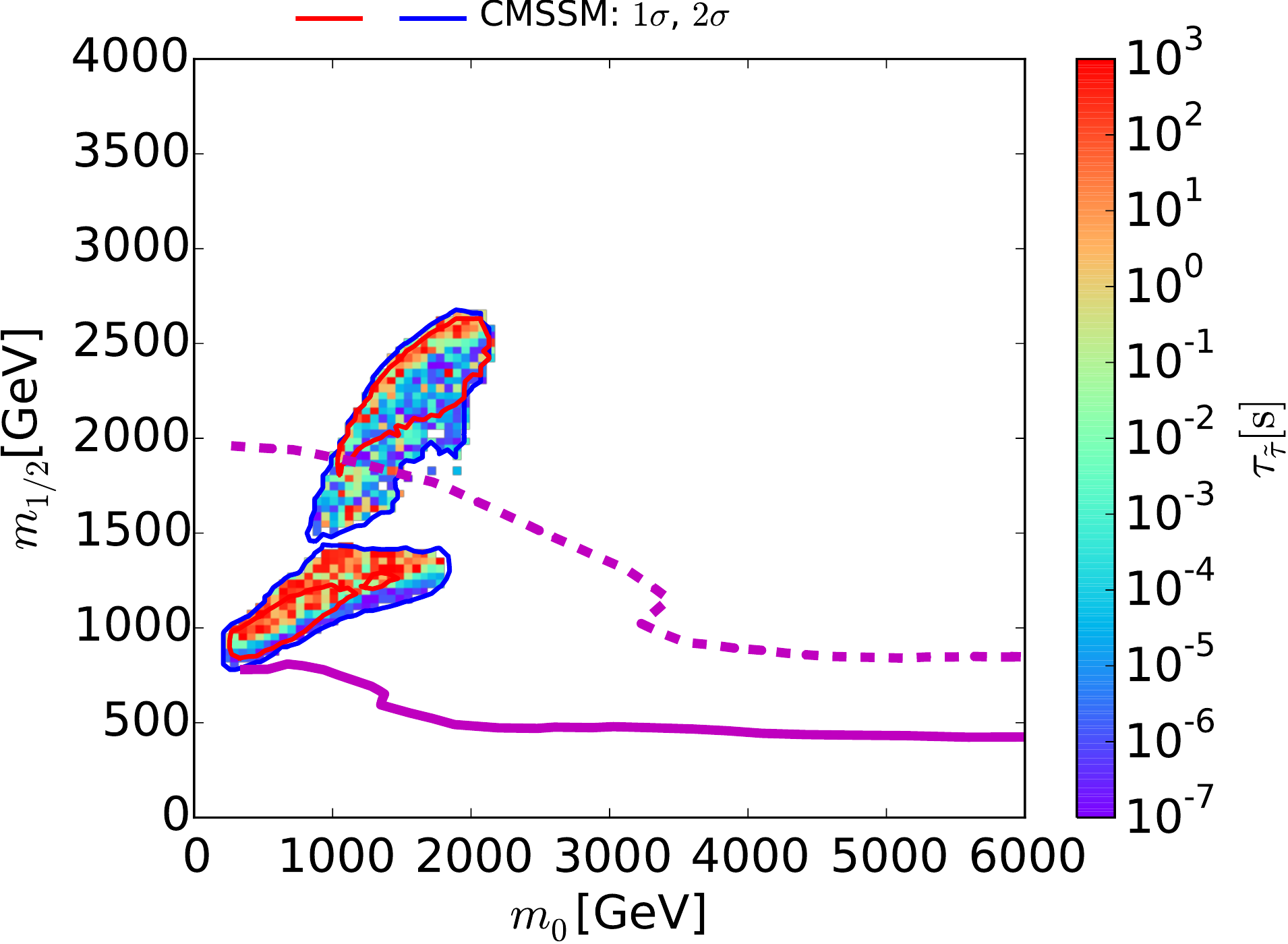}}
\resizebox{7.5cm}{!}{\includegraphics{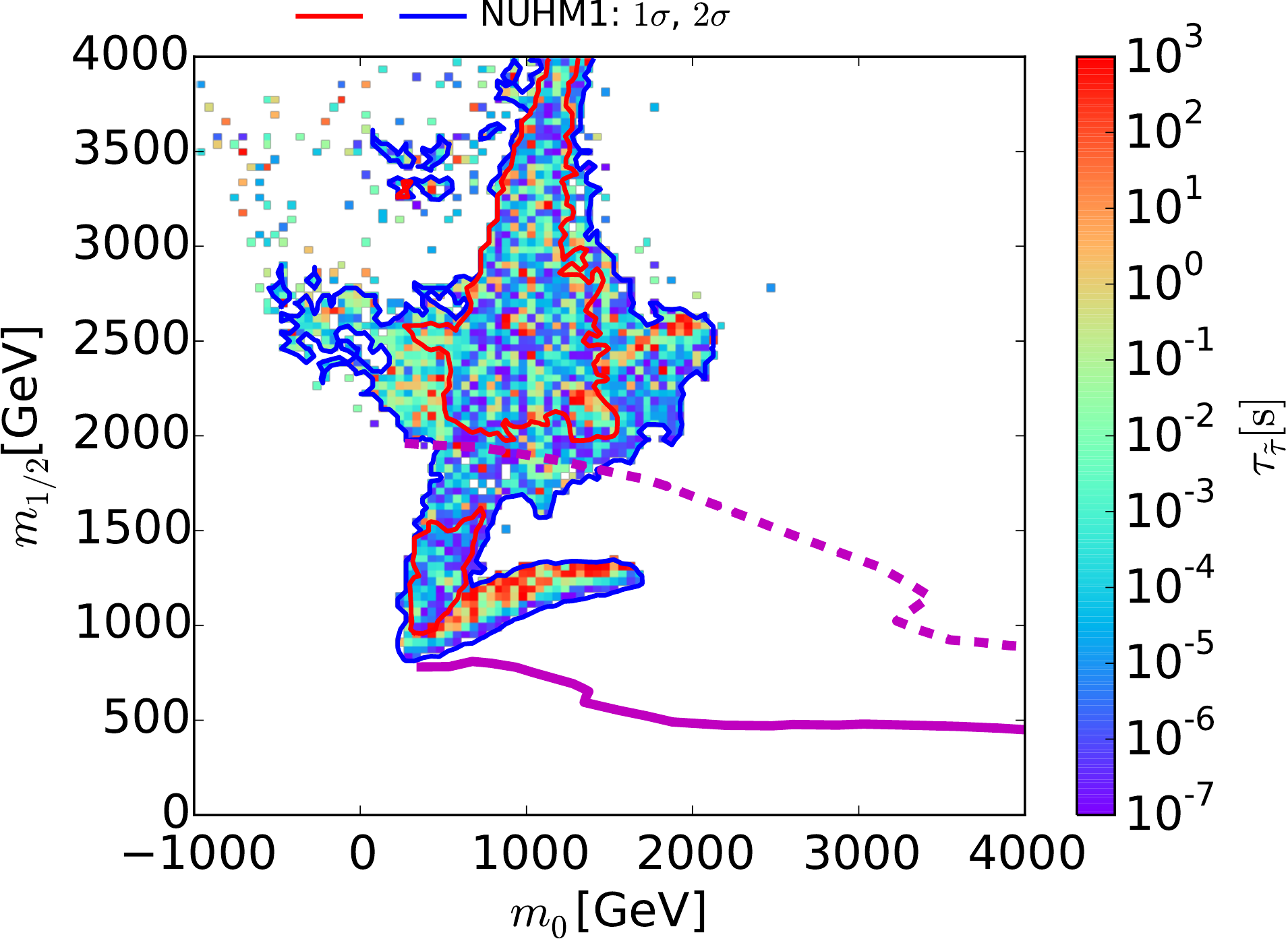}}
\end{center}
\vspace{-0.5cm}
\caption{\it
The $(m_0, m_{1/2})$ planes in the CMSSM (left panel) and the NUHM1 (right panel), showing (colour coded)
the lifetime of the lighter stau~\protect\cite{Citron:2012fg} for the best fit at each point in the plane~\protect\cite{MCDM}.
The red, blue, and purple contours have the same significances as in Fig.~\protect\ref{fig:SUSY1}.
}
\label{fig:SUSY2}
\end{figure*}

A corollary of the importance of coannihilation is that in many scenarios the next-to-lightest 
supersymmetric particle (NLSP) may have a mass only slightly greater than that of the LSP,
in which case it may have a long lifetime, opening up the possibility of signatures from
displaced vertices and/or massive metastable charged particles passing through the detector~\cite{MCDM}.
For example, in the CMSSM, NUHM1, and NUHM2 one can find that 
$m_{\tilde \tau_1} - m_{LSP} < m_\tau$, in which case the ${\tilde \tau_1}$ lifetime can be
very long~\cite{Citron:2012fg}, as seen in Fig.~\ref{fig:SUSY2}~\cite{MCDM}, which displays in colour code the lifetime of
the ${\tilde \tau_1}$ at the best fit point for each pair of $(m_0, m_{1/2})$ values in the CMSSM
(left panel) and the NUHM1 (right panel). We see that a long-lived stau may be a distinctive signature
in the regions of these models that can be explored in future runs of the LHC.
Long-lived NLSP signatures also appear in other models of SUSY breaking, e.g., minimal
anomaly-mediated SUSY breaking, in which the appropriate DM density is obtained by
coannihilation of the LSP with a nearly degenerate long-lived wino.

{\it It is, therefore, desirable to consider other possible signatures of DM models, such as the appearance of long-lived particles.}
(See also the discussion of pseudo-Dirac DM in the previous Section.)

\begin{figure*}[htb!]
\resizebox{8.0cm}{!}{\includegraphics{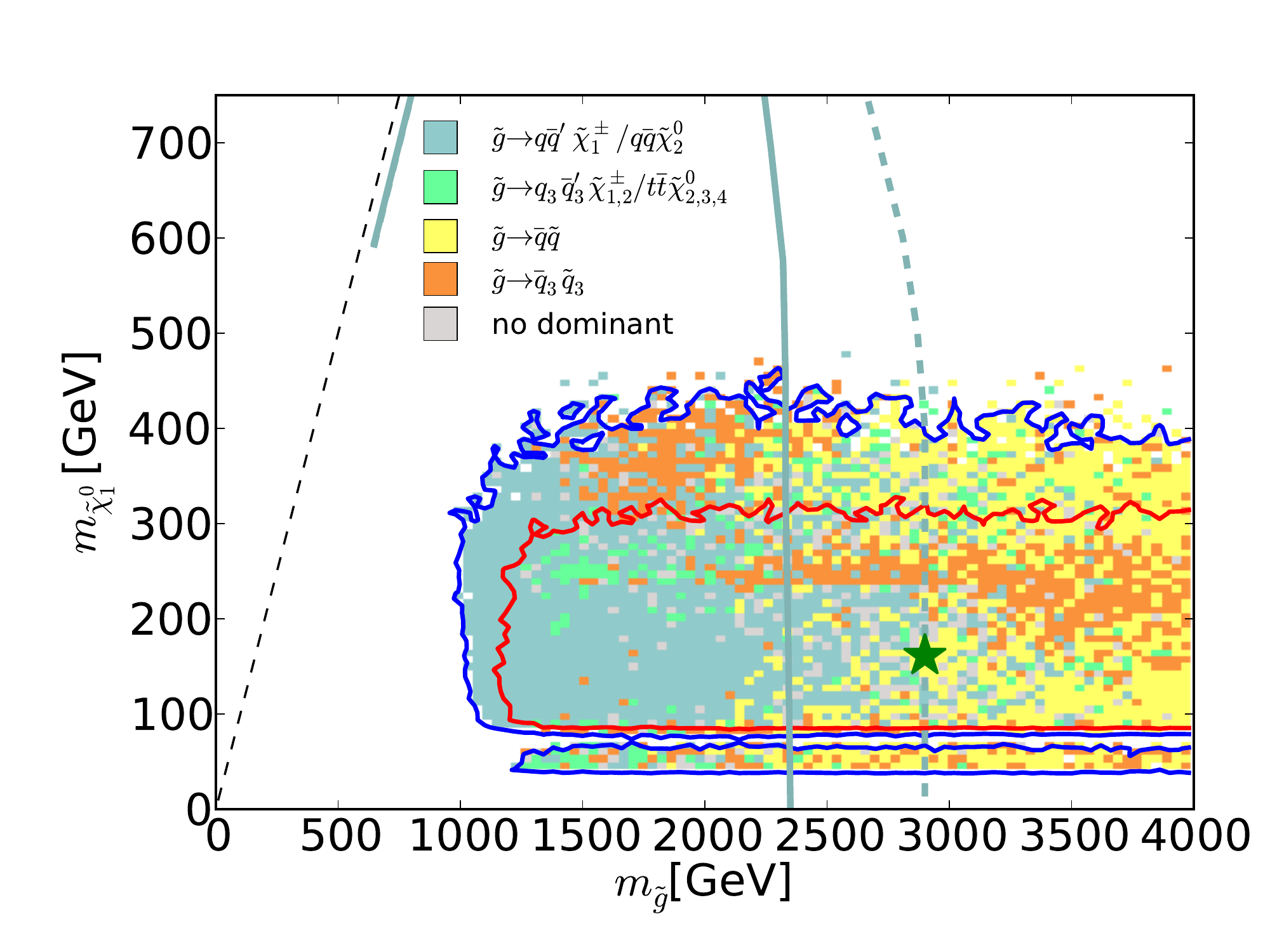}}
\resizebox{8.0cm}{!}{\includegraphics{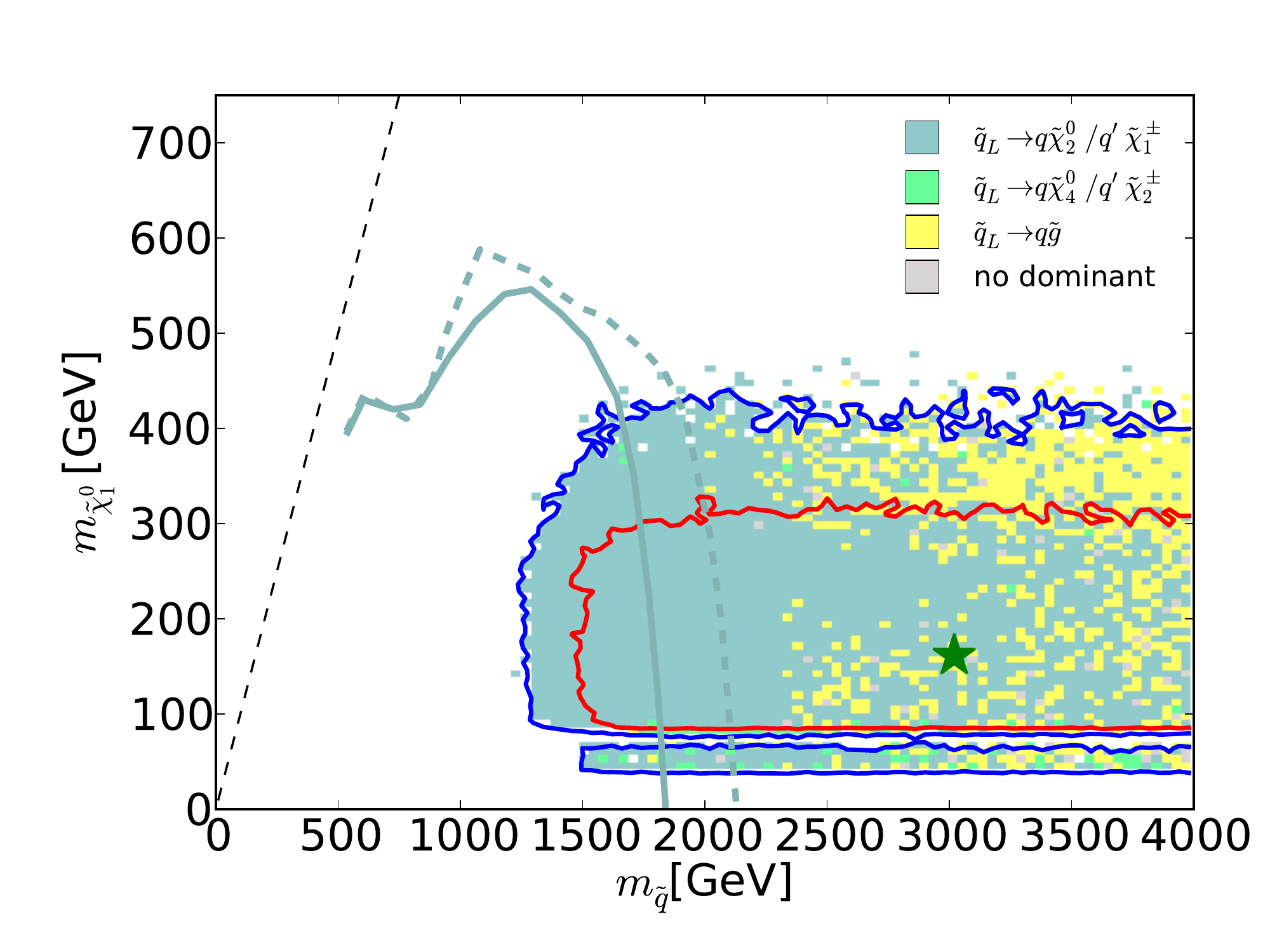}}  \\[1em]
\vspace{-0.5cm}
\caption{\it Illustration of the dominant ${\tilde g}$ decays (left panel)
and ${\tilde q}$ decays (right panel) in the pMSSM10~\protect\cite{pMSSM10}.
The pale blue solid (dashed) lines show the estimated LHC sensitivities
with 300~\infb (3000~\infb). 
}
\vspace{1em}
\label{fig:SUSY3}
\end{figure*}

Furthermore, simplified DM models typically do not take into account the complexity
of many mechanisms of producing DM particles. In SUSY, as well as other frameworks
such as extra dimensions, most DM particles are not produced directly at the LHC, but
appear at the final stage of cascade decays of heavier SUSY particles. Typically, strongly interacting
particles (e.g. squarks and gluinos in SUSY models) are produced and then decay via many
possible intermediate particles into the DM particle (e.g., the LSP in SUSY models).
Figure~\ref{fig:SUSY3} illustrates the important possible decays of gluinos and squarks in the
pMSSM10, colour-coded according to the dominant decay for the best fit parameter set at each
point in the displayed plane~\cite{pMSSM10}. A comprehensive study of SUSY models should take these decays
and their branching fractions into account; assuming that one particular decay mode is dominant
is likely to lead to an over-estimated exclusion of realistic models.

\begin{figure*}[htb!]
\resizebox{8.0cm}{!}{\includegraphics{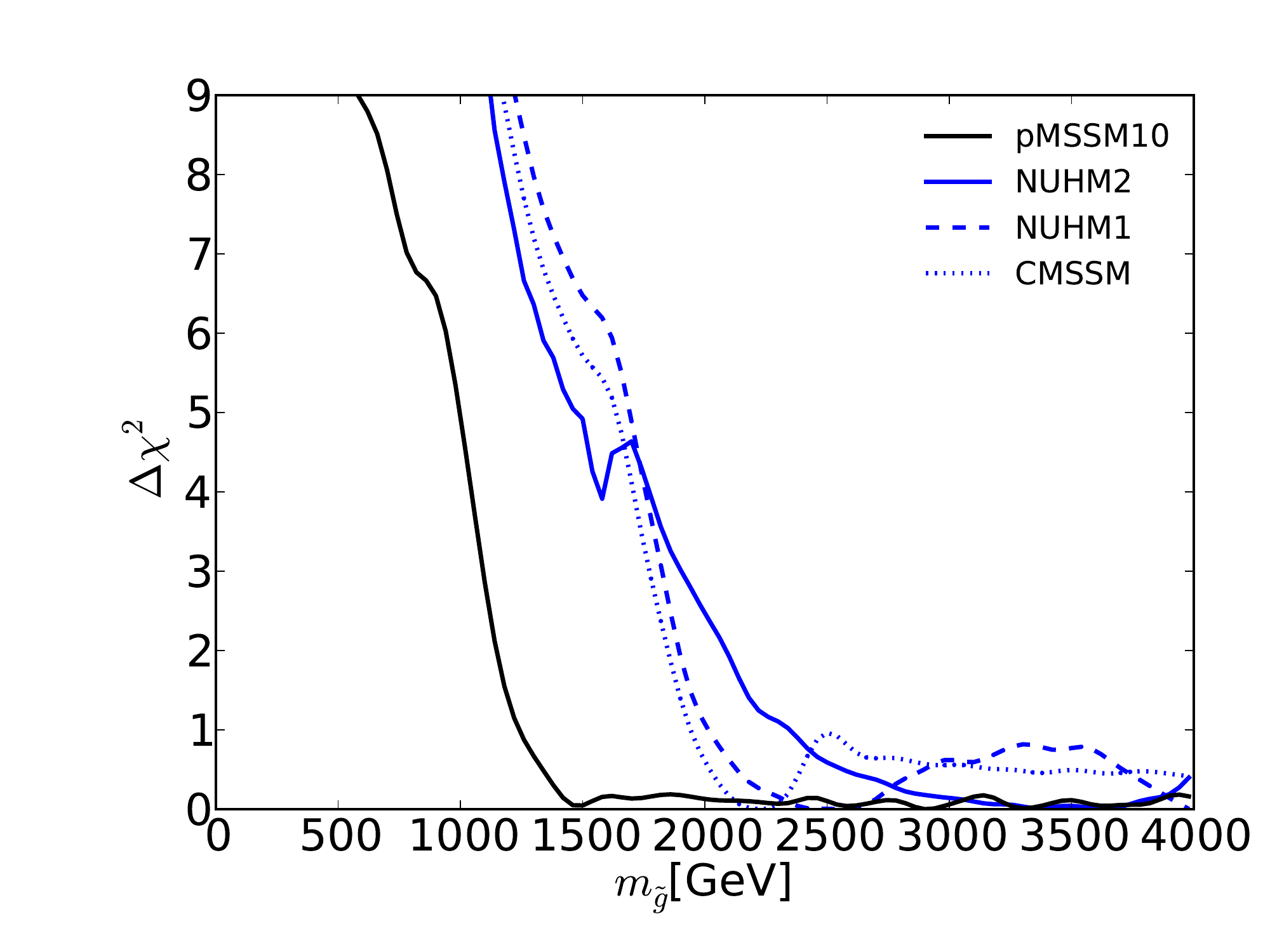}}
\resizebox{8.0cm}{!}{\includegraphics{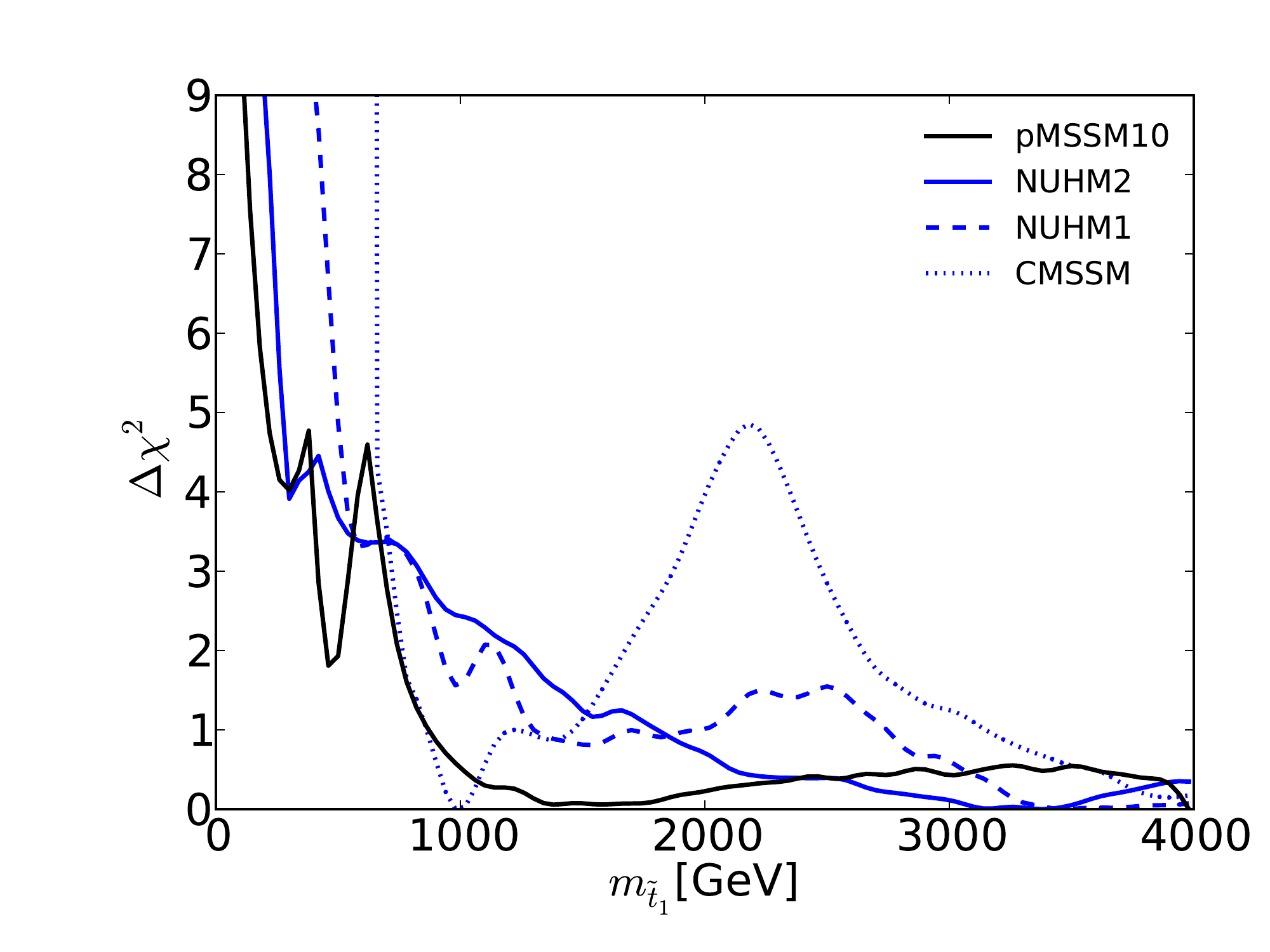}}
\caption{\it One-dimensional profile likelihood functions
for \mgl\ and \mstop{1}:
the solid black lines are for the pMSSM10, the solid blue lines for the NUHM2,
the dashed blue lines for the NUHM1, and the dotted blue lines for the
CMSSM~\protect\cite{pMSSM10}.}
\label{fig:SUSY4}
\end{figure*}

The importance of these considerations is illustrated in Fig.~\ref{fig:SUSY4}, that displays
the $\chi^2$ likelihood functions for the gluino mass (left panel) and the lighter stop mass
(right panel) in the pMSSM10 (solid black lines), the NUHM2 (solid blue lines), the NUHM1
(dashed blue lines), and the CMSSM (dashed blue lines)~\cite{pMSSM10}. In each model, careful attention has
been paid to the implementation of the LHC Run 1 constraints on a variety of different SUSY
production and decay channels and their respective branching fractions. Two important
points are worth noting. In the case of the gluino, the lower limit on its mass from LHC Run 1
is significantly weaker than for the other models, reflecting the importance of
taking into account the complexity of possible SUSY
cascade decay channels seen in Fig.~\ref{fig:SUSY3}. In the stop case, the pMSSM10
features a compressed stop region with $\Delta \chi^2 \lesssim 2$ that is not visible in the
NUHM2, NUHM1, and CMSSM cases. Understanding the interplay between several different 
production and decay mechanisms is essential to estimate correctly the LHC reach in this
region.

Therefore, we conclude that care must be taken in interpreting simplified models: {\it many competing decay modes are possible in
realistic models, which are not likely to feature simple decay chains.}

\subsection{Interplay of Collider and Direct Detection Searches in SUSY}

\begin{figure*}[htb!]
\begin{center}
\resizebox{7.5cm}{!}{\includegraphics{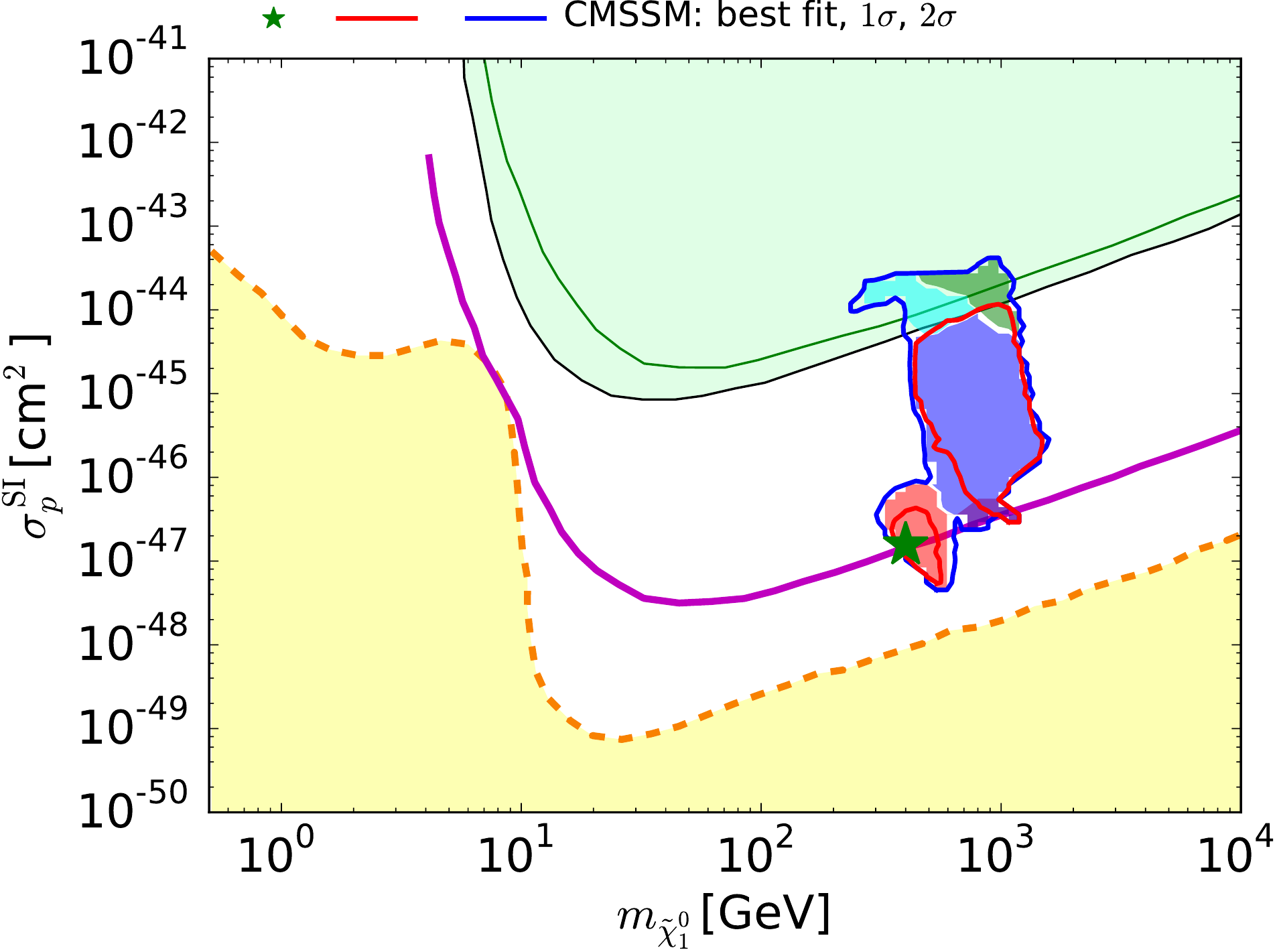}}
\resizebox{7.5cm}{!}{\includegraphics{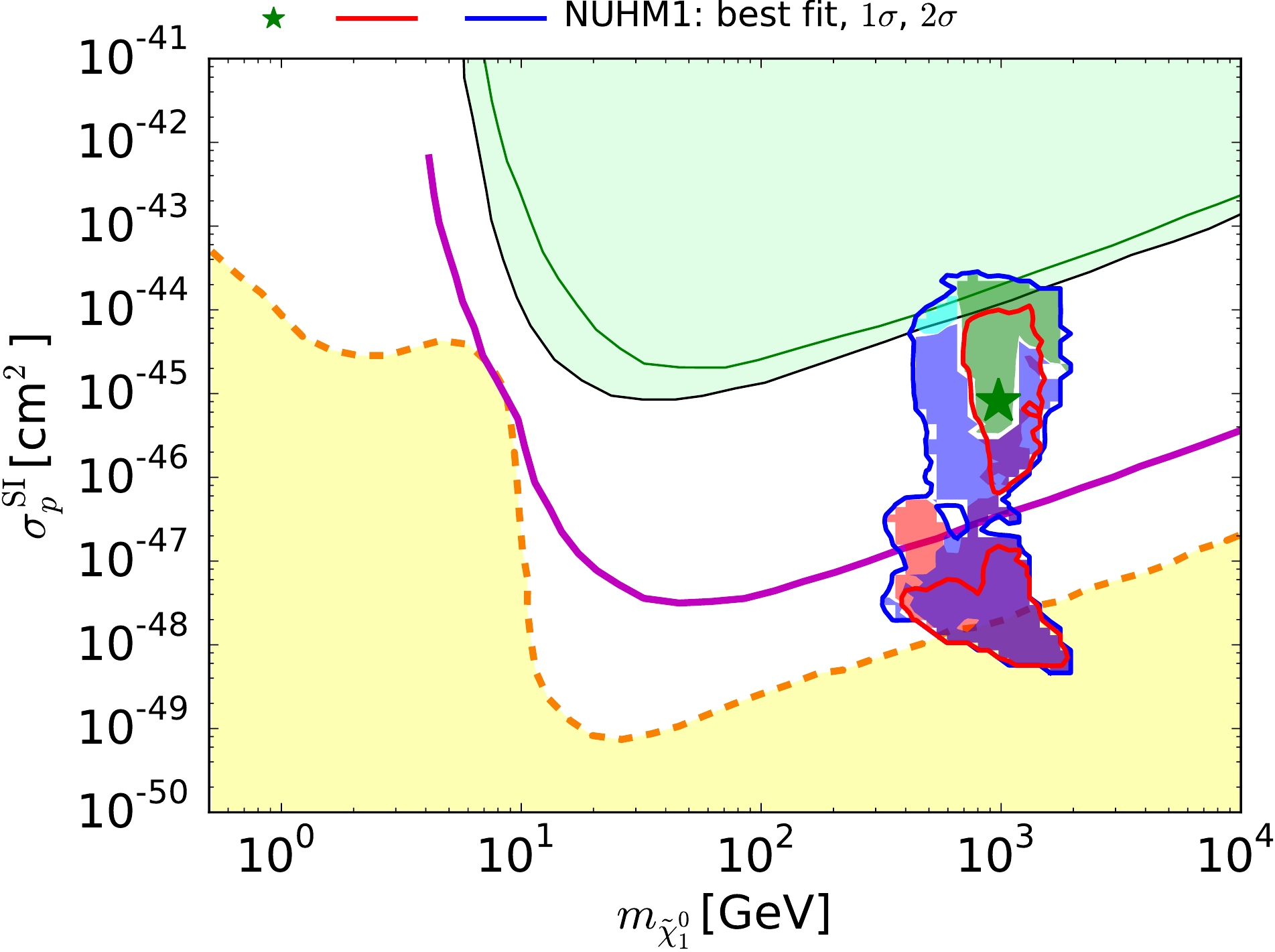}}\\[1em]
\resizebox{7.5cm}{!}{\includegraphics{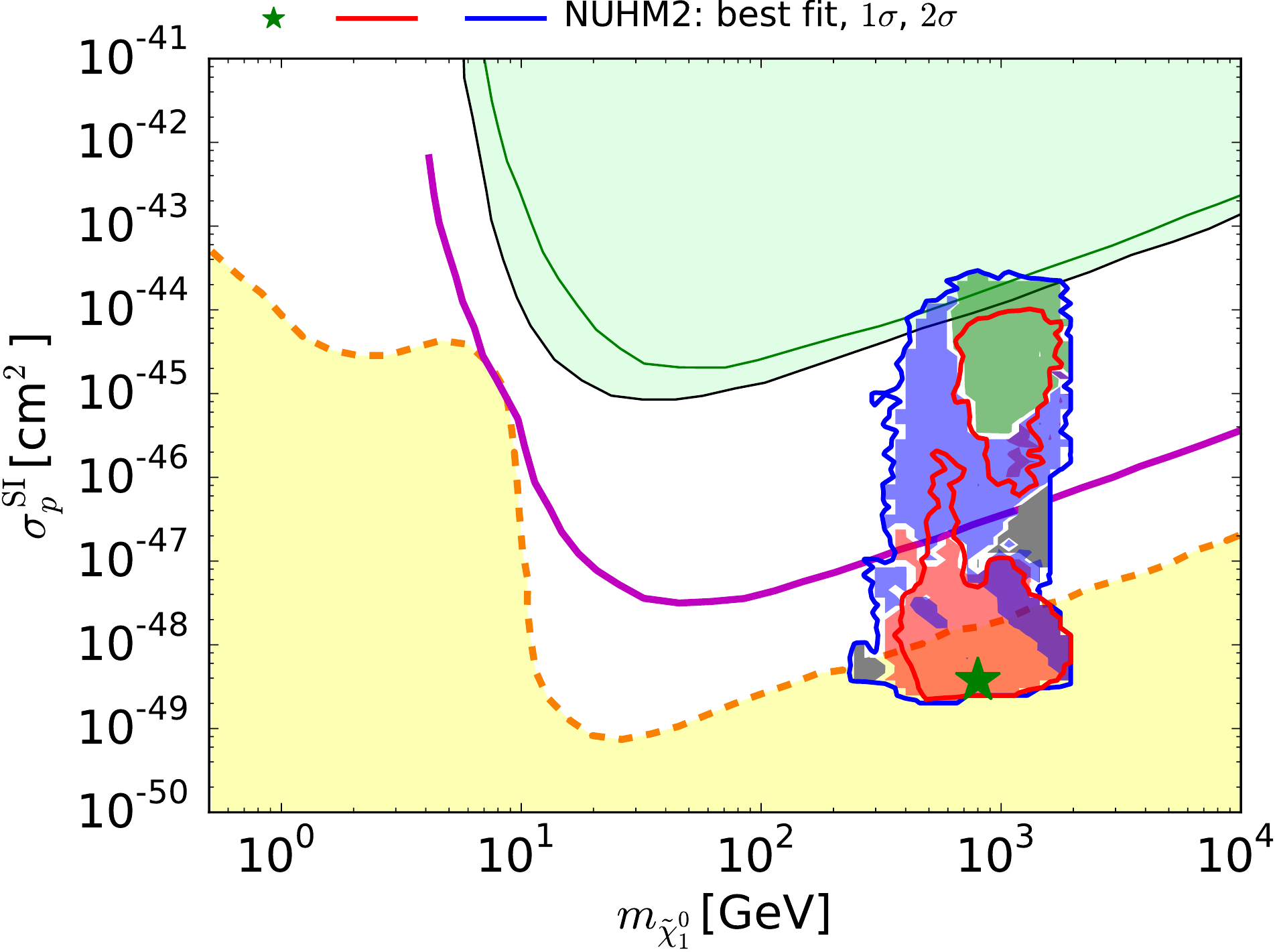}}
\resizebox{7.5cm}{!}{\includegraphics{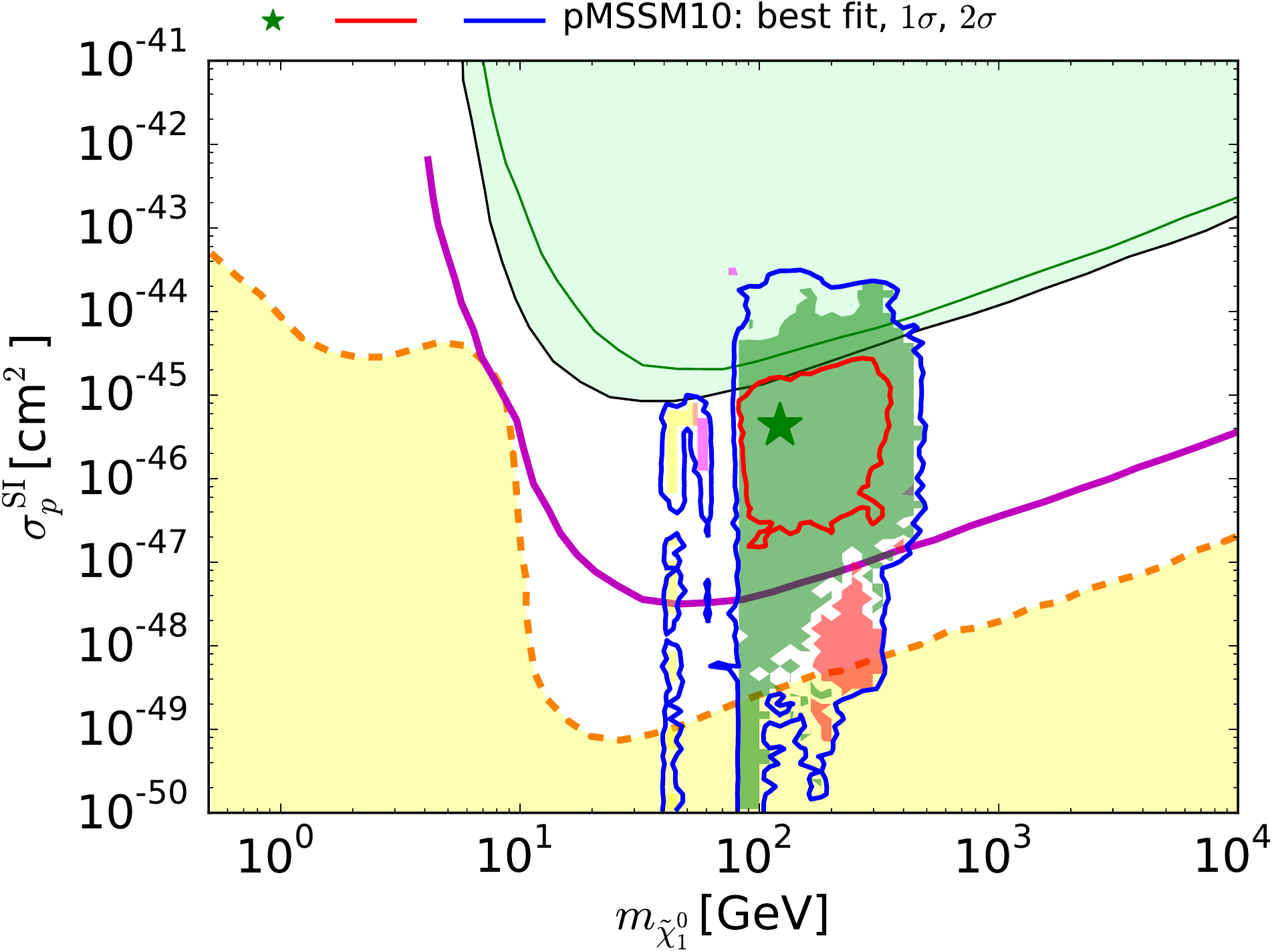}}\\
\resizebox{15cm}{!}{\includegraphics{n12c_dm_legend}}
\end{center}
\vspace{-0.5cm}
\caption{\it The $(\mneu1, \ssi)$ planes in the CMSSM (upper left),
the NUHM1 (upper right), the NUHM2 (lower left), and the pMSSM10 (lower right)~\protect\cite{MCDM}.
The red and blue solid lines are  the $\Delta \chi^2 = 2.30$ and 5.99 contours,
and the solid purple lines show the projected 95\% exclusion sensitivity of the LZ experiment~\protect\cite{LZ}. The green and black lines show the current sensitivities of the XENON100~\protect\cite{Xe100}
and LUX~\protect\cite{LUX} experiments,
respectively, and the dashed orange line shows the
astrophysical neutrino `floor'~\protect\cite{Billard:2013qya, Snowmass}, below which astrophysical neutrino backgrounds dominate (yellow region).
}
\label{fig:SUSY5}
\end{figure*}

Figure~\ref{fig:SUSY5} emphasizes that the interplay between LHC and direct DM
searches is quite different in different SUSY models~\cite{MCDM}. The detectability of a specific model
depends on the dominant mechanism for fixing the DM density via its spin-independent DM scattering
cross section \ssi, as can be seen in each of the panels.
For example, in the CMSSM, the stop coannihilation regions lie very close to the current 
LUX exclusion, whereas the $H/A$ annihilation region likely lies within the future reach
of the LZ experiment~\cite{LZ} and the stau coannihilation region may require a more sensitive experiment.
On the other hand, in the pMSSM10 the chargino coannihilation region apparently lies
mainly within reach of LZ, whereas portions of the chargino coannihilation region,
the stau coannihilation region, and the $h$ and $Z$ funnels may lie below the neutrino `floor'
where there is an irreducible neutrino background. Overall assessments of the LHC and
direct search sensitivities for these models, the NUHM1 and the NUHM2, are given in
Table~\ref{tab:SUSY1}.

\begin{table*}[htb!]
	\begin{center}
	\caption{\it
		\label{tab:SUSY1}
	Summary of SUSY detectability in the CMSSM, NUHM1, NUHM2, and pMSSM10
	models at the LHC in searches for $\ETslash$ events, long-lived charged particles (LL), and heavy $A/H$
	Higgs bosons, and in direct DM search experiments, depending on the dominant mechanism for
	bringing the DM density into the cosmological range~\protect\cite{MCDM}. The symbols $\checkmark$, ($\checkmark$) and
	$\times$ indicate good prospects, interesting possibilities and poorer prospects, respectively.
	The symbol -- indicates that a DM mechanism is not important for the corresponding model.
	}
	\vspace{5mm}
	{
	\begin{tabular}{ | c || c || c | c | c | c|}
		\hline
		DM & Exp't & \multicolumn{4}{c|}{Models} \\ 
		mechanism & & CMSSM & NUHM1 & NUHM2 & pMSSM10 \\ \hline
		${\staue}$ & LHC & {$\checkmark$ $\ETslash$, $\checkmark$ LL} & ($\checkmark$ $\ETslash$, $\checkmark$ LL) & ($\checkmark$ $\ETslash$, $\checkmark$ LL) & ($\checkmark$ $\ETslash$), $\times$ LL \\ 
		coann. & DM & ($\checkmark$) & ($\checkmark$) & $\times$ & $\times$ \\ \hline
		$\cha{1}$ & LHC & -- & $\times$ &  $\times$ & ($\checkmark$ $\ETslash$)  \\ 
		coann. & DM & -- & $\checkmark$ & $\checkmark$ & ($\checkmark$)  \\ \hline
		${\tilde t_1}$ & LHC & -- & -- & $\checkmark$ $\ETslash$ & --  \\ 
		coann. & DM & -- & -- & $\times$ & --  \\ \hline
		$A/H$ & LHC & $\checkmark$ $A/H$ & ($\checkmark$ $A/H$) & ($\checkmark$ $A/H$) & --  \\
		funnel & DM & $\checkmark$ & $\checkmark$ & ($\checkmark$) & -- \\ \hline
		Focus & LHC & ($\checkmark$ $\ETslash$) & -- & -- & --  \\
		point & DM & $\checkmark$ & -- & -- & -- \\ \hline
		$h,Z$ & LHC & -- & -- & -- & ($\checkmark$ $\ETslash$) \\
		funnels & DM & -- & -- & -- & ($\checkmark$) \\ \hline
			\end{tabular}}
			\end{center}
\end{table*}

Based on these findings, we see that {\it a detailed consideration of the relevant DM mechanisms is as important for direct searches as it is for LHC searches}, and needs to be taken into account in assessing the interplay between these search strategies. 


\subsection{Lessons from SUSY for simplified DM models}

In this Section we have discussed the lessons we can learn for the development for simplified models from a complete theory like SUSY. This is important to identify potential oversimplification of simplified models and how this can be overcome. For example, one should check that a simplified model can reasonably be extended to yield an acceptable DM density, remembering that there are many different mechanisms for bringing the DM density into the cosmological range. In addition to the conventional
annihilation and freezeout, one should consider extending the
simplified model approach to include other possibilities such as coannihilation with some other, 
almost degenerate particle (e.g., the stau, stop, wino in SUSY), as well as the possibility of
rapid annihilation via direct channel resonances. One should keep in mind possible 
non-\ETslash\ final-state signatures such as displaced vertices
and/or massive long-lived particles in coannihilation scenarios. One should also remember that DM
particles appear typically at the ends of cascade decays of heavier particles, and it may be misleading
to assume that any particular production or decay channel dominates. The sensitivities of both the LHC and direct
DM detection experiments are quite dependent on these features, and it is desirable for
simplified models to be extended to take at least some of these possibilities into account.


 \section{Summary and Recommendations}
 \label{summary}
 
In this White Paper we have summarised the discussions and corresponding follow-up studies of the brainstorming meeting ``Next generation of simplified Dark Matter models" held at the Imperial College, London on May 6, 2016 \cite{Agenda}. Based on this work we have defined a short list of recommendations, which we think will be important for defining both short-term and long-term strategies for the evolution of simplified Dark Matter models. This White Paper is an input to the 
ongoing discussion within the experimental and theoretical community about the extension and refinement of simplified Dark Matter models.

In Section 2 we studied in detail the extension of SDMMs with a scalar mediator, as currently used by ATLAS and CMS, to include mixing with the SM 
Higgs boson. We conclude that including mixing provides a more realistic description of the underlying kinematic properties that a complete physics model would possess. The addition of the mixing with the 
Higgs also provides the opportunity to interpret this class of models in the context of LHC Higgs 
measurements, as these results constrain the required mixing angle in these models. Furthermore, 
the scalar mixing model also provides the option to compare and combine consistently searches targeting different experimental signatures. For example, in this model a consistent interpretation of missing transverse energy searches, such as monojet, mono-$V$, and VBF-tagged analyses, which are sensitive to different production modes --- gluon fusion, associated, and VBF production, respectively --- is possible. 
Therefore, connecting the missing energy DM searches with 
other LHC measurements of properties of SM final states will result in a more complete and rigorous interpretation.
{We recommend that the class of SDMMs with scalar mediator and mixing with the 
Higgs boson should become part of the portfolio of simplified models studied by the LHC experiments.}

 Using the example of the recently observed excess in the high-mass diphoton searches in ATLAS and CMS for definiteness, we have discussed in Section~3 how a hypothetical signal for
 production of a new mediator can be connected to DM using simplified models. This exercise was intended as an example of a case 
 study of how to correlate
 searches with different experimental signatures in order to characterise the properties of a newly discovered particle in the context of DM.  This study highlights that within the framework of simplified models, possibly combined with effective couplings, it is rather straightforward to connect a new physics signal observed in a visible channel with DM searches or vice versa.
 Using the simplified DM model with scalar (pseudoscalar) mediator and extending it using effective couplings, we have shown that it is possible to not only connect the diphoton excess with different important visible signatures such as $\gamma Z$, $ZZ$, $WW$, and dijets, but also with generic DM signatures such as the monojet search. Therefore, this pragmatic ansatz enables one to link a potential signal in one channel with searches for other experimental signatures, which then can be used to verify/falsify potential signal models and to study their underlying nature. We believe that exploring these links 
is vital for guiding the experimental search programme in case of a discovery of a new particle. {\ We recommend that the development of discovery-oriented simplified models that manifest themselves in a variety of experimental signatures, such as the one used in the example to characterise the $750 \, {\rm GeV}$ diphoton excess, should be an important part of future activities of the LHC DM working group.}

 We highlighted in Section 4 the importance of $t$-channel and spin-2 mediator SDMMs, as well as models in which the properties of the DM candidate are different from the currently canonically assumed Dirac fermion, such as pseudo-Dirac DM. {\ We recommend that SDMMs with $t$-channel exchange and other properties like a  spin-2 mediator or different DM candidates should be studied with higher priority in the future}.

Last but not least, we have discussed in Section 5 important properties of SUSY DM and how these could aid the development of new simplified DM models that possess more realistic mechanisms for bringing the DM density into the cosmological range. In addition to the conventional annihilation and freezeout, SDMMs could be extended to include other possibilities such as coannihilation with an almost degenerate particle, as well as the possibility of rapid annihilation via direct-channel resonances. {We recommend that properties of complete models, such as SUSY and its DM sector, should become a stronger guide 
for the development of more realistic SDMMs in the future}.

 \subsection{Recommendations of the White Paper in a nutshell}

 \begin{itemize}
 \item  {\it We recommend that the class of SDMMs with scalar mediator and mixing with the 
 Higgs boson should become part of the official portfolio of simplified models studied by the LHC experiments. }
 \item {\it We recommend the development of discovery-oriented SDMMs that manifest themselves in a variety of experimental signatures, such as the one used to characterise the $750 \, {\rm GeV}$ diphoton excess, should be an important part of future activities of the LHC DM working group.} 
 \item {\it  We recommend that SDMMs with $t$-channel exchange and other properties such as a spin-2 mediator or different DM candidates should be studied with higher priority in the future}.
 \item  {\it We recommend that properties of complete models, such as SUSY and its DM sector, should become a stronger guide for the development of more realistic SDMMs}.
 \end{itemize}

\section{Acknowledgements}
The work of JE was partially supported by STFC Grant ST/L000326/1, that of GL by DOE Award DE-SC0010010, and that of KH by DOE Award DE-SC0015973. UH is grateful to the CERN Theoretical  Physics Department and the MITP in Mainz for their hospitality and their partial support during the completion of this work.

\bibliography{ref}
\bibliographystyle{JHEP}

\end{document}